\documentclass[%
preprint,
onecolumn,
titlepage,
tightenlines,
12pt
]
{revtex4-2}

\usepackage{dcolumn}
\usepackage{Paper}

\usepackage[export]{adjustbox}

\newcommand{\chord}{\ensuremath{c}}
\newcommand{\aoa}{\ensuremath{\alpha}}
\newcommand{\velocityscale}{\ensuremath{U_{\infty}}} 
\newcommand{\presscale}{\ensuremath{P_{\infty}}} 
\newcommand{\densityf}{\ensuremath{\rho_{f}}}
\newcommand{\reynolds}{\ensuremath{Re}} 
\newcommand{\visc}{\ensuremath{\nu}} 

\newcommand{\pres}{\ensuremath{p}} 
 
\newcommand{\presdim}{\ensuremath{P}} 
\newcommand{\body}{\ensuremath{\Psi}} 
\newcommand{\vel}{\ensuremath{\bm{u}}}
\newcommand{\velbs}{\ensuremath{U_\infty}}
\newcommand{\stress}{\ensuremath{\bm{f}}} 
\newcommand{\bodypoint}{\ensuremath{\bm{\chi}}} 
\newcommand{\surface}{\ensuremath{s}} 
\newcommand{\flowpoint}{\ensuremath{\bm{x}}} 
\newcommand{\torsbody}{\ensuremath{\Psi_f}} 
\newcommand{\torsbodypoint}{\ensuremath{\bm{\chi}_{f}}}
\newcommand{\hinge}{\ensuremath{\bm{\chi}^{0}_{f}}} 
\newcommand{\fluiddomain}{\ensuremath{\Omega}} 
\newcommand{\rigbodypoint}{\ensuremath{\bm{\chi}_{a}}}

\newcommand{\timet}{\ensuremath{t}}
\newcommand{\periodt}{\ensuremath{t/T}}
\newcommand{\phase}{\ensuremath{\phi}}
\newcommand{\angvel}{\ensuremath{\omega}}
\newcommand{\pangvel}{\ensuremath{\Omega}}

\newcommand{\defl}{\ensuremath{\beta}}
\newcommand{\stiff}{\ensuremath{k_{\defl}}}
\newcommand{\inert}{\ensuremath{i_{\defl}}}
\newcommand{\mass}{\ensuremath{m_{\defl}}}
\newcommand{\stiffdim}{\ensuremath{K_{\defl}}}
\newcommand{\inertdim}{\ensuremath{I_{\defl}}}
\newcommand{\deflmean}{\ensuremath{\overline{\defl}}}
\newcommand{\fifty}{\ensuremath{50\%}}
\newcommand{\sixty}{\ensuremath{60\%}}
\newcommand{\twenty}{\ensuremath{20\%}}
\newcommand{\addedinert}{\ensuremath{i_{a}}}
\newcommand{\addedinertdim}{\ensuremath{I_{a}}}

\newcommand{\slev}{\ensuremath{\Gamma_{SLEV}}}
\newcommand{\lev}{\ensuremath{\Gamma_{LEV}}}
\newcommand{\plev}{\ensuremath{\Gamma_{PLEV}}}
\newcommand{\tev}{\ensuremath{\Gamma_{TEV}}}

\newcommand{\lift}{\ensuremath{C_l}}

\newcommand{\liftimp}{\ensuremath{\Delta \overline{C}_l}}
\newcommand{\Cp}{\ensuremath{C_p}}

\newcommand{\gapratio}{\ensuremath{r_{g}}}
\newcommand{\gap}{\ensuremath{l_{g}}}
\newcommand{\gaplarge}{\ensuremath{l_{G}}}
\newcommand{\flapratio}{\ensuremath{r_{f}}}
\newcommand{\flapvortex}{\ensuremath{l_{v}}}
\newcommand{\flap}{\ensuremath{l_{f}}}
\newcommand{\flapdist}{\ensuremath{l_{\defl}}}
\newcommand{\flapdistdim}{\ensuremath{L_{\defl}}}

\begin{document}

\preprint{APS/123-QED}


\title{Fluid-structure interaction of a bio-inspired passively deployable flap for lift enhancement}

\author{Nirmal J. Nair}
 \email{njn2@illinois.edu}
\affiliation{%
 Department of Aerospace Engineering, University of Illinois at Urbana-Champaign, IL 61801, USA
}%

\author{Andres Goza}%
\affiliation{%
 Department of Aerospace Engineering, University of Illinois at Urbana-Champaign, IL 61801, USA
}%


\begin{abstract}

Birds have a remarkable ability to perform complex maneuvers at post-stall angles of attack such as landing, take-off, hovering and perching. The passive deployment of self-actuating covert feathers in response to unsteady flow separation while performing such maneuvers provides a passive, adaptive flow control paradigm for these aerodynamic capabilities. Most studies involving covert-feathers-inspired passive flow control have modeled the feathers as a rigidly attached or a freely moving flap on a wing. A flap mounted via a torsional spring enables a configuration more emblematic of the finite stiffness associated with the covert-feather dynamics (the free-flap case is the zero-stiffness limit of this more general torsional spring configuration). The performance benefits and flow physics associated with this more general case remain largely unexplored. In this work, we model covert feathers as a passively deployable, torsionally hinged flap on the suction surface of a stationary airfoil. We numerically investigate this airfoil-flap system at a low Reynolds number of $Re=1{,}000$ and angle of attack of $20^\circ$ by performing high-fidelity nonlinear simulations using a projection-based immersed boundary method. A parametric study performed by varying the stiffness of the spring, mass of the flap and location of the hinge yielded lift improvements as high as 27\% relative to the baseline flap-less case and revealed two dominant flow behavioral regimes. A detailed analysis revealed that the stiffness-dependent mean flap deflection and inertia-dependent amplitude and phase of flap oscillations altered the dominant flow characteristics in both the regimes. Of special interest for performance benefits were the flap parameters that enhanced the lift-conducive leading-edge vortex while weakening the trailing-edge vortex and associated detrimental effect of upstream propagation of reverse flow. These parameters also yielded a favorable temporal synchronization of flap oscillations with the vortex-shedding process in both regimes.

\end{abstract}

\maketitle

\section{Introduction}

Flow control is a process of manipulating the flow to achieve desirable performance of enhanced lift, reduced drag, delayed flow separation and stall mitigation \cite{gad2007flow}. In the past few decades, significant advances have been made in the area of passive flow control. Passive flow control techniques involve design modifications and passive actuations that do not require external power such as vortex generators \cite{lin2002review}, Gurney flaps \cite{wang2008gurney} and roughness elements \cite{saric2011passive}. Despite these advances, the control aim of maintaining aerodynamic performance in the presence of significantly separated flow and vortex shedding at post-stall angles of attack remains a challenge due to its complex, unsteady and nonlinear nature. Such flow conditions are routinely encountered in micro- and unmanned-aerial vehicles (MAVs and UAVs, respectively) that perform agile and complex maneuvers.
Birds, on the other hand, have a remarkable ability to fly under adverse flow conditions and sudden gusts owing to their natural flow control strategies \cite{choi2012biomimetic}. To incorporate the use of these natural strategies into UAVs and MAVs, various bio-inspired flow control devices have been developed \cite{gencc2020traditional}. Devices that involve geometrical modifications to the wings and surface morphology include owl-wings-inspired leading-edge serrations \cite{rao2017owl} and morphing wings that adaptively adjust the camber to incoming flow conditions \cite{hu2008flexible,gamble2020load}. 
Off-surface actuation and flow manipulation via alula \cite{kim2015function,walker2010deformable,ito2019function,linehan2020scaling} and covert feathers \cite{carruthers2007automatic,videler2006avian} provide further avenues for passively attaining aerodynamic benefits. Covert feathers are a set of self-actuating feathers located on the upper surface of the wings.
During unsteady flow separation at large angles of attack, these feathers can passively deploy and interact with the separated flow. These self-actuating effectors can be exploited in bio-inspired MAVs and UAVs for attaining high agility and maneuverability \cite{jiakun2020review,bechert2000fluid,duan2018design}. In this work, we focus on the fluid-structure interaction mechanisms of covert-feather-inspired passive flow control strategies.

Covert feathers are generally modeled as a rigidly attached or a freely moving flap on the upper surface of a wing. Several aerodynamic benefits of utilizing covert-feather-inspired flaps have been reported. At post-stall angles of attack, lift improvements ranging from 15\% to 50\% have been attained experimentally at Reynolds number of $Re=10^4$--$10^6$ \cite{kernstine2008initial,schluter2010lift}. On airfoils that exhibit soft-stall characteristics, the flaps were able to delay the stall angle of attack while for sharp-stall airfoils, a reduction in the drop in post-stall lift without delaying stall was reported \cite{duan2021covert}. In another study, a delay in stall was achieved without substantial increase in maximum lift \cite{johnston2012investigation}. Besides stall delay, flaps were able to mitigate lift breakdown at stall, resulting in gentler flight characteristics \cite{schluter2010lift}. The flaps were found to be beneficial during the occurrence of both trailing- and leading-edge stall behavior \cite{altman2016post}. At small angles of attack, flaps made using actual bird feathers attached on the pressure surface of the airfoil provided lift and lift-to drag ratio improvements as high as 186\% and 70\%, respectively \cite{wang2019experimental}. A reduction in lift fluctuations occurring due to severe vortex shedding at a low $Re=1{,}000$ was achieved when flexible flaps were affixed at specific deployment angles \cite{fang2019passive}. 

Several physical mechanisms have been identified to contribute to the aerodynamic benefits provided by the passively deployable flap. A ``pressure dam'' effect, where the flap acted as a dam in maintaining a lower suction pressure upstream of the flap, was found to be a dominant phenomenon yielding significant lift enhancements \cite{bramesfeld2002experimental}. This effect manifested as a pressure discontinuity on the suction surface of both soft- and sharp-stall airfoils \cite{duan2021covert}. In one of the studies, the optimum flap deflection was found to be the one where the flap nearly touched the boundary of the separated region or the shear layer \cite{meyer2007separation}. Such a flap configuration formed a barrier to progression of reverse flow from the trailing to the leading edge \cite{izquierdo2021experimental}, causing the separation point to move aft on the airfoil \cite{johnston2012investigation}. Since the flow was attached over larger portions of the wing in the presence of the flap, the wing could sustain higher angles of attack without stalling \cite{schluter2010lift}. The location of the flap was also found to contribute differently to the aerodynamic performance. In some studies, flaps at upstream locations near the mid-chord or the leading edge were found to be more beneficial \cite{kernstine2008initial}, due in part to the flap delaying the shedding of the leading-edge vortex during a dynamic ramp-up motion \cite{izquierdo2021experimental}. In contrast, the pressure dam effect had a stronger effect on lift when the flap was located more downstream near the trailing edge \cite{bramesfeld2002experimental}. Additionally, the flap near the trailing edge divided the separated recirculation zone into two regions which was linked to delay flow separation \cite{fang2019passive}. 

While it is clear that covert-inspired flaps modeled using freely moving and static flaps are beneficial for aerodynamic performance, the fully coupled dynamics---and possible additional benefits---associated with a flap mounted via a torsional spring remain largely unexplored. This more general model is more emblematic of the finite stiffness of bird feathers, and includes the free flap as a sub-case in the zero stiffness limit. In \citet{rosti2018numerical}, a numerical parametric study was performed by varying the length and moment of inertia of the flap and stiffness of the spring. Maximal lift was attained when the natural frequency of the flap matched the frequency of vortex shedding. These benefits were attributed to a delay in the release of the leading-edge vortex (LEV) during a dynamic ramp-up motion \cite{rosti2018passive} and reduction of the trailing-edge vortex (TEV) circulation due to a blowing-jet-like effect when the flap oscillated downwards \cite{rosti2018numerical}. These studies largely focused on a flap location of $70\%$ chord from the leading edge, and flap stiffness/inertia properties designed to assess the effects of aligning the vacuum-scaled resonant frequency with that of the underlying vortex-shedding behavior. \citet{nair2022effects} subsequently considered a wider range of flap locations and structural parameters, and identified two flow behavioral regimes conducive to aerodynamic performance. Although the pressure dam effect was the primary contributor to lift in both regimes, the mechanisms by which the associated low pressure regions were attained were distinct in these regimes. However, a detailed analysis of how the stiffness, inertia and flap location contributed to the interaction of the flap with the separated flow and vortex shedding was not performed. \citet{nair2022numerical} also found that a single flap was more beneficial to lift than multiple torsional flaps at low $Re=1{,}000$, for the multi-flap parameters considered in that study. A more extensive study of the flow physics and the fluid-structure interaction mechanisms of a single passively deployable, torsionally attached flap on an airfoil is required and addressed here.

In this work, we perform two-dimensional (2D) high-fidelity, strongly-coupled fluid-structure interaction simulations of flow past a stationary airfoil with a passively deployable, torsionally mounted flap on the suction surface. The Reynolds number of the flow based on the chord length is set to $1{,}000$ and the post-stall  angle of attack is fixed at $20^\circ$. A wide range of moment of inertia of the flap and stiffness of the spring, varying several orders of magnitude, is considered here. We also consider mounting the flap at distinct locations along the airfoil. A detailed analysis of the distinct flow regimes the airfoil-flap system exhibits is performed by critically investigating the flow processes modulated by the hinge stiffness and flap inertia. We quantify the effects of flap parameters on the pressure dam effect, barrier to the upstream propagation of reverse flow, strengths of the leading- and trailing-edge vortices, and temporal synchronization of flap oscillations with the vortex-shedding process. 


\section{Problem setup and numerical methodology}
\label{problem}

\subsection{Problem setup}
\label{problemsetup}

\begin{figure}
\centering
\includegraphics[scale=1]{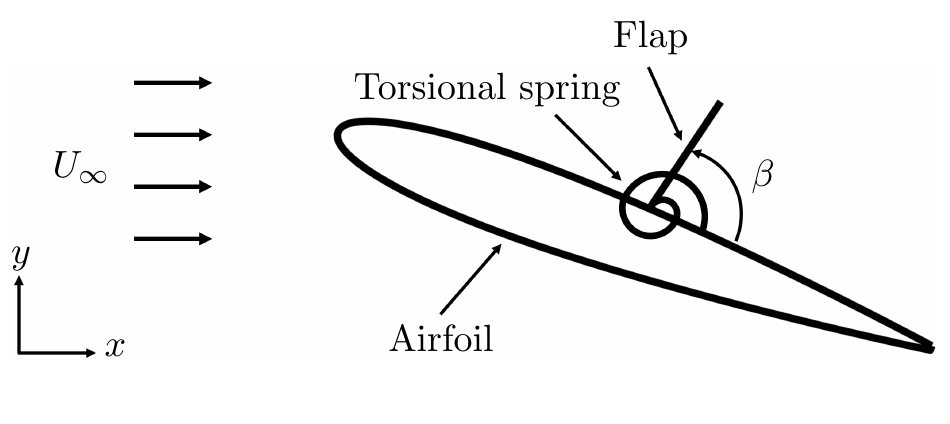}
\caption{Schematic of the system of a passively deployable flap on an airfoil}
\label{schematicairfoil}
\end{figure}

The schematic of the problem setup is shown in Fig.~\ref{schematicairfoil}. It consists of a stationary NACA0012 airfoil of chord length $c$ at a post-stall angle of attack of $\aoa = 20^\circ$ in a flow with freestream velocity $\velocityscale$. The Reynolds number based on the chord length, $\reynolds=\velocityscale\chord/\visc$, is $\reynolds = 1 {,}000$, where $\visc$ is the kinematic viscosity of the fluid. A flap of length $0.2c$ is hinged on the upper surface of the airfoil via a torsional spring, where the instantaneous deployment angle of the flap is given by $\defl$. Initially, the flap is rested at an angle of $5^\circ$ from the airfoil surface, which is taken as the undeformed (zero stress) deflection angle. As the vortex-shedding process occurs at the high angle of attack of $20^\circ$, the flap passively deploys and interacts with the flow.

The varying parameters in this study are  the  dimensionless moment of inertia of the flap, $\inert$, stiffness of the  torsional spring, $\stiff$ and chordwise distance of the flap from the leading edge, $\flapdist$, defined as,
\begin{equation}
\inert= \frac{\inertdim}{\densityf \chord^4}, \quad \quad 
\stiff=\frac{\stiffdim}{\densityf \velocityscale^2 \chord^2}, \quad \quad
\flapdist=\frac{\flapdistdim}{\chord}.
\label{eqn:param_def}
\end{equation}
Here, $\inertdim$, $\stiffdim$ and $\flapdistdim$ are the dimensional analogs of $\inert$, $\stiff$ and $\flapdist$; and $\densityf$ is the density of the fluid. The parameter $\flapdist$ is reported as a percentage of the chord length from the leading edge in this manuscript. Inertia is varied as $\inert \in \{10^{-5}, 10^{-4}, 10^{-3}, 10^{-2}\}$ while stiffness is varied in the interval $\stiff \in [10^{-4}, 10^{-1}]$. Three flap locations of $\twenty$, $\fifty$ and $\sixty$ of the chord length from the leading edge are considered. 

While the parameters in (\ref{eqn:param_def}) are sufficient to fully characterize the system dynamics, to facilitate a more physically intuitive interpretation we recast the torsional inertia $\inert$ as mass ratio $\mass = \rho_s h/\densityf l_f$, where $h$ and $l_f$ are the thickness and length of the flap, respectively. This mass ratio can be understood as a ratio between the dimensional moment of inertia, $\inertdim=\rho_s h l_f^3/3$, and the theoretical estimate of the added torsional inertia of the flap, $\addedinertdim=9\pi\densityf l_f^4/128$ \cite{gracey1941additional} (ignoring the $\mathcal{O}(1)$ constant factor). A mass ratio of one, therefore, implies that the flap is approximately as resistant to an applied torque as the surrounding volume of fluid displaced by the oscillating flap. The results in the subsequent sections will confirm this interpretation. Now, on converting $\inert$ to $\mass$, the corresponding mass ratio is varied as $\mass \in 1.875\times \{10^{-2}, 10^{-1}, 10^{0}, 10^{1}\}$. That is, in this work we consider mass ratios where the flap is much lighter than, commensurate with, and much heavier than the flow inertia.

Finally, the aerodynamic performance of the airfoil-flap system will be often analyzed in terms of coefficients of lift and pressure defined as,
\begin{equation}
\lift = \frac{F_y}{\frac{1}{2}\densityf \velocityscale^2 \chord}, \quad \quad
\Cp = \frac{\presdim-\presscale}{\frac{1}{2}\densityf \velocityscale^2},
\end{equation}
where $F_y$ is the dimensional integrated force along the baseline airfoil (not including the flap) in the $y$ direction, $\presdim$ is the dimensional pressure variable and $\presscale$ is the freestream pressure.

\subsection{Numerical methodology}

The strongly-coupled projection-based immersed boundary method of \citet{goza2017strongly}, adapted for problems involving torsionally mounted flaps by \citet{nair2022strongly}, has been utilized to perform the simulations in this work. In this method, the following dimensionless governing equations are solved,
\begin{equation}
\frac{\partial \vel}{\partial \timet} + \vel \cdot \nabla \vel = -\nabla \pres + \frac{1}{\reynolds}  \nabla^2 \vel + \int_{\body} \bm{\stress}(\bodypoint(\surface,\timet)) \delta(\bodypoint(\surface,\timet)-\flowpoint)d\surface
\label{ns}
\end{equation}
\begin{equation}
\nabla \cdot \vel = 0 
\label{continuity}
\end{equation}
\begin{equation}
\inert \frac{\partial^2 \defl}{\partial \timet^2}  + \stiff \defl = - \int_{\torsbody} (\torsbodypoint-\hinge) \times \stress(\torsbodypoint) d\torsbodypoint 
\label{newtons}
\end{equation}
\begin{equation}
\int_{\fluiddomain} \vel(\flowpoint) \delta(\flowpoint-\rigbodypoint)d\flowpoint = 0
\label{bcr}
\end{equation}
\begin{equation}
\int_{\fluiddomain} \vel(\flowpoint) \delta(\flowpoint-\torsbodypoint)d\flowpoint = \frac{\partial \defl}{\partial t} \hat{\bm{e}}^i \times (\torsbodypoint-\hinge)
\label{bct}
\end{equation}
In immersed boundary method, two separate grids---one fixed and other moving---are used to represent the flow domain and body surface, respectively. Accordingly, $\flowpoint$ denotes the fixed Eulerian coordinate representing a position in the fluid domain $\fluiddomain$ while $\bodypoint(\surface, \timet)$ denotes the moving Lagrangian coordinate attached to the bodies (airfoil and flap) in the set $\body$. The surfaces of all the bodies are parametrized by $\surface$. For non-dimensionalizing the Navier-Stokes \eqref{ns} and continuity \eqref{continuity} equations, the characteristic length scale used is the airfoil chord $\chord$ and velocity scale is the freestream velocity, $\velocityscale$. Accordingly, $\timet$ is non-dimensionalized by $\chord/\velocityscale$ while the surface stress imposed on the fluid by the body $\stress$ and pressure $\pres$ were nondimensionalized by $\densityf \velocityscale^2$, where $\densityf$ is the fluid density. The rotational Newton's equation of motion of the torsional flap $\torsbody$ is given by Eq.~\eqref{newtons} where $\torsbodypoint$ is the Lagrangian coordinate of $\torsbody$. This equation consists of the inertial and stiffness terms on the left and aerodynamic torque on the flap about its hinge $\hinge$ due to the surface stress imposed by the fluid (therefore, the negative sign in front) on the right hand side. Eq. \eqref{bcr} and \eqref{bct} are the no-slip boundary conditions enforced on the airfoil and flap, respectively, where $\hat{\bm{e}}^i$ is a unit vector denoting the direction of the angular velocity of the flap and $\rigbodypoint$ is the Lagrangian coordinate of the airfoil. These no-slip constraints provide closure to the governing equations by allowing us to solve for the surface stress term $\stress(\bodypoint)$ that enforces the boundary condition on the respective bodies.

The flow equations \eqref{ns}--\eqref{continuity} are reformulated in a streamfunction-vorticity formulation and are spatially discretized using the standard second-order finite difference method. For time-discretization of Eq. \eqref{ns}, an Adams-Bashforth scheme is used for the nonlinear term while a Crank-Nicolson method is employed for the diffusive term. Eq.~\eqref{newtons} is discretized using an implicit Newmark scheme. To ensure strong fluid-structure coupling and stability of the method for flaps with a wide range of inertia and stiffness, the boundary condition constraints \eqref{bcr}--\eqref{bct} and the surface stress term in Eq. \eqref{ns} are implicitly treated at the current time step. 
For enforcing far-field Dirichlet boundary conditions of zero vorticity, the multi-domain approach of \citet{colonius2008fast} is utilized. After the full discretization of the equations followed by a block-LU decomposition, the resulting system of equations are iterated using Newton's method with a convergence criteria on the flap deflection angle.

The spatial grid and time step sizes based on a previously performed grid-convergence study \cite{nair2022strongly} are set to be $\Delta x/\chord=0.00349$ and $\Delta t/(\chord/\velocityscale) = 0.0004375$, respectively. Following \citet{goza2017strongly}, the immersed boundary spacing is set to be twice as that of the flow grid spacing of the finest grid. The convergence criteria on the flap deflection angle is $\| \Delta \defl \|_{\infty} \le 10^{-7}$. For the multi-domain approach for far-field boundary conditions, five grids of increasing coarseness are used where the finest and coarsest grid levels are $[-0.5,2.5]\chord \times [-1.5, 1.5]\chord$ and $[-23,25]\chord \times [-24, 24]\chord$, respectively. 

\section{Results}
\label{results}

\subsection{Qualitative flow features}
\label{qualitative}

\begin{figure}
\begin{adjustwidth}{}{0.5cm} 
\centering
\centering
\begin{subfigure}[t]{0.283\textwidth}
\centering
\includegraphics[scale=1]{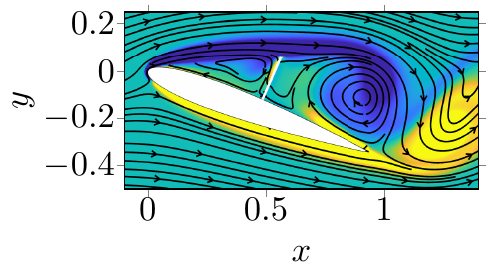}
\vspace{-0.7cm}
\caption{$t=0 \ T$ }
\label{generalvorticity1}
\end{subfigure}
\begin{subfigure}[t]{0.22\textwidth}
\centering
\includegraphics[scale=1]{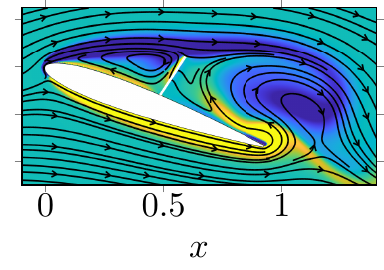}
\vspace{-0.7cm}
\caption{$t=0.27 \ T$ }
\label{generalvorticity2}
\end{subfigure}
\begin{subfigure}[t]{0.22\textwidth}
\centering
\includegraphics[scale=1]{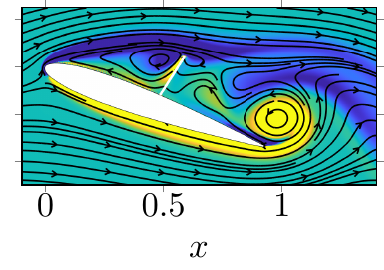}
\vspace{-0.7cm}
\caption{$t=0.55 \ T$ }
\label{generalvorticity3}
\end{subfigure}
\begin{subfigure}[t]{0.22\textwidth}
\centering
\includegraphics[scale=1]{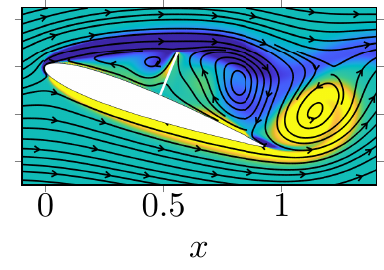}
\vspace{-0.7cm}
\caption{$t=0.82 \ T$ }
\label{generalvorticity4}
\end{subfigure}
\centering
\begin{subfigure}[t]{0.283\textwidth}
\centering
\includegraphics[scale=1]{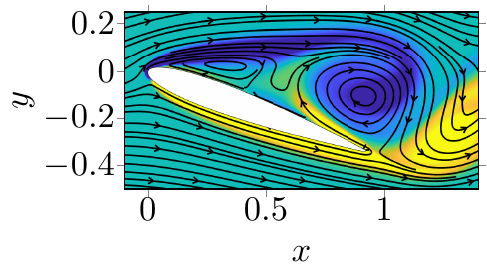}
\vspace{-0.7cm}
\caption{$t=0 \ T$ }
\label{generalvorticity5}
\end{subfigure}
\begin{subfigure}[t]{0.22\textwidth}
\centering
\includegraphics[scale=1]{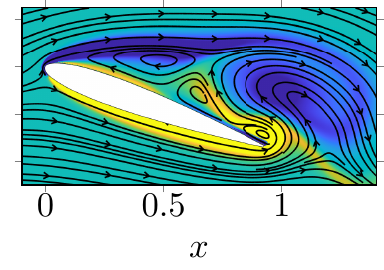}
\vspace{-0.7cm}
\caption{$t=0.27 \ T$ }
\label{generalvorticity6}
\end{subfigure}
\begin{subfigure}[t]{0.22\textwidth}
\centering
\includegraphics[scale=1]{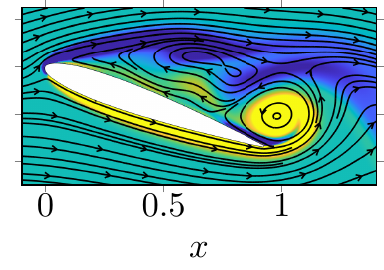}
\vspace{-0.7cm}
\caption{$t=0.55 \ T$ }
\label{generalvorticity7}
\end{subfigure}
\begin{subfigure}[t]{0.22\textwidth}
\centering
\includegraphics[scale=1]{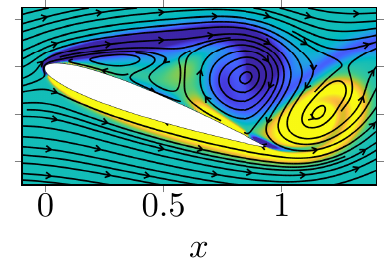}
\vspace{-0.7cm}
\caption{$t=0.82 \ T$ }
\label{generalvorticity8}
\end{subfigure}
\end{adjustwidth}
\caption{Vorticity contours depicting vortex shedding from the leading and trailing edges of the airfoil for a representative case of $\stiff=0.001$, $\mass=1.875$ and $\fifty$ location (top row) and the flap-less baseline airfoil case (bottom row) at four time instants in a lift cycle.}
\label{generalvorticity}
\end{figure}
\begin{figure}
    \centering
    \includegraphics[scale=1]{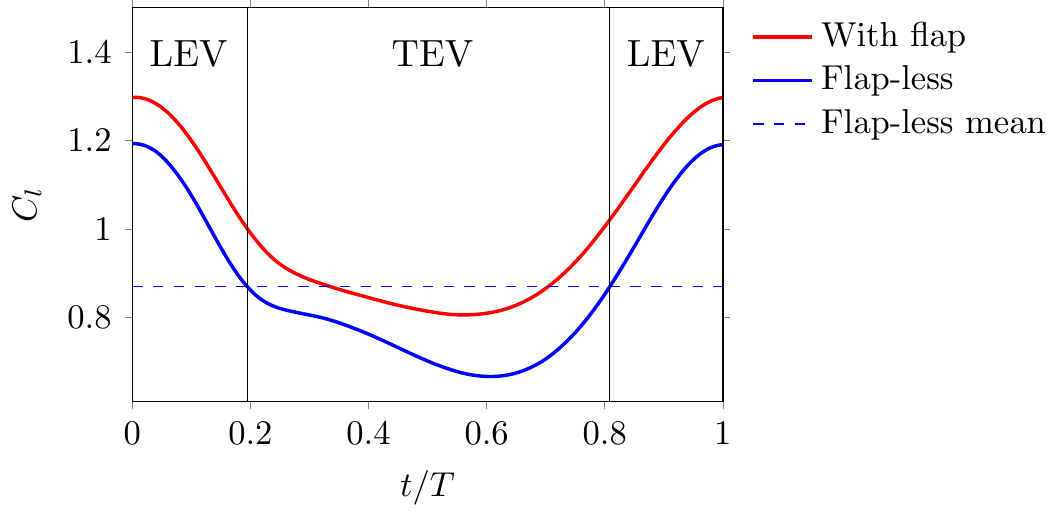}
    \caption{Coefficient of lift in one time period of the lift cycle for the flap ($\fifty$ location, $\stiff=0.001$, $\mass=1.875$) and flap-less cases. The time period is divided into two portions dominated by the LEV and TEV where the time instants of demarcation are $\periodt\approx 0.2$ and $\periodt\approx 0.8$.}
    \label{generalcl}
\end{figure}

The flap initially begins from its undeformed configuration on the airfoil surface, $\defl=0^\circ$, and then passively deploys into the flow as flow separation on the suction surface of the airfoil and vortex-shedding process occurs.
After a transient period, for all parameters considered, the airfoil-flap system enters limit-cycle oscillation (LCO) where the flap deflection and airfoil lift undergo oscillations with largely constant amplitude about a stationary mean value.
In this work, we focus on the flow physics in the LCO regime, occurring in our simulations for $\timet>20$, associated with periodic vortex shedding from the leading and trailing edges of the airfoil. The alternating shedding of the leading- and trailing-edge vortices, LEV and TEV respectively, can be visualized via vorticity contours for a representative case of the flap at $\fifty$ location, $\stiff=0.001$ and $\mass=1.875$ in Fig.~\ref{generalvorticity1}--\ref{generalvorticity4}. These vorticity snapshots are plotted at four time instants in one lift cycle in the LCO regime, where the periodic formation, shedding and interaction of the LEV and TEV can be clearly observed. In this manuscript, one lift cycle is defined between two consecutive peaks of $\lift$. For demonstration, one period of the lift cycle for the case corresponding to the vorticity contours is plotted in Fig.~\ref{generalcl}. From Fig.~\ref{generalvorticity1} it can be observed that at $\periodt=0$, the downstream advecting LEV shed from the previous lift cycle is above the trailing edge. The low pressure and clockwise circulation of the fully rolled-up LEV above the airfoil suction surface provides maximum lift, as seen from $\lift$ at $\periodt=0$ in Fig.~\ref{generalcl}. The downstream advecting LEV then induces the formation of the TEV at $\periodt \approx 0.27$ as shown in Fig.~\ref{generalvorticity2}, which then continues to roll up at $\periodt \approx 0.55$ as shown in Fig.~\ref{generalvorticity3}. The growing strength of the TEV reduces the lift producing clockwise circulation around the airfoil, decreases the pressure on the lower surface, and thereby reduces the lift, which is also seen from Fig.~\ref{generalcl}, where a trough in $\lift$ is attained around $\periodt \approx 0.6$. While the TEV grows in strength, a new LEV also simultaneously begins to form and advect downstream. As the TEV sheds away from the trailing edge at $\periodt=0.81$ in Fig.~\ref{generalvorticity4}, this newly formed LEV continues to gain in circulation strength as it advects downstream, thereby increasing the lift as noted from Fig.~\ref{generalcl}. Eventually, the rolled-up LEV reaches the trailing edge, maximum lift is attained at $\periodt=1$ and the cycle is repeated. During this entire period, the flap continues to remain deployed and oscillate about a mean deflection angle. 

Similar vortex-shedding phenomena and associated lift dynamics are also observed in the case of an airfoil without a flap, as depicted in Fig.~\ref{generalvorticity5}--\ref{generalvorticity8} and Fig.~\ref{generalcl}, respectively. Therefore, we emphasize that vortex shedding is ubiquitous in all the flap and flap-less cases and is the dominant flow characteristic in the large-angle-of-attack airfoil-flap problem considered in this work. However, there are intricate mechanisms through which the passively deployed flap interacts with the separated flow and vortex shedding which will be shown to favorably improve the lift of the airfoil. To facilitate a systematic discussion of the flap-vortex interactions in the later sections of this manuscript, we divide one time period of flap oscillations into two temporal regions dominated by the TEV and LEV as highlighted in Fig.~\ref{generalcl}. Here, the starting and ending portions of the lift cycle from $\periodt=0$ to $\periodt\approx 0.2$ and $\periodt \approx 0.8$ to $\periodt = 1$, respectively, correspond to the LEV-dominant portion while the TEV is dominant between $\periodt\approx0.2$ and $\periodt \approx 0.8$, which are also coherent with the above discussion of the vortex-shedding process. The time instants of $\periodt \approx0.2$ and $\periodt \approx0.8$ demarcating the LEV and TEV regions are identified using the baseline flap-less airfoil case since the qualitative features of the vortex-shedding process are consistent across all the flap and flap-less airfoil systems. Specifically, these times correspond to the instants when $\lift$ of the flap-less case crosses its mean value, which is also demonstrated in Fig.~\ref{generalcl} by plotting the mean lift of the flap-less case.

In the following sections, we perform a systematic parametric study of the airfoil-flap system. We characterize the qualitatively distinct regimes that the fully-coupled FSI system exhibits, and provide detailed insights into the interplay between the flap dynamics and the formation and interaction of vortical structures for parameters yielding lift benefits. 

\subsection{Parametric study}
\label{parametricstudy}

\begin{figure}
\centering
\begin{subfigure}[t]{0.45\textwidth}
\centering
\includegraphics[scale=1]{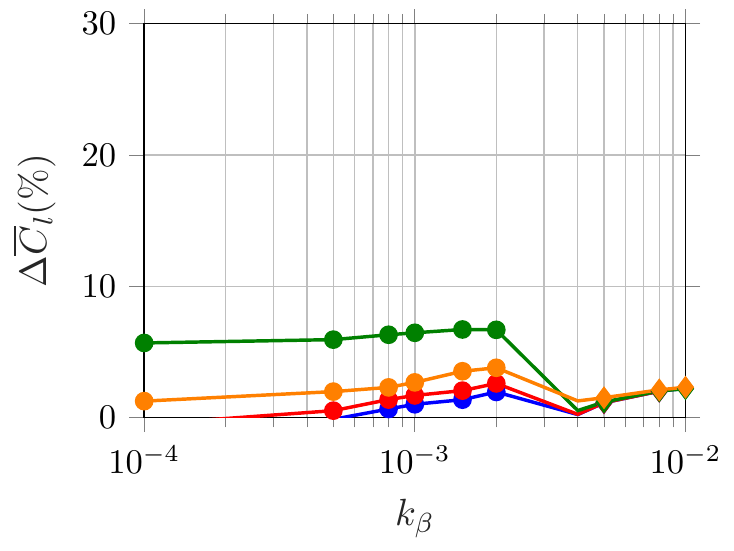}
\caption{Flap hinged at $\twenty$ chord.}
\label{parametric20}
\end{subfigure}
\begin{subfigure}[t]{0.45\textwidth}
\centering
\includegraphics[scale=1]{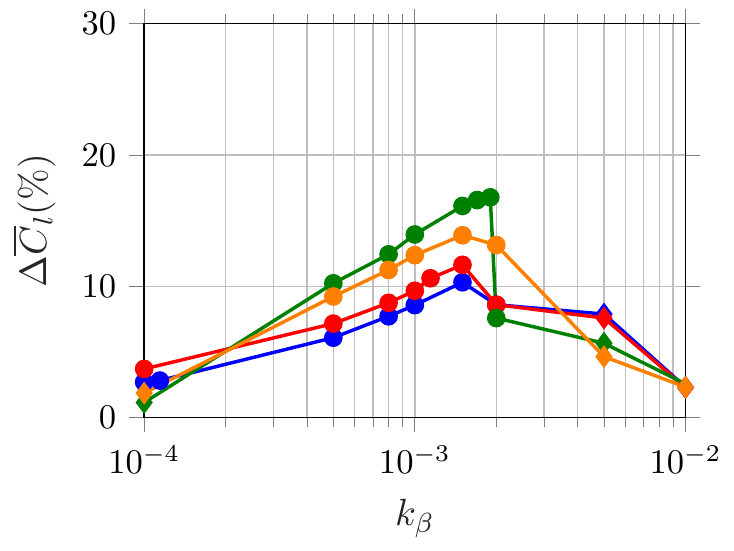}
\caption{Flap hinged at $\fifty$ chord.}
\label{parametric50}
\end{subfigure}
\begin{subfigure}[t]{0.9\textwidth}
\centering
\includegraphics[scale=1]{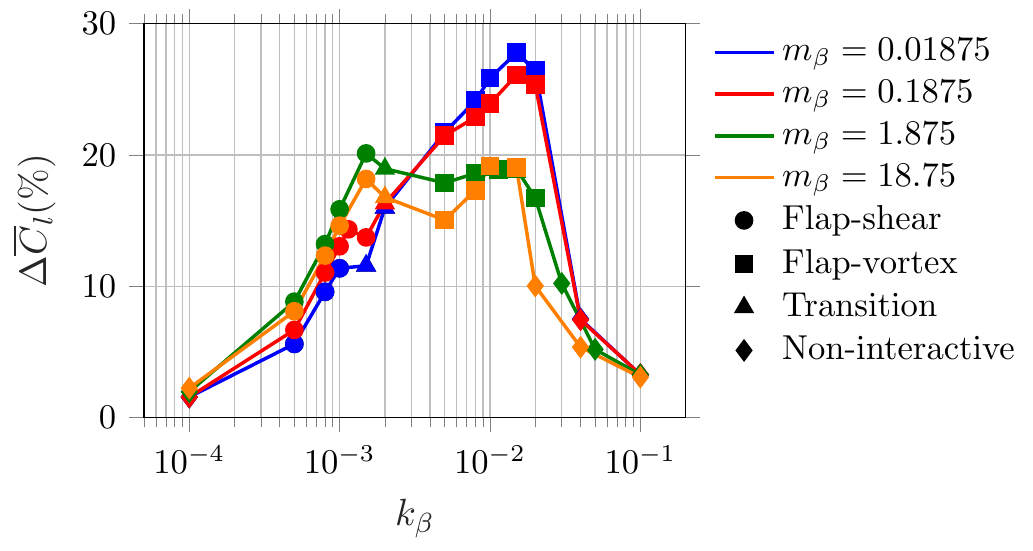}
\caption{Flap hinged at $\sixty$ chord}
\label{parametric60}
\end{subfigure}
\caption{Performance plots showing percentage change in mean lift for the various cases of mass ratio, stiffness and hinge location. The marker symbols denote the flow regimes identified using a flow classification algorithm described in Appendix~\ref{classification}.}
\label{parametric}
\end{figure}

A parametric study is performed by varying the stiffness of the hinge, mass of the flap and location of the hinge. The results of this study are presented in Fig.~\ref{parametric}. Here, improvements in mean lift of the airfoil with a torsionally mounted flap compared to the flap-less airfoil case in percentage, $\liftimp$, are plotted against stiffness for various mass ratios and flap locations. All the mean quantities are evaluated in the LCO regime ($\timet>20$). Lift improvements as high as $27\%$ are attained using a passively deployable flap. However, such significant benefits are realized only at certain flap parameters. 
To systematically understand the role of various flap parameters on performance, we analyze the variations in performance trends due to each parameter starting with the flap location. 

The flaps located at $\sixty$ and $\twenty$ provided the best and worst lift improvements, respectively, implying that the downstream flap locations are more beneficial for performance than upstream located flaps. In contrast to the quantitative differences in lift improvements across flap locations, there are certain qualitative similarities in performance trends. For instance, in Fig.~\ref{parametric20} and \ref{parametric50} for the $\twenty$ and $\fifty$ locations, respectively, a single peak in performance is observed at approximately $\stiff \approx 0.0015$, largely independent of the mass ratio. The flap at $\sixty$ location also exhibits a performance peak at a similar stiffness of $\stiff\approx0.0015$ for large mass ratios. However, an important qualitative distinction is observed between these locations, where the $\sixty$ case exhibits an additional peak at an order of magnitude higher stiffness of $\stiff \approx 0.015$ while the corresponding lift improvements for the $\twenty$ and $\fifty$ cases are not considerable.  

\begin{figure}
\centering
\begin{subfigure}[t]{0.45\textwidth}
\centering
\includegraphics[scale=1]{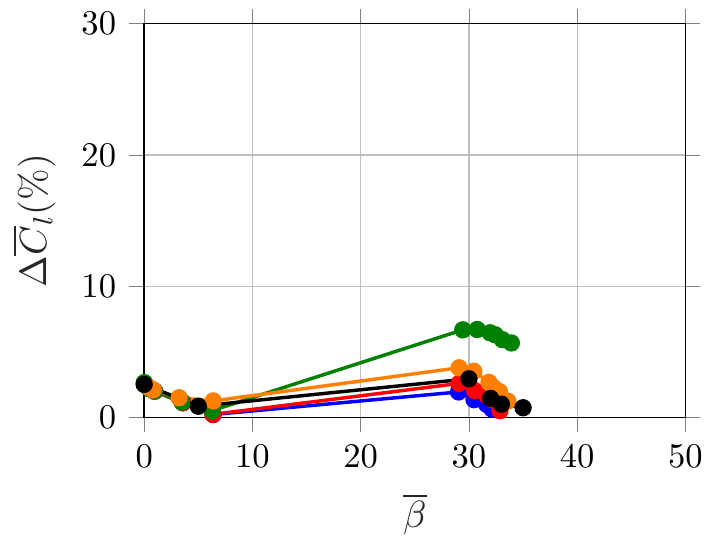}
\caption{Flap hinged at $\twenty$ chord.}
\end{subfigure}
\begin{subfigure}[t]{0.45\textwidth}
\centering
\includegraphics[scale=1]{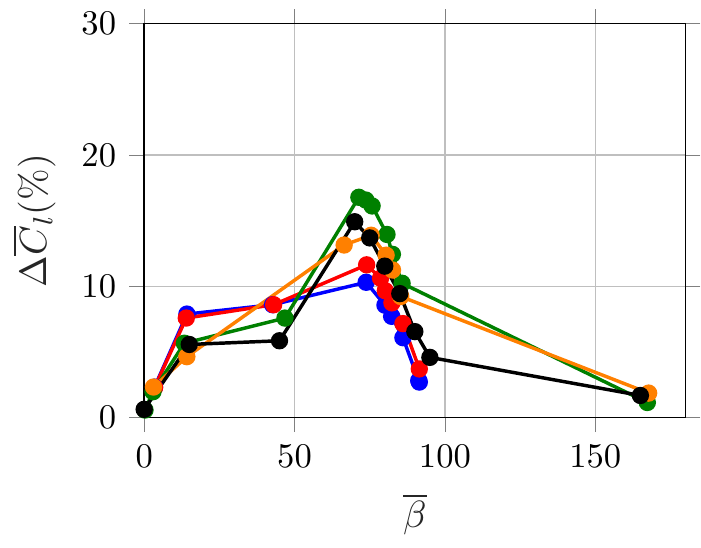}
\caption{Flap hinged at $\fifty$ chord.}
\end{subfigure}
\begin{subfigure}[t]{0.9\textwidth}
\centering
\includegraphics[scale=1]{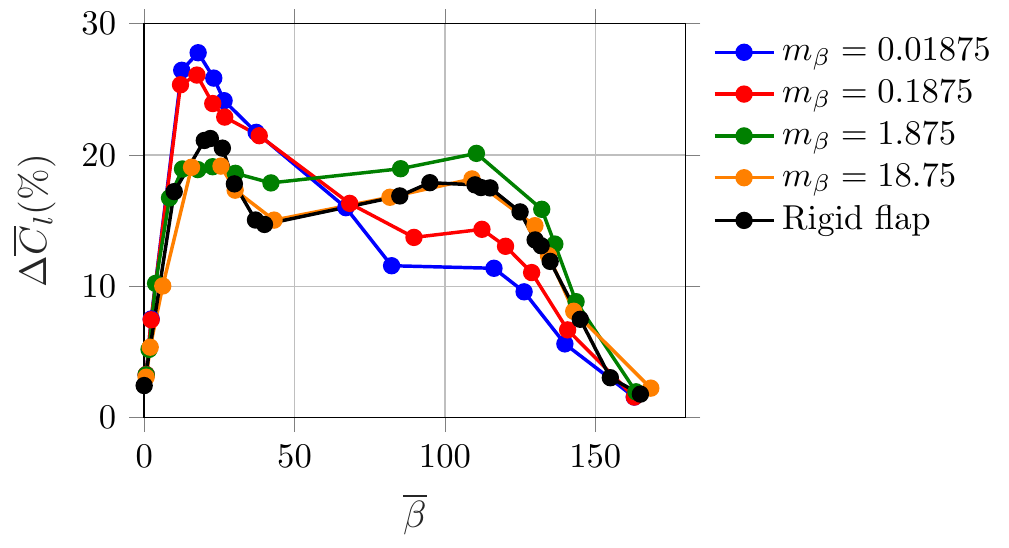}
\caption{Flap hinged at $\sixty$ chord.}
\end{subfigure}
\caption{Performance plots showing percentage change in mean lift for the various cases of inertia, stiffness-dependent mean flap deflection and hinge location. The rigid flap cases correspond to a stationary rigid flap affixed at various deployment angles.}
\label{parametricbeta}
\end{figure}

Since stiffness appears to be a more influential parameter than mass ratio in setting the qualitative performance trends, we look at the role of stiffness next.
One of the dominant effects of stiffness is to set a nominal mean deflection angle about which the flap oscillates \cite{nair2022effects}. Indeed, one might imagine that the flap over-deploys for small stiffness values and under-deploys for large stiffness values. We probe the impact of stiffness on mean deflection angle, and in turn on lift, by re-plotting the performance plots from Fig.~\ref{parametric} with $\deflmean$ along the $x$-axis in Fig.~\ref{parametricbeta}. It can be clearly seen that the cases corresponding to extreme flap configurations of $\deflmean \approx 150^\circ$ and $\deflmean \approx 0^\circ$, corresponding to very low and large stiffness, respectively, do not provide considerable performance benefits regardless of inertia. However, intermediate flap deployments corresponding to moderate values of stiffness provide significant lift improvements. 

We also note that across four orders of magnitude of mass ratio,  peak lift not only occurs at the same stiffness (\emph{c.f.} Fig.~\ref{parametric}) but also at the same mean flap deflection angle (\emph{c.f.} Fig.~\ref{parametricbeta}). In fact, the general trend in the mean lift improvement of the moving flap follows the trend of a rigid flap as shown by the black line in Fig.~\ref{parametricbeta}. Here, the data for the rigid flap were obtained via rigid-body simulations of the airfoil-flap system with a stationary rigid flap affixed at various deployment angles. This qualitative similarity in trend suggests that (i) the role of stiffness is to set the mean deflection angle in a manner largely agnostic to the mass ratio, and (ii) this mean deflection angle is the primary parameter in setting the qualitative flow regime and associated aerodynamic performance.

\begin{figure}
\begin{adjustwidth}{}{0.5cm} 
\centering
\centering
\begin{subfigure}[t]{0.283\textwidth}
\centering
\includegraphics[scale=1]{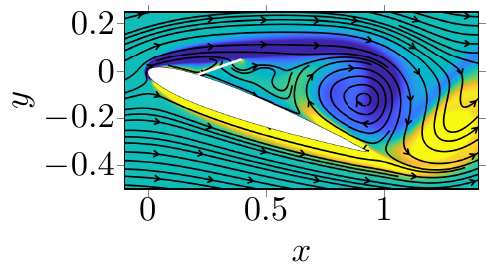}
\vspace{-0.7cm}
\caption{$\twenty$ location, \linebreak $\stiff=0.0015$}
\label{flapconfig1}
\end{subfigure}
\begin{subfigure}[t]{0.22\textwidth}
\centering
\includegraphics[scale=1]{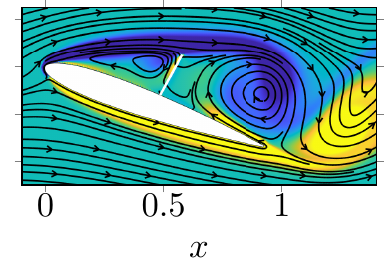}
\vspace{-0.7cm}
\caption{$\fifty$ location, $\stiff=0.0015$}
\label{flapconfig2}
\end{subfigure}
\begin{subfigure}[t]{0.22\textwidth}
\centering
\includegraphics[scale=1]{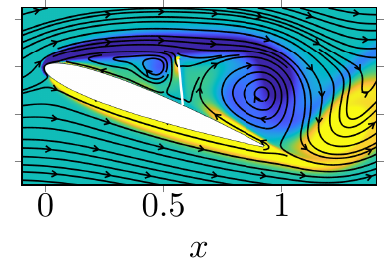}
\vspace{-0.7cm}
\caption{$\sixty$ location, $\stiff=0.0015$}
\label{flapconfig3}
\end{subfigure}
\begin{subfigure}[t]{0.22\textwidth}
\centering
\includegraphics[scale=1]{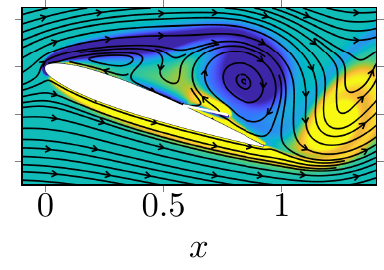}
\vspace{-0.7cm}
\caption{$\sixty$ location, $\stiff=0.015$}
\label{flapconfig4}
\end{subfigure}
\end{adjustwidth}
\caption{Vorticity contours demonstrating flap configurations in the flap-shear (subplots (a), (b) and (c)) and flap-vortex (subplot (d)) interaction regimes. The contours are plotted at a representative time instant of $\periodt=0$ and $\mass=1.875$.}
\label{flapconfig}
\end{figure}

One might also hypothesize that the stiffness-dependent mean flap configuration can be used to identify qualitatively different regimes; e.g., one might expect qualitatively distinct flow physics associated with the local peaks at very different mean deflection angles in Fig.~\ref{parametricbeta}. In this vein, we utilize  a k-means clustering based algorithm to classify the dynamics associated with various mean deflection angles. The details of the classification methodology, which utilizes two meaningfully chosen lengthscales representative of the shear layer separating the bulk and near-body flow dynamics and the proximity of the flap to shedding of the leading edge vortex, are provided in Appendix~\ref{classification}. The results of this flow classification procedure are indicated by the markers in Fig.~\ref{parametric}. It can be observed from Fig.~\ref{parametric} that the cases with stiffness in the vicinity of $\stiff=0.0015$ belong to what we term the \emph{flap-shear interaction regime} for all the $\twenty$, $\fifty$ and $\sixty$ locations. To visualize the associated flap configurations, vorticity contours at a representative time instant of $\periodt=0$, for $\mass=1.875$, $\stiff=0.0015$ and all three locations are plotted in Fig.~\ref{flapconfig1}, \ref{flapconfig2} and \ref{flapconfig3}, respectively. It can be seen that the relatively large flap deflection is such that the flap tip lies in close vicinity of the high momentum shear layer. All parameters classified within this regime share the prominently (but not overly) deployed flap that has a mean deployment near the shear layer separating the bulk and near-body flow, which underscores the chosen terminology. 

On the other hand, the cases around $\stiff=0.015$, which correspond to the second peak for the flap at $\sixty$ location (\emph{c.f.} Fig.~\ref{parametric60}), are assigned to the \emph{flap-vortex interaction regime}. The vorticity contour at $\periodt=0$ for $\mass=1.875$, $\stiff=0.015$ and $\sixty$ location is plotted in Fig.~\ref{flapconfig4}. The relatively low flap deflection is such that the flap can strongly interact with the vortex-shedding process. These coupled dynamics will be discussed in further detail in the sections below. Note that none of the cases at the $\twenty$ and $\fifty$ locations fall under the flap-vortex regime because the flap length of $0.2c$ is not long enough to interact with the vortices when the flap is located relatively upstream. Some parametric cases between the flap-shear and flap-vortex interaction regimes in Fig.~\ref{parametric60} are assigned to the \emph{transition regime}, where interplay with both the shear layer and bulk vortex-shedding processes are observed. Finally, cases corresponding to extreme stiffness values and flap deflections are grouped together into the \emph{non-interactive regime}, since they hardly interact with either the shear layer or vortices and therefore do not provide considerable changes to lift in comparison to the interactive regimes.

We now discuss the effect of inertia or mass ratio on lift performance. While across four orders of magnitude of mass ratio the general trend in the mean lift improvement remains similar, there are detailed differences across mass ratios indicated in Fig.~\ref{parametric}. Specifically, the effect of mass ratio is not uniformly beneficial or detrimental for lift performance. For example, for the cases in the flap-shear interaction regime near $\stiff=0.0015$, the lighter flaps perform more poorly than the heavier ones, with maximal performance attained at an intermediate mass ratio of $\mass=1.875$. In contrast, for the cases in the flap-vortex interaction regime near $\stiff=0.015$, the lighter flaps provide larger mean lift improvements. It will be shown in the later sections that one of the dominant effects of mass ratio is to set the amplitude of flap oscillations. The lowest mass flaps oscillate with largest amplitude which decreases with increasing mass. For a very large mass ratio of $\mass=18.75$, the flap response is quasi-static and therefore, the associated performance in Fig.~\ref{parametricbeta} (orange line) strongly mimics that of a rigidly affixed flap (black line). The time-dependent flap dynamics encoded in mass ratio, therefore, have a secondary effect in establishing the FSI dynamics and corresponding lift behavior.

To summarize, these parametric studies suggest that the role of stiffness is to set the mean flap deflection angle, and a k-means classification algorithm demonstrates that there are meaningful categorical distinctions that can be established based on this mean deflection angle. The mass ratio or inertia alters the amplitude and phase of flap oscillations which has a secondary effect on lift performance. In the following sections, we describe in detail the physical mechanisms that yield the high lift benefits in the flap-shear and flap-vortex interaction regimes. Within each regime, we analyze the effects of the static mean deflection angle (driven by variations in stiffness) and of dynamic flap motion about this mean angle (encoded in mass ratio) on the vortex-shedding process and how this in turn modulates performance.


\subsection{Flap-shear interaction regime}
\label{flapshear}

\begin{figure}
\begin{adjustwidth}{}{0.5cm} 
\centering
\begin{subfigure}[t]{0.283\textwidth}
\centering
\includegraphics[scale=1]{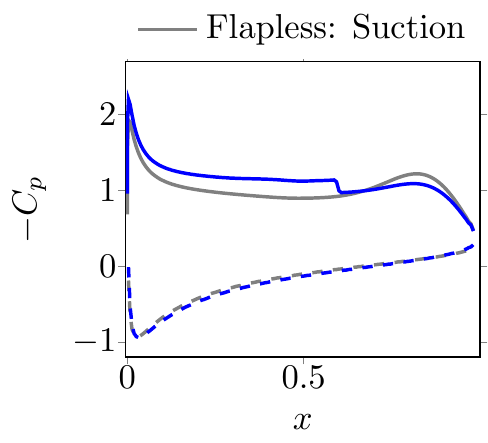}
\vspace{-0.7cm}
\caption{$t=0 \ T$ }
\label{fscp1}
\end{subfigure}
\begin{subfigure}[t]{0.22\textwidth}
\centering
\includegraphics[scale=1]{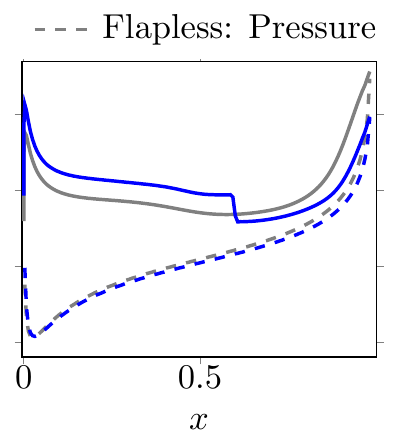}
\vspace{-0.7cm}
\caption{$t=0.27 \ T$ }
\label{fscp2}
\end{subfigure}
\begin{subfigure}[t]{0.22\textwidth}
\centering
\includegraphics[scale=1]{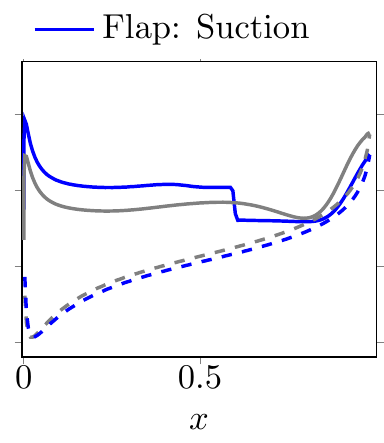}
\vspace{-0.7cm}
\caption{$t=0.55 \ T$ }
\label{fscp3}
\end{subfigure}
\begin{subfigure}[t]{0.22\textwidth}
\centering
\includegraphics[scale=1]{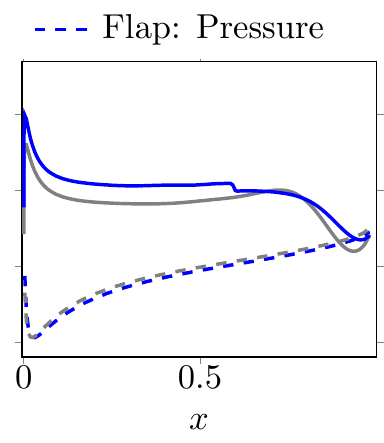}
\vspace{-0.7cm}
\caption{$t=0.82 \ T$ }
\label{fscp4}
\end{subfigure}
\centering
\begin{subfigure}[t]{0.283\textwidth}
\centering
\includegraphics[scale=1]{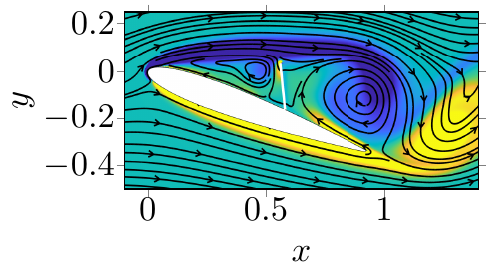}
\vspace{-0.7cm}
\caption{$t=0 \ T$ }
\label{fs1}
\end{subfigure}
\begin{subfigure}[t]{0.22\textwidth}
\centering
\includegraphics[scale=1]{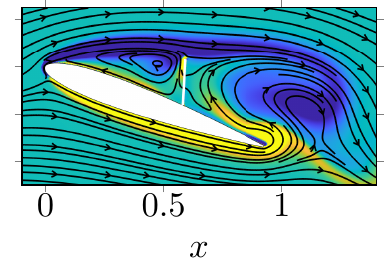}
\vspace{-0.7cm}
\caption{$t=0.27 \ T$ }
\label{fs2}
\end{subfigure}
\begin{subfigure}[t]{0.22\textwidth}
\centering
\includegraphics[scale=1]{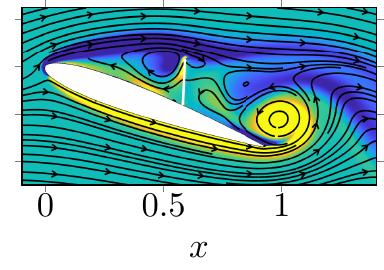}
\vspace{-0.7cm}
\caption{$t=0.55 \ T$ }
\label{fs3}
\end{subfigure}
\begin{subfigure}[t]{0.22\textwidth}
\centering
\includegraphics[scale=1]{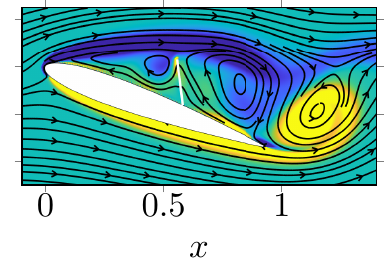}
\vspace{-0.7cm}
\caption{$t=0.82 \ T$ }
\label{fs4}
\end{subfigure}
\end{adjustwidth}
\caption{$\Cp$ distribution (top row) demonstrating the pressure dam effect and vorticity contours with superimposed streamlines (bottom row) demonstrating barrier to upstream propagation of reverse flow imposed by the flap for a representative case of $\stiff=0.0015$, $\mass=1.875$ and $\sixty$ location in the flap-shear interaction regime at four time instants in a lift cycle. $\Cp$ distributions of the baseline flap-less case are also provided for comparison in the top row. Refer Fig.~\ref{generalvorticity5}--\ref{generalvorticity8} for vorticity contours of the baseline case.}
\label{fs}
\end{figure}

We consider a representative case within this regime, with the flap fixed at $\sixty$ location, $\stiff=0.0015$ and $\mass=1.875$ to discuss the lift enhancement mechanisms. For this case, the $\Cp$ distribution on the airfoil surface at four time instants in one time period of the lift cycle are plotted in Fig.~\ref{fscp1}--\ref{fscp4}. For comparison, $\Cp$ on the flap-less airfoil is also plotted in the same figure. In each plot, a step discontinuity in $\Cp$ on the suction surface of the airfoil at the location of the flap is formed. Upstream of this discontinuity, a lower pressure zone compared to the flap-less case is created. Since the flap acts as a dam in maintaining a low pressure region upstream of the hinge, the ensuing step discontinuity is identified to be the manifestation of the pressure dam effect \cite{bramesfeld2002experimental}. 

To understand the pressure dam effect in detail, vorticity contours with streamlines at the same four time instants are plotted in Fig.~\ref{fs1}--\ref{fs4}. It can be observed that the flap is significantly extended towards the shear layer at all times. At such large flap deployment angles, the flap acts as a barrier in dividing the large separated flow region into two smaller regions. This barrier prevents the mixing of regions of lower and higher pressure near the leading and trailing edges, respectively, allowing for lower pressure to be maintained upstream of the flap hinge. In the absence of the flap as shown in Fig.~\ref{generalvorticity5}--\ref{generalvorticity8}, the mixing mainly occurs due to the upstream propagation of reverse flow occurring downstream near the trailing edge. This upstream propagation of reverse flow is more pronounced during the growth and shedding of the TEV. For visualization, multiple upstream moving streamlines from the trailing to the leading edge can be observed in the flap-less vorticity contours at time instants of $\periodt=0.27$ and $\periodt=0.55$ in Fig.~\ref{generalvorticity6} and \ref{generalvorticity7}, respectively. On the other hand, the highly extended flap in the flap-shear interaction regime significantly blocks these streamlines in Fig.~\ref{fs2} and \ref{fs3}. Accordingly, the step discontinuities in Fig.~\ref{fscp2} and Fig. \ref{fscp3} are stronger during this TEV-dominant portion of the lift cycle. 

We note that even in the presence of the flap, some streamlines are observed to pass through the flap in Fig.~\ref{fs}. We emphasize that this phenomenon does not imply a permeable flap. Instead, the flap motion sets a non-zero flow velocity via the no-slip boundary condition due to which the streamlines appear to pass through the flap. A closer inspection will also reveal that the streamlines are aligned in the flap-normal direction in the vicinity of the flap, as is physically necessary. Therefore, despite the slight upstream flow induced by the flap's motion, the velocity of the associated reverse flow is modulated and reduced by the slowly moving flap.

In the following sections, we decompose the effects of the flap as those due to the mean location of the flap and the dynamics about the mean respectively. Following the discussion in Sec.~\ref{parametricstudy}, the first of these effects is dominated by the torsional stiffness, and the latter by the mass ratio. We systematically investigate the effect of these parameters on the flow physics next.

\subsubsection{Static behavior: effect of varying stiffness}
\label{fs_static}

To systematically decouple and study the effects of stiffness-dependent mean flap deflection without incorporating the effects of mass-dependent flap dynamics, in this section we focus on the higher mass cases, $\mass>1$, due to their relatively low amplitude and therefore, better approximation towards exhibiting a static behavior. The additional performance-affecting physical mechanisms due to increasing flap dynamics and amplitude with decreasing mass are separately analyzed in the next section.

Within the flap-shear regime in the stiffness range from $\stiff=0.0005$ to $0.002$, the maximal lift is around $\stiff=0.0015$ (\emph{c.f.} Fig.~\ref{parametric50}). Maximal lift occurs at this stiffness value due to an optimal balance between two dominant competing effects---a ``secondary'' LEV trapped by the flap and the magnitude of flow velocity in the shear layer above the airfoil. These competing effects are described next.

\begin{figure}
\begin{adjustwidth}{}{0.3cm} 
\centering
\begin{subfigure}[t]{0.3\textwidth}
\centering
\includegraphics[scale=1]{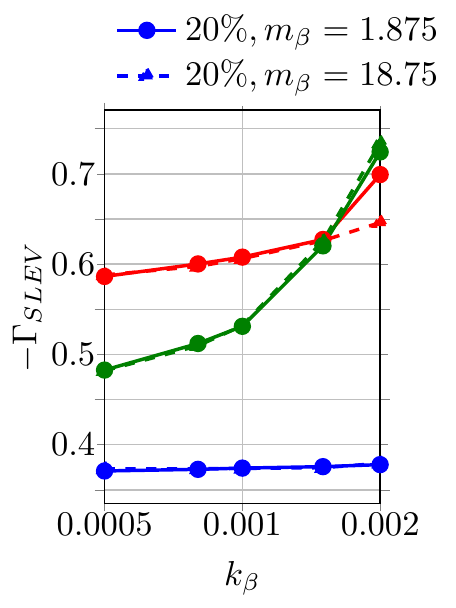}
\caption{\centering Mean SLEV strength.}
\label{fs_static_trends1}
\end{subfigure}
\begin{subfigure}[t]{0.3\textwidth}
\centering
\includegraphics[scale=1]{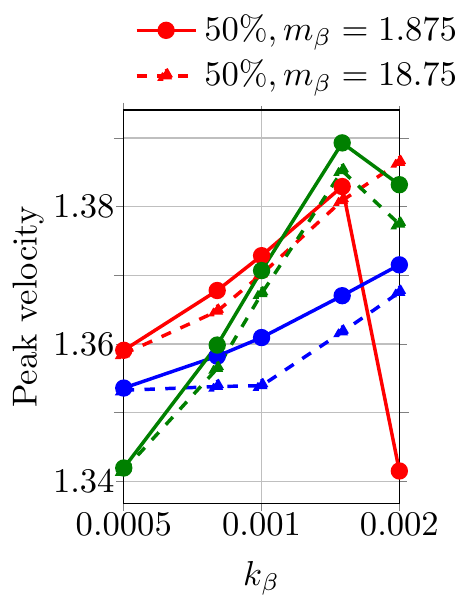}
\caption{\centering Mean peak velocity attained at $x=0.2$.}
\label{fs_static_trends2}
\end{subfigure}
\begin{subfigure}[t]{0.3\textwidth}
\centering
\includegraphics[scale=1]{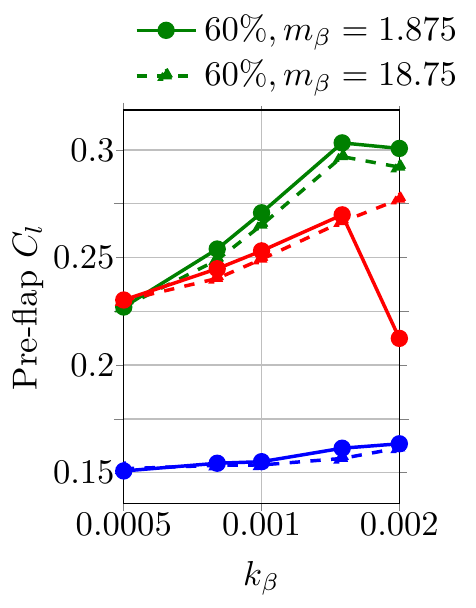}
\caption{\centering Mean pre-flap suction lift.}
\label{fs_static_trends3}
\end{subfigure}
\end{adjustwidth}
\caption{Plots of various physical quantities characteristic of static behavior in the flap-shear regime averaged over a time period for $\mass>1$ cases.}
\label{fs_static_trends}
\end{figure}

From the vorticity contours in Fig.~\ref{fs}, we observe that the flap divides the large separated flow region into two zones. In the upstream zone of the flap, the trapped flow continues to recirculate owing to the high momentum shear layer bounding from above. We refer to this trapped recirculated flow as the secondary LEV (SLEV) since it forms a part of the clockwise rotating vortex generated at the leading edge. 
A larger SLEV  augments the existing low pressure region of the pressure dam effect. The magnitude of the SLEV is proportional to the recirculation area upstream of the flap---the larger the area, the larger the SLEV circulation strength, $\slev$. The area is in turn dependent on the flap deployment angle which is largely set by stiffness. Therefore, the larger the stiffness, the lower the deployment angle, resulting in a larger upstream recirculation area, eventually yielding a stronger SLEV. The methodology for quantifying $\slev$ is provided in Appendix~\ref{quantvortstrength}. Now, $\slev$ for the various cases of stiffness (in the flap-shear regime), mass ratio ($\mass>1$) and hinge locations are plotted in Fig.~\ref{fs_static_trends1}. As expected, $\slev$ increases with increasing stiffness for all mass and locations. This increase in $\slev$ is the first dominant mechanism which contributes to improved performance, specifically the pre-flap suction lift, which is also plotted in Fig.~\ref{fs_static_trends3}. Here, the pre-flap suction lift is the lift contribution by the upper surface of the airfoil upstream of the flap (\emph{i.e.} before the discontinuity in the $\Cp$ plot). We plot the pre-flap lift instead of the total lift of the upper surface since the effect of the SLEV is prominent in the region upstream of the flap. 

\begin{figure}
\centering
\begin{subfigure}[t]{0.45\textwidth}
\centering
\includegraphics[scale=1]{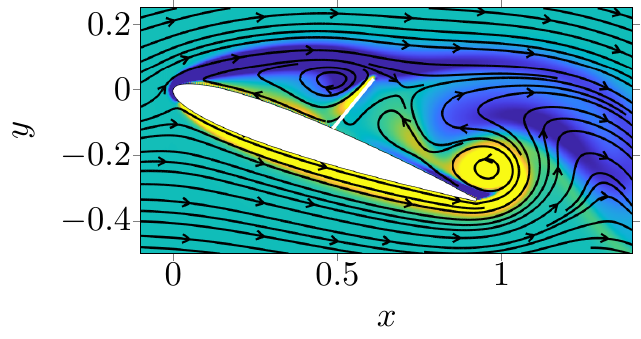}
\caption{Increased blocking at $\stiff=0.0015$.}
\label{fs_static_reverse1}
\end{subfigure}
\begin{subfigure}[t]{0.4\textwidth}
\centering
\includegraphics[scale=1]{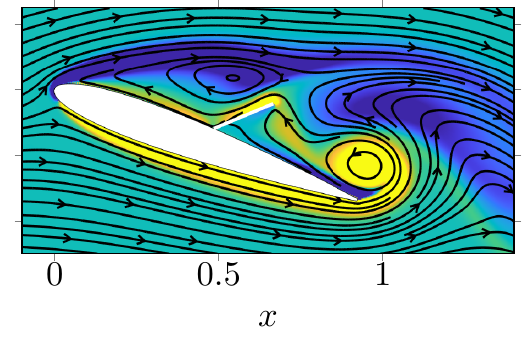}
\caption{Decreased blocking at $\stiff=0.002$.}
\label{fs_static_reverse2}
\end{subfigure}
\caption{Demonstration of the stiffness-dependent flap deployment angle dictating the magnitude of blocking of propagation of TEV-induced reverse flow via vorticiy contours and superimposed streamlines at $\periodt=0.55$ for $\mass=1.875$ and $\fifty$ location.}
\label{fs_static_reverse}
\end{figure}

We also note from Fig.~\ref{fs_static_trends3} that for some cases, as the stiffness is increased to $\stiff=0.002$, the pre-flap lift decreases even though the corresponding $\slev$ continues to increase in Fig.~\ref{fs_static_trends1}. This trend is observed for $\mass=1.875$ case at $\fifty$ flap location and both mass ratio cases at $\sixty$ location. This decrease in lift with increasing stiffness is associated with the continued decrease in flap deployment to the point where the flap detaches from the shear layer. This detachment weakens the barrier imposed by the flap and allows increased mixing between the upstream and downstream flow regions separated by the flap. A noticeable example of increased mixing due to the shear-layer-detached flap is demonstrated in Fig.~\ref{fs_static_reverse}, where vorticity contour plots of $\stiff=0.0015$ and $\stiff=0.002$ cases for the $\fifty$ location and $\mass=1.875$ at a representative time instant of $\periodt=0.45$ are plotted. For $\stiff=0.0015$ in Fig.~\ref{fs_static_reverse1}, the flap extended towards the shear layer effectively blocks the TEV-induced reverse flow. However, when the stiffness is increased to $\stiff=0.002$ in Fig.~\ref{fs_static_reverse2}, the flap is observed to detach away from the shear layer which causes an increased upstream propagation of TEV-induced reverse flow. We emphasize that the amplitude of flap oscillations in the flap-shear regime, specifically for $\mass>1$, is not significant which can also be visualized from the temporal vorticity plots of related cases in Fig.~\ref{fs1}--\ref{fs4} and Fig.~\ref{generalvorticity1}--\ref{generalvorticity4}. Therefore, the shear-layer-extended and -detached flap configurations for $\stiff=0.0015$ and $\stiff=0.002$ cases, respectively, and associated mixing characteristics are valid at all time instants. 
We note that the increased reverse flow in the latter further strengthens the TEV, which is detrimental to lift since the TEV reduces the clockwise circulation around the airfoil. This lower clockwise circulation in turn reduces the downstream moving flow velocity magnitude in the shear layer above the suction surface of the airfoil near the leading edge. This lower flow velocity in the shear layer is unfavorable to the low pressure region upstream of the flap. 
As a proxy for the flow velocity within the shear layer, the mean of peak $x$-component of flow velocity attained at a representative location of $x=0.2$ is plotted in Fig.~\ref{fs_static_trends2}. The peak velocity is taken as the maximal value across all $y$-locations, at $x=0.2$, which is then averaged over a lift cycle. We note that the chosen $x$-location is suitable for this analysis because our aim is purely to have a representative measure of the flow momentum in the shear layer. It can be observed that the cases beyond $\stiff=0.0015$ that exhibit a decrease in the pre-flap lift with increasing stiffness in Fig.~\ref{fs_static_trends3} also exhibit a decrease in the peak velocity in the shear layer above the suction surface upstream of the flap. This reduction in velocity, which primarily occurs due to the detachment of the flap from the shear layer, is the second dominant mechanism that deleteriously affects lift. 

\begin{figure}
    \centering
    \includegraphics[scale=1]{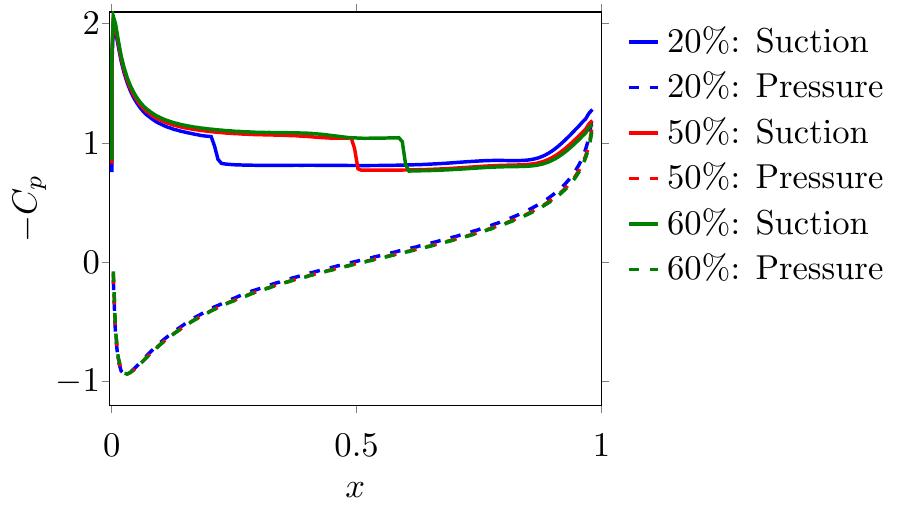}
    \caption{Comparison of the pressure dam effect at different flap locations via mean $\Cp$ distribution for $\mass=1.875,\stiff=0.0015$.}
    \label{fs_static_cpcomp}
\end{figure}

Finally, from Fig.~\ref{parametric} we note that the flaps at downstream locations have a higher maximal lift compared to those at upstream locations. Specifically, the peak performance in the flap-shear interaction regime for $\sixty$ case is the highest while that of $\twenty$ is lowest. This is because the flaps at locations further downstream allow a larger pre-flap suction surface to benefit from regions of low pressure created by the pressure dam effect. This increased exposure to the pressure dam effect can be visualized in Fig.~\ref{fs_static_cpcomp}, where the mean $\Cp$ distribution for three flap locations with $\mass=1.875$ and $\stiff=0.0015$ are plotted. The $\sixty$ case has the largest region of pre-flap suction pressure as compared to the flaps located further upstream.

To summarize, maximal performance in the flap-shear interaction regime is observed around $\stiff\approx 0.0015$ due to the competing effects of (advantageous) increasing SLEV strength and (disadvantageous) decreasing flow velocity in the shear layer above the suction surface due to the detachment of the flap from the shear layer as the stiffness is increased. In this section, we analyzed the lift enhancement mechanisms associated with the mean flap deflection through the lens of higher mass cases of $\mass>1$ to focus on the (quasi-)static effect of the flap on the flow. In the next section, we will investigate how flap dynamics, largely governed by mass ratio, play a role in favorably or adversely altering this quasi-static behavior. 

\subsubsection{Dynamic behavior: effect of varying inertia}
\label{fs_dynamic}

In Sec.~\ref{parametricstudy}, we noted that in the flap-shear regime, the lowest mass ratio yielded the least lift improvement whereas peak performance for a given stiffness was attained at $\mass=1.875$. In this section, we discuss the dominant mechanisms through which the flap dynamics (encoded in mass ratio) modulate the TEV-induced reverse flow and LEV strength, which in turn affect lift performance.

\begin{figure}
    \centering
    \includegraphics[scale=1]{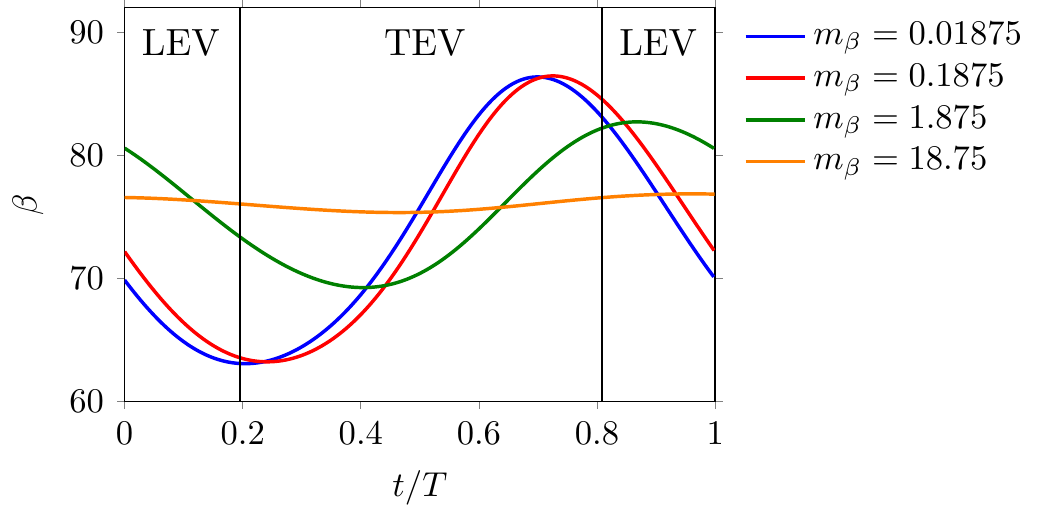}
    \caption{Flap deflection in one time period of the lift cycle for $\fifty$ location and $\stiff=0.0015$. The time period is divided into two portions dominated by the LEV and TEV where the time instants of demarcation are $\periodt\approx 0.2$ and $\periodt\approx 0.8$.}
    \label{deflfs}
\end{figure}

To explain the effect of mass ratio on the flap dynamics consisting of amplitude and phase, we plot the flap deflection in one time period of the limit-cycle oscillation (LCO) for a representative case of $\stiff=0.0015$, $\fifty$ location and different mass ratios in Fig.~\ref{deflfs}. Firstly, the amplitude of the flap oscillation decreases as the mass is increased and for a large mass ratio of $\mass=18.75$, the flap is almost stationary. Secondly, the flap oscillations also have a different phase relative to each other (and the lift dynamics), especially discernible for the lowest three cases of mass ratio. In order to understand the effects of mass-dependent varying amplitude and phase on the flap-fluid interactions and associated performance, we divide the lift cycle into two parts---the LEV- and TEV-dominant portions---as shown in Fig.~\ref{deflfs}. This temporal segregation is described in Sec.~\ref{qualitative} and Fig.~\ref{generalcl}, where the baseline case is used as the reference due to the consistent nature of vortex shedding across all the flap and flap-less airfoil cases. This baseline case also provides a meaningful reference for the phase of the flap dynamics: we define the phase of flap oscillations, $\phase$, to be the time delay (in radians) between the observed peak of $\beta$ and the starting time instant of the TEV-portion ($\periodt=0.2$): $\phase = 2\pi(\underset{\periodt}{\text{argmax}}(\defl)-0.2)$. The reason for using the start of the TEV-portion of the signal to define the phase is explained in Appendix~\ref{quantvortstrength}.
In the following, we analyze the effects of flap dynamics on performance in  two parts---the TEV- and LEV-dominated portions of the lift cycle. In each portion of the cycle, we identify the physical mechanisms that favorably or detrimentally contribute to lift. 

\begin{figure}
\centering
\begin{subfigure}[t]{0.45\textwidth}
\centering
\includegraphics[scale=1]{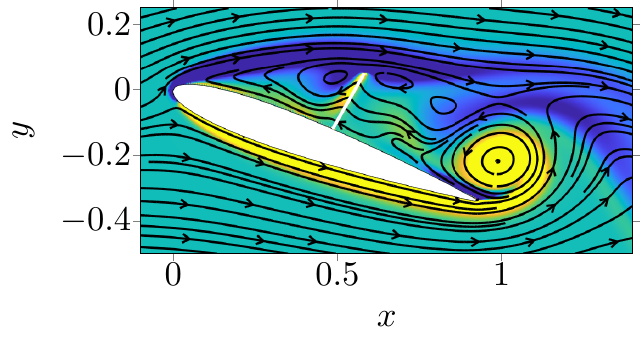}
\caption{Increased reverse flow at $\mass=0.01875$.}
\label{fs_dynamic_reverse1}
\end{subfigure}
\begin{subfigure}[t]{0.4\textwidth}
\centering
\includegraphics[scale=1]{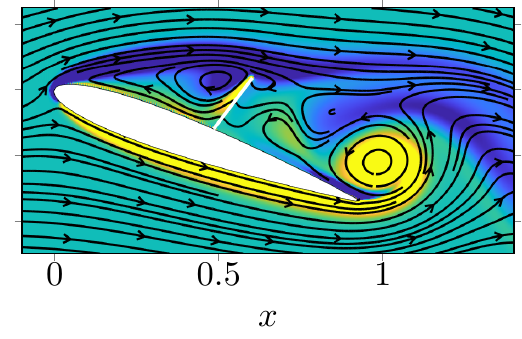}
\caption{Decreased reverse flow at $\mass=1.875$.}
\label{fs_dynamic_reverse2}
\end{subfigure}
\caption{Demonstration of propagation of TEV-induced reverse flow facilitated by the upstream oscillation of the flap via vorticiy contours and superimposed streamlines at $\periodt=0.55$ for $\stiff=0.0015$ and $\fifty$ location.}
\label{fs_dynamic_reverse}
\end{figure}

\begin{figure}
    \centering
    \includegraphics[scale=1]{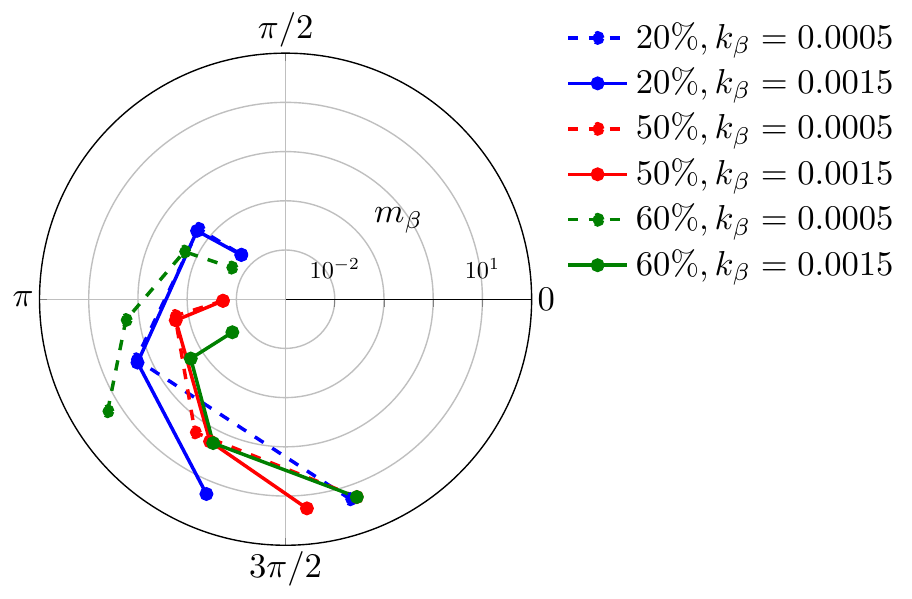}
    \caption{Phase difference of flap oscillations v/s mass ratio (along the radial axis) for various stiffness and locations.}
    \label{fs_dynamic_phase}
\end{figure}

First, we analyze the TEV-dominant portion and identify a separate physical mechanism between the downstream ($\fifty$ and $\sixty$) and upstream ($\twenty$) flap locations. This distinction will be connected to the effects of the TEV and SLEV, whose presence are dominantly experienced downstream near the trailing edge and upstream near the leading edge, respectively. 
%
For the flaps at the downstream locations of $\fifty$ and $\sixty$, the ability of the flap to block the TEV-induced reverse flow dictates whether there are performance benefits or detriments during this TEV-portion of the cycle. For a given stiffness-dependent mean deflection angle, the ability to maximize blocking is linked to the timing (or phase) and amplitude of the flap dynamics relative to the TEV formation process. First, to motivate the importance of timing, we recall that during the TEV-dominant portion from $\periodt\approx0.2$ to $\periodt\approx0.8$, the TEV induces reverse flow that propagates upstream from the trailing to the leading edge. If the flap motion is such that it also oscillates in the upstream direction (towards the leading edge) during this TEV-dominant portion of the lift cycle, then the flap facilitates the upstream propagation of TEV-induced reverse flow via the no-slip constraint imposed by the flap. 
This flap motion corresponds to a phase of $\phase\approx\pi$ (according to the above-mentioned definition of phase) and is detrimental to lift. 
For demonstration, we consider the flap dynamics of the lighter flaps ($\mass<1$) in Fig.~\ref{deflfs} where these flaps move in the upstream direction (increasing deflection) at the initiation of the TEV-portion of the cycle, $\periodt \approx 0.2$. To visualize how this upstream oscillating flap facilitates TEV-induced reverse flow, vorticity contour for the lightest flap, $\mass=0.01875$, at $\fifty$ location and $\stiff=0.0015$ at a time instant of $\periodt=0.55$ when the TEV is approximately at its maximum strength is plotted in Fig.~\ref{fs_dynamic_reverse1}. Several upstream moving streamlines passing through the upstream oscillating flap can be observed. On the other hand, in the comparative vorticity contour for the larger mass ratio of $\mass=1.875$ in Fig.~\ref{fs_dynamic_reverse2}, only fewer streamlines pass through the flap. This upstream propagation of TEV-induced reverse flow experienced by light flaps reduces the pressure difference across the flap and thereby the lift-enhancing pressure-dam effect. Motivated by the observed importance of timing of flap oscillations relative to the growth of TEV, we now probe the phase for all mass ratios, flap locations and stiffness values of $\stiff=0.0005$ and $\stiff=0.0015$ in Fig.~\ref{fs_dynamic_phase}. The figure demonstrates that for the downstream locations of $\fifty$ and $\sixty$, lower flap masses indeed yield a phase closer to $\pi$ than zero, implying that the flap oscillations of lower-mass flaps promote TEV-induced reverse flow and associated lift detriments.

\begin{figure}
\begin{adjustwidth}{}{0.3cm} 
\centering
\begin{subfigure}[t]{0.24\textwidth}
\centering
\includegraphics[scale=1]{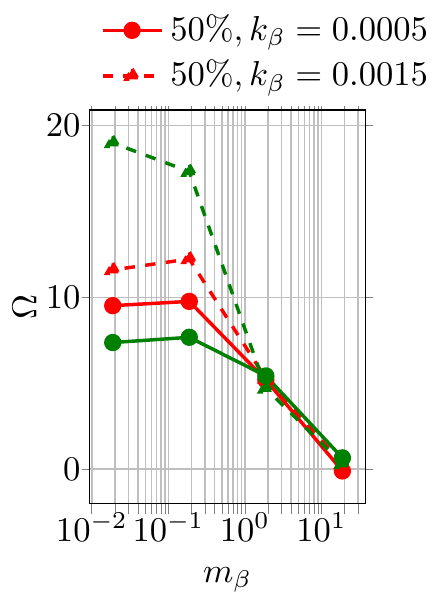}
\caption{\centering Integrated angular velocity of the flap.}
\label{fs_dynamic_trends1_1}
\end{subfigure}
\begin{subfigure}[t]{0.24\textwidth}
\centering
\includegraphics[scale=1]{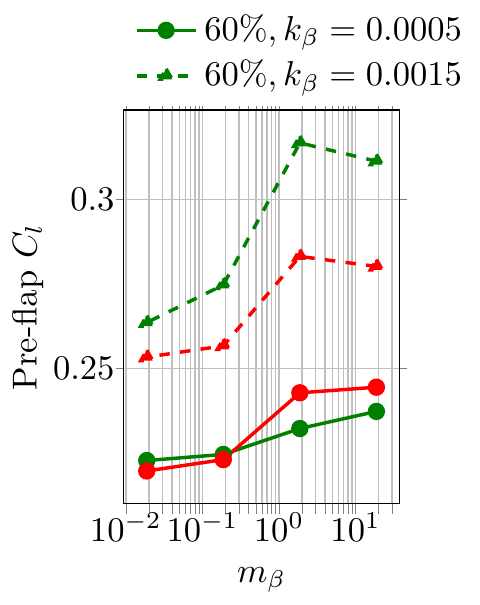}
\caption{\centering Mean pre-flap suction lift.}
\label{fs_dynamic_trends1_2}
\end{subfigure}
\begin{subfigure}[t]{0.24\textwidth}
\centering
\includegraphics[scale=1]{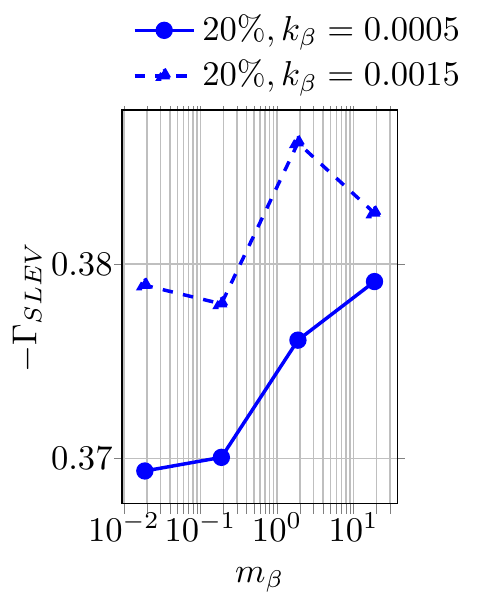}
\caption{\centering Mean negative SLEV strength.}
\label{fs_dynamic_trends1_3}
\end{subfigure}
\begin{subfigure}[t]{0.24\textwidth}
\centering
\includegraphics[scale=1]{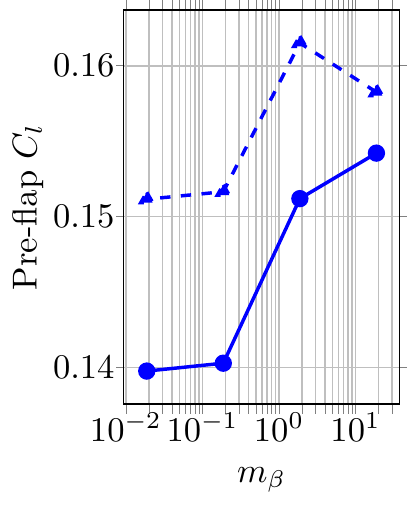}
\caption{\centering Mean pre-flap suction lift.}
\label{fs_dynamic_trends1_4}
\end{subfigure}
\end{adjustwidth}
\caption{Plots of various physical quantities characteristic of dynamic behavior in the flap-shear interaction regime averaged or integrated over the TEV-dominant portion of the lift cycle for different locations, mass ratios and stiffness.}
\label{fs_dynamic_trends1}
\end{figure}

The impact of the phase difference between the flap and the underlying shedding dynamics depends on the amplitude of the flap oscillations. For example, one might expect that for sufficiently small flap amplitudes, the dynamic effect of the flap is negligible irrespective of the phase with respect to the underlying flow behavior. We therefore use the angular velocity of the flap to quantify the magnitude of TEV-induced reverse flow. We note that this measure is a reasonable proxy for TEV-induced reversed flow since the flap tip is sufficiently close to the shear layer in the flap-shear regime that, to reasonable approximation, we can neglect any reverse flow occurring between the flap tip and the shear layer. Therefore, the amount of reverse flow is estimated by integrating the angular velocity ($\angvel$) of the flap during the TEV-dominant portion of the lift cycle: $\pangvel = \int_{TEV} \angvel d(\periodt)$. This integrated angular velocity is plotted for the various cases of mass ratio, stiffness and locations of $\fifty$ and $\sixty$ in Fig.~\ref{fs_dynamic_trends1_1}. As expected, it can be seen that the lower-mass flaps have a larger positive angular velocity due to their large-amplitude oscillations compared to the quasi-static $\mass=18.75$ case. This large-amplitude motion is disadvantageous to lift because of the phase relationship of these flap dynamics highlighted earlier. As a proxy for performance, in Fig.~\ref{fs_dynamic_trends1_2} we also plot the pre-flap lift improvement for the same cases as those in Fig.~\ref{fs_dynamic_trends1_1}. The cumulative effect of the phase difference of approximately $\pi$ and large flap velocity results in the lowest mass flaps providing the least pre-flap lift improvements in the TEV-portion of the lift cycle in Fig.~\ref{fs_dynamic_trends1_2}.

\begin{figure}
\centering
\begin{subfigure}[t]{0.45\textwidth}
\centering
\includegraphics[scale=1]{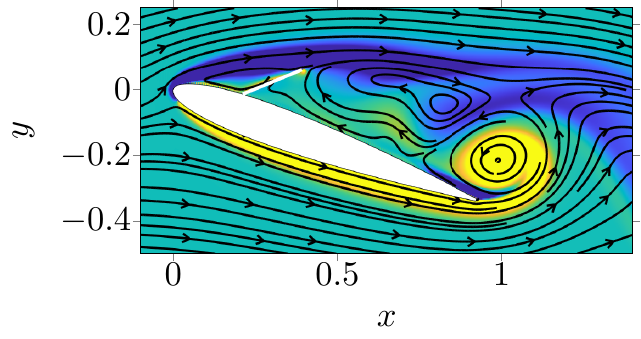}
\caption{$\mass=0.01875$}
\label{fs_dynamic_reverse3}
\end{subfigure}
\begin{subfigure}[t]{0.4\textwidth}
\centering
\includegraphics[scale=1]{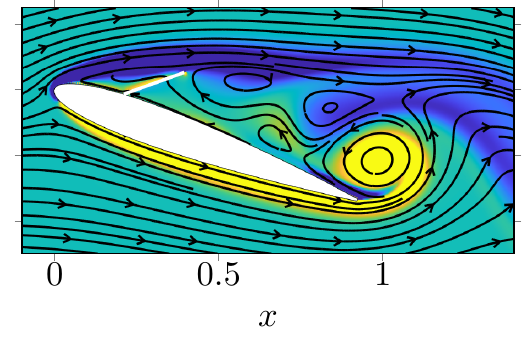}
\caption{$\mass=1.875$}
\label{fs_dynamic_reverse4}
\end{subfigure}
\caption{Demonstration of insubstantial variations in TEV-induced reverse flow for upstream $\twenty$ flap location and $\stiff=0.0015$ via vorticiy contours and superimposed streamlines at $\periodt=0.55$.}
\label{fs_dynamic_reverse20}
\end{figure}

We continue our discussion of the TEV-portion of the lift cycle by now considering the flaps at $\twenty$ location. Here, due to the farther upstream location of the flap from the trailing edge, the role of the flap aiding the TEV-induced reverse flow is not significant as compared to the downstream flap locations discussed above. For demonstration, consider the vorticity contours at $\periodt=0.55$ when the TEV is approximately at its maximum strength for two different mass ratios of $\mass=0.01875$ and $\mass=1.875$, $\stiff=0.0015$ and $\twenty$ location in Fig.~\ref{fs_dynamic_reverse20}. In both cases, the reverse flow has propagated upstream with significantly less hindrance than the similar flaps located at $\fifty$ in Fig.~\ref{fs_dynamic_reverse}. Therefore, any additional facilitation or blocking of the TEV-induced reverse flow has insubstantial effect on lift.

Instead, the SLEV formed near the leading edge is found to be more influential to performance at the upstream location of $\twenty$. 
Recall the following two facts---(a) the SLEV strength is higher when the flap deflection is lower (\emph{c.f.}, Sec.~\ref{fs_static}) and (b) the mean flap deflection is largely agnostic to mass ratio at a given stiffness (\emph{c.f.,}, Sec. \ref{parametricstudy}). Based on these two statements, for a fixed stiffness, the mass ratio that yields a flap deflection that is lower than the (nearly) constant mean for the majority of the TEV-dominant portion of the lift cycle, will yield a larger SLEV strength during this portion.
In other words, the SLEV is enhanced when the phase of mass-dependent flap dynamics is $\phase \approx -\pi/2$, while $\phase \approx \pi/2$ is least conducive to the growth of the SLEV. Now, the phase of flap oscillations as well as the mean SLEV circulation strength, $\slev$, in the TEV-portion of the lift cycle for the $\twenty$ location, and varying flap parameters are plotted in Fig.~\ref{fs_dynamic_phase} and Fig.~\ref{fs_dynamic_trends1_3}, respectively. It can be seen in Fig.~\ref{fs_dynamic_phase} that the lower masses are associated with a phase (away from $-\pi/2$) that is detrimental to SLEV growth compared to the larger masses. Accordingly, the mean $\slev$ plotted in Fig.~\ref{fs_dynamic_trends1_3} is lowest for the lower-mass flaps. The effect of reduced $\slev$ is manifested as reduced pre-flap suction lift for lower-mass flaps plotted in Fig.~\ref{fs_dynamic_trends1_4}.  

Our prior discussion of the TEV-portion of the lift cycle clarified the detrimental role of light flaps at both the upstream and downstream flap locations. It is also clear from the associated figures (\emph{e.g.}, the pre-flap stress in Fig.~\ref{fs_dynamic_trends1_2}) that intermediate masses yielded the highest lift benefits. To clarify the beneficial role of intermediate mass values, we now analyze the mechanisms in the LEV-dominant portion of the cycle, for times of approximately $\periodt\in [0, 0.2] \cup [0.8,1]$. 
In the LEV-portion of the lift cycle, the LEV is the dominant contributor to lift as evidenced by the peak lift attained in the LEV-portion in Fig.~\ref{generalcl}. Whether the LEV is enhanced or deteriorated is dictated by the flap-LEV interactions, which in turn affects performance. Therefore, we next focus on these flap-LEV interactions and  demonstrate that the dynamics of these interactions in the LEV-portion can be divided into two parts---formation of the primary LEV (PLEV) from the SLEV and advection of the PLEV into the airfoil wake. 

\begin{figure}
    \centering
    \includegraphics[scale=1]{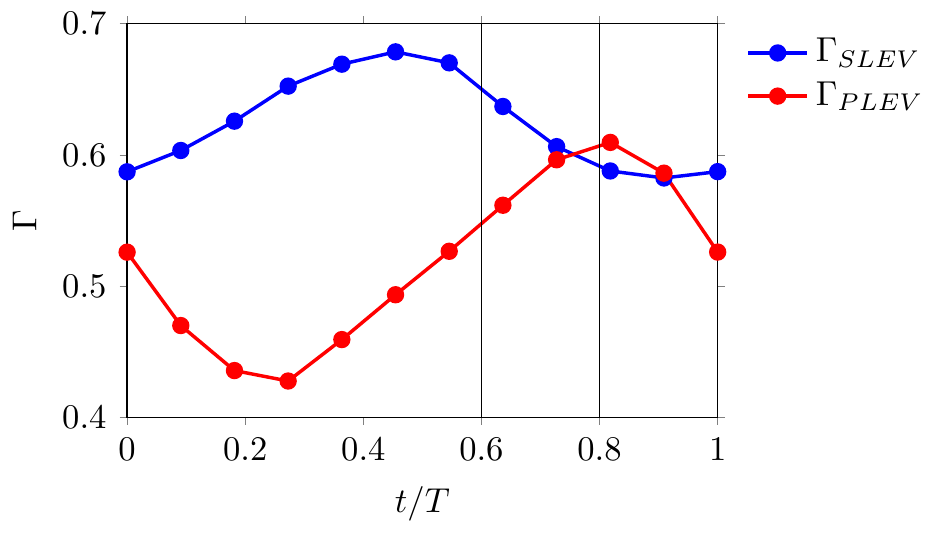}
    \caption{Vortex circulation strengths in one time period of the lift cycle for $\fifty$ location, $\mass=1.875$ and $\stiff=0.0015$. The region between the black vertical lines, $\periodt\in[0.6,0.8]$, denotes the SLEV-to-PLEV transfer process.}
    \label{slevplev}
\end{figure}

In the first part, the PLEV is formed when a portion of the circulating flow in the SLEV trapped upstream of the flap advects downstream. This process can be visualized from the vorticity contours in Fig.~\ref{fs}, where in going from $\periodt=0.55$ in Fig.~\ref{fs3} to $\periodt=0.82$ in Fig.~\ref{fs4}, the SLEV (upstream of the flap) is found to decrease with a simultaneous formation of the PLEV (downstream of the flap). We term this process as the SLEV-to-PLEV transfer and it occurs roughly towards the end of the TEV-portion at $\periodt\in[0.6,0.8]$. To make this process evident, the SLEV ($\slev$) and PLEV ($\plev$) circulation strengths in one lift cycle for the case of $\mass=1.875$, $\stiff=0.0015$ and $\fifty$ location are plotted in Fig.~\ref{slevplev}. It can be seen that in $\periodt\in[0.6,0.8]$ (highlighted by the vertical black lines), $\slev$ decreases while $\plev$ increases. Now, the flap motion that synchronizes favorably with the SLEV-to-PLEV transfer process is expected to enhance the lift-conducive PLEV. For example, the flap with $\mass=1.875$ oscillates upstream (increasing deflection) during the SLEV-to-PLEV transfer process (see the green line in Fig.~\ref{deflfs} in $\periodt\in[0.6,0.8]$). Therefore, the decreasing circulation region for the SLEV (due to increasing flap deflection) is synchronized with the decreasing SLEV strength (\emph{c.f.} Fig.~\ref{slevplev}) as the SLEV is transferred downstream to form the PLEV. This synchronization makes the SLEV-to-PLEV transfer process more efficient than for cases where such a synchronization is not present, such as the (quasi-static) larger mass ratio case of $\mass=18.75$. In fact in the latter setting, the flap roughly acts as a rigid barrier to the efficient advection of the LEV. 

\begin{figure}
\centering
\begin{subfigure}[t]{0.45\textwidth}
\centering
\includegraphics[scale=1]{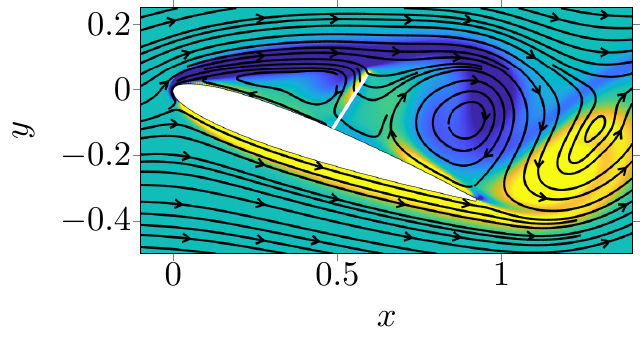}
\caption{Increased PLEV displacement at $\mass=0.01875$.}
\label{fs_dynamic_levdisplace1}
\end{subfigure}
\begin{subfigure}[t]{0.4\textwidth}
\centering
\includegraphics[scale=1]{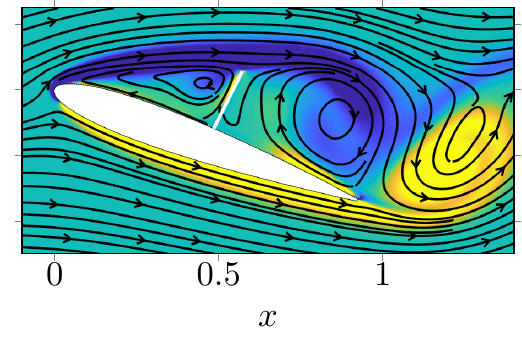}
\caption{Decreased PLEV displacement at $\mass=1.875$.}
\label{fs_dynamic_levdisplace2}
\end{subfigure}
\caption{Demonstration of downstream displacement of PLEV facilitated by the downward oscillation of the flap via vorticity contours and superimposed streamlines at $\periodt=0.91$ for $\stiff=0.0015$ and $\fifty$ location.}
\label{fs_dynamic_levdisplace}
\end{figure}

The second part of the flap-LEV interaction is concerned with the speed of advection of the PLEV once the SLEV-to-PLEV transfer is completed. The flap that quickly displaces the PLEV downstream is expected to be more detrimental to lift than the flap that slows the PLEV-advection process, allowing the associated low-pressure benefits to persist for longer. More specifically, the flap that oscillates downwards (decreasing deflection) with higher velocity in $\periodt=[0.8,1]$ will impart a higher momentum to displace the PLEV downstream. For instance in Fig.~\ref{deflfs}, consider the lower-mass flaps ($\mass<1.875$) which have a phase of $\phase\approx\pi$ (\emph{c.f.} Fig.~\ref{fs_dynamic_phase}) and accordingly oscillate downwards in $\periodt=[0.8,1]$. To visualize the effect of this flap motion on the PLEV, a vorticity contour at a representative time instant of $\periodt=0.91$, $\mass=0.01875$, $\stiff=0.0015$ and $\fifty$ location is plotted in Fig.~\ref{fs_dynamic_levdisplace1}. Here, the streamlines emerging through the downward-oscillating flap displaces the PLEV further downstream. For comparison, we have also plotted a vorticity contour for the corresponding intermediate mass $\mass=1.875$ case in Fig.~\ref{fs_dynamic_levdisplace2}. It can be observed that the $\mass=1.875$ flap does not accelerate the PLEV advection process in Fig.~\ref{fs_dynamic_levdisplace2} as much as the lower-mass flap counterpart due to the its delayed phase (\emph{c.f.} Fig.~\ref{fs_dynamic_phase}) in addition to its lower flap velocity. 

\begin{figure}
\begin{adjustwidth}{}{0.3cm} 
\centering
\begin{subfigure}[t]{0.4\textwidth}
\centering
\includegraphics[scale=1]{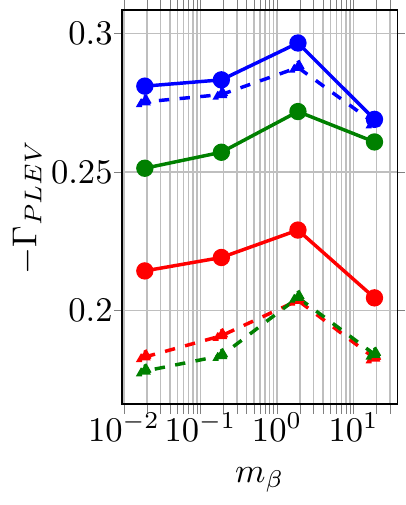}
\caption{\centering Mean negative PLEV strength.}
\label{fs_dynamic_trends2_1}
\end{subfigure}
\begin{subfigure}[t]{0.45\textwidth}
\centering
\includegraphics[scale=1]{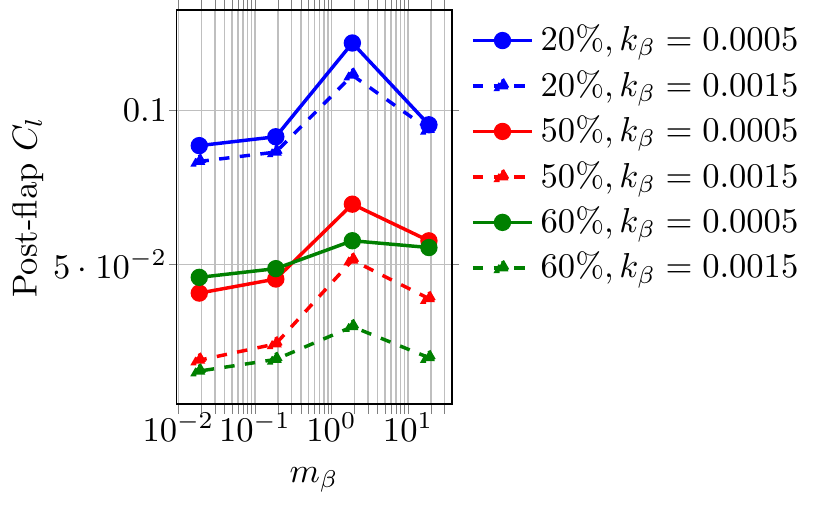}
\caption{\centering Mean post-flap suction lift.}
\label{fs_dynamic_trends2_2}
\end{subfigure}
\end{adjustwidth}
\caption{Plots of various physical quantities characteristic of dynamic behavior in the flap-shear interaction regime averaged or integrated over the LEV-dominant portion of the lift cycle for different locations, mass ratios and stiffness.}
\label{fs_dynamic_trends2}
\end{figure}

To generalize the effect of phase on the PLEV-advection process for all mass ratios, flap locations and stiffness values of $\stiff=0.0005$ and $\stiff=0.0015$, we note from Fig.~\ref{fs_dynamic_phase} that the lower flap masses indeed yield a phase closer to $\pi$ than the phase of $\phase\approx-\pi/2$ of the larger-mass counterparts. This implies that the lower-mass flaps accelerate the PLEV advection and yield associated lift detriments.  
Finally, to quantify the cumulative effect of the SLEV-to-PLEV transfer and the PLEV advection for these cases, we plot the PLEV strength ($\plev$) and post-flap suction lift averaged over the LEV-portion of the lift cycle in Fig.~\ref{fs_dynamic_trends2}. Post-flap suction lift is the lift contribution by the upper surface of the airfoil downstream of the flap (\emph{i.e.} after the discontinuity in the $\Cp$ plot). We plot the post-flap lift since the effect of the PLEV is prominent in the region downstream of the flap. From Fig.~\ref{fs_dynamic_trends2_1}, the largest $\plev$ is attained at an intermediate mass ratio of $\mass=1.875$ which also yields the highest post-flap suction lift in Fig.~\ref{fs_dynamic_trends2_2}. This intermediate mass ratio is found to be maximal because (a) an efficient SLEV-to-PLEV transfer is enabled by the moderate flap oscillations of $\mass=1.875$, which is non-existent in the quasi-static larger-mass case of $\mass=18.75$ and (b) the advection of the PLEV is relatively decelerated as compared to the lower-mass cases of $\mass<1.875$ due to a favorable phase of $\phase=-\pi/2$ of flap oscillations and lower angular velocity.

To summarize the effects of dynamic behavior of the flap on flow physics in the flap-shear regime, the mass ratio of $\mass=1.875$ is observed to provide the best performance improvement with the lowest mass cases providing the least benefits in the flap-shear regime owing to several fluid-structure interaction mechanisms. For the downstream flap locations of $\fifty$ and $\sixty$, this mass ratio yields a phase and amplitude of flap dynamics that mitigate (disadvantageous) upstream-propagating reverse flow during the TEV-portion of the cycle and facilitate a more efficient SLEV-to-PLEV transfer as well as a slower PLEV advection during the LEV-portion. For the upstream flap location of $\twenty$, the same benefits arise during the LEV-portion of the cycle, and for the TEV-portion the benefits come not from mitigating reverse flow (as in the downstream flap locations) but from maximizing the lift-producing SLEV. 
In the next section, we identify the lift enhancing mechanisms of the flap-vortex interaction regime and perform similar analysis of the effect of varying stiffness and mass ratio on aerodynamic performance.

\subsection{Flap-vortex interaction regime}
\label{flapvortex}

\begin{figure}
\begin{adjustwidth}{}{0.5cm} 
\centering
\begin{subfigure}[t]{0.283\textwidth}
\centering
\includegraphics[scale=1]{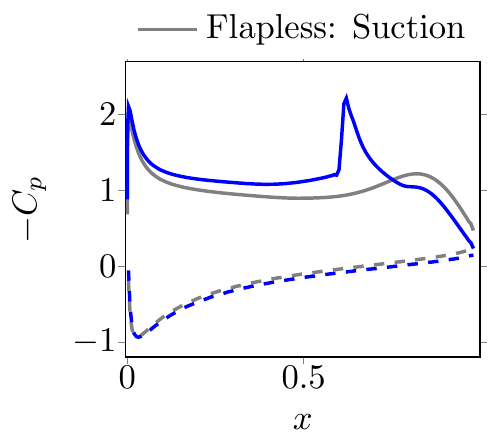}
\vspace{-0.7cm}
\caption{$t=0 \ T$ }
\label{fvcp1}
\end{subfigure}
\begin{subfigure}[t]{0.22\textwidth}
\centering
\includegraphics[scale=1]{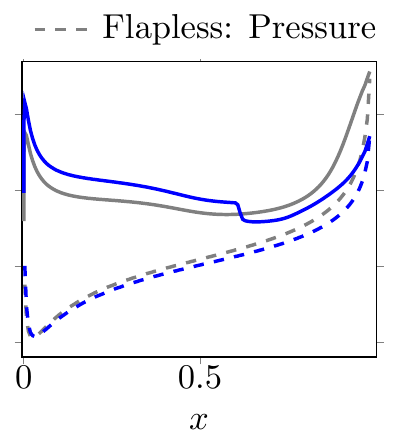}
\vspace{-0.7cm}
\caption{$t=0.27 \ T$ }
\label{fvcp2}
\end{subfigure}
\begin{subfigure}[t]{0.22\textwidth}
\centering
\includegraphics[scale=1]{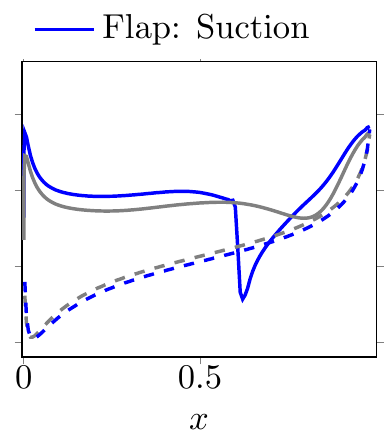}
\vspace{-0.7cm}
\caption{$t=0.55 \ T$ }
\label{fvcp3}
\end{subfigure}
\begin{subfigure}[t]{0.22\textwidth}
\centering
\includegraphics[scale=1]{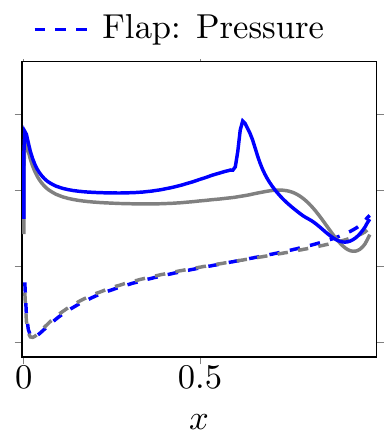}
\vspace{-0.7cm}
\caption{$t=0.82 \ T$ }
\label{fvcp4}
\end{subfigure}
\centering
\begin{subfigure}[t]{0.283\textwidth}
\centering
\includegraphics[scale=1]{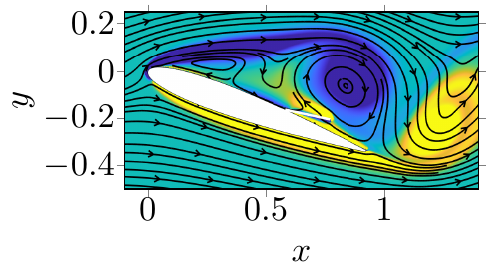}
\vspace{-0.7cm}
\caption{$t=0 \ T$ }
\label{fv1}
\end{subfigure}
\begin{subfigure}[t]{0.22\textwidth}
\centering
\includegraphics[scale=1]{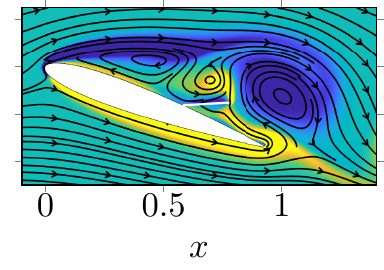}
\vspace{-0.7cm}
\caption{$t=0.27 \ T$ }
\label{fv2}
\end{subfigure}
\begin{subfigure}[t]{0.22\textwidth}
\centering
\includegraphics[scale=1]{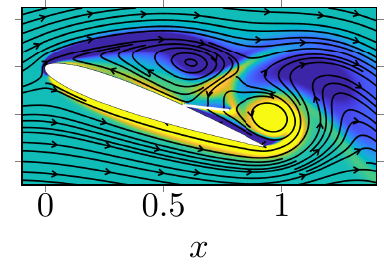}
\vspace{-0.7cm}
\caption{$t=0.55 \ T$ }
\label{fv3}
\end{subfigure}
\begin{subfigure}[t]{0.22\textwidth}
\centering
\includegraphics[scale=1]{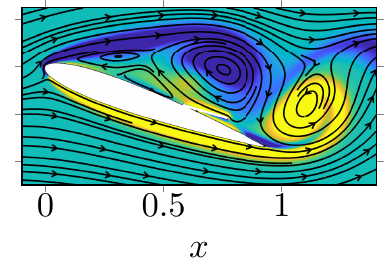}
\vspace{-0.7cm}
\caption{$t=0.82 \ T$ }
\label{fv4}
\end{subfigure}
\end{adjustwidth}
\caption{$\Cp$ distribution (top row) demonstrating the pressure dam effect and vorticity contours with superimposed streamlines (bottom row) demonstrating barrier to upstream propagation of reverse flow imposed by the flap and LEV for a representative case of $\stiff=0.015$, $\mass=1.875$ and $\sixty$ location in the flap-vortex interaction regime at four time instants in a lift cycle. $\Cp$ distributions of the baseline flap-less case are also provided for comparison in the top row. Refer Fig.~\ref{generalvorticity5}--\ref{generalvorticity8} for vorticity contours of the baseline case.}
\label{fv}
\end{figure}

We consider a representative case of the flap fixed at $\sixty$ location, $\stiff=0.015$ and $\mass=1.875$ to discuss the lift enhancing mechanisms in this regime. For this case, the $\Cp$ distribution on the airfoil surface and vorticity contours at four time instants in one time period of the lift cycle are plotted in Fig.~\ref{fvcp1}--\ref{fvcp4} and \ref{fv1}--\ref{fv4}, respectively. There are instances where a pressure dam effect, similar to that in the flap-shear interaction regime are observed (\emph{c.f.}, the low pressure zone maintained upstream of the flap on the suction surface followed by a step discontinuity in Fig.~\ref{fvcp2} with any of the $C_p$ plots from Fig.~\ref{fs}). The vorticity contour at the associated time instant, Fig.~\ref{fv2}, also reveals a similar blockage of the upstream-propagating reverse flow induced by the TEV. However, the $C_p$ plots and vorticity snapshots at the other time instances indicate a distinct mechanism from the flap-shear regime, where flap oscillations occur nearer to the airfoil surface and have more direct interaction with the vortex-shedding process. This observation justifies the k-means categorical distinction between this regime and the flap-shear regime, and also suggests why this flap-vortex regime is not observed for upstream flap locations where such flap-vortex interplay is not possible. We characterize the distinct mechanisms for this new regime in this section.

We observe that the flap fully blocks the TEV-induced reverse flow during the initial growth of the TEV at $\periodt=0.27$ in Fig.~\ref{fv2}, despite its reduced deflection. However, as the TEV further grows in strength, the reduced flap deflection is unable to entirely block the reverse flow at $\periodt=0.55$, as interpreted from the streamlines originating from the TEV and propagating above the flap in Fig.~\ref{fv3}. While this reduced flap deflection is not conducive to the blocking of TEV-induced reverse flow, it allows the LEV to efficiently advect downstream through the region between the shear layer and the flap with little hindrance. At these instances when the TEV strength is near-maximum, the high-momentum carrying LEV which has approached near the flap provides additional blocking of the TEV-induced reverse flow. Therefore, the upstream-propagating streamlines above the flap are restricted only to the flap hinge and do not progress towards the leading edge as seen in Fig.~\ref{fv3}. This type of aggregate blocking by the flap and the LEV, which is only observed when the flap strongly interacts with the vortex shedding, yields a pressure-dam like effect.

In the following sections, analogous to the flap-shear regime, we first present the dominant effects of the mean flap location, which is largely driven by a balance of the mean flow forces on the flap and the internal flap stress determined by stiffness. We then identify how dynamics about this mean state, encoded in flap mass, further modulate the effects of the flap on flow and performance.

\subsubsection{Static behavior: effect of varying stiffness}
\label{fv_static}

\begin{figure}
\centering
\begin{subfigure}[t]{0.45\textwidth}
\centering
\includegraphics[scale=1]{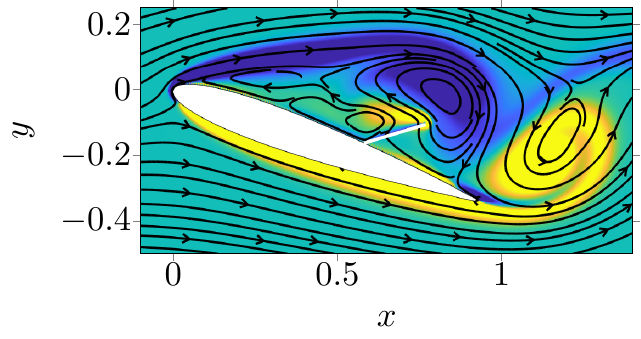}
\caption{LEV distortion at $\stiff=0.005$.}
\label{fv_tlev1}
\end{subfigure}
\begin{subfigure}[t]{0.4\textwidth}
\centering
\includegraphics[scale=1]{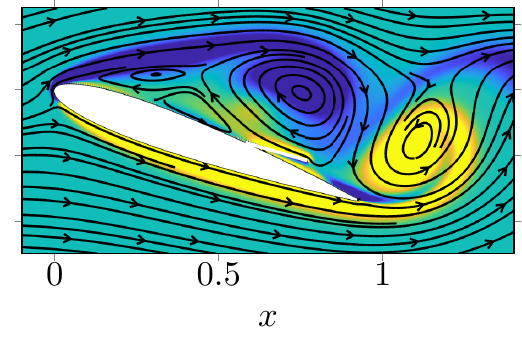}
\caption{LEV enhancement at $\stiff=0.015$.}
\label{fv_tlev2}
\end{subfigure}
\caption{Demonstration of LEV deformation as it advects above the flap at $\periodt=0.82$ for $\mass=1.875$ and $\sixty$ location via vorticity contours.}
\label{fv_tlev}
\end{figure}

Similar to the static-behavior analysis of the flap-shear interaction regime in Sec.~\ref{fs_static}, in this section, we discuss the effect of stiffness-dependent mean flap configuration on lift performance in the flap-vortex regime through the lens of higher mass ratios cases of $\mass>1$.
Within the flap-vortex regime for large mass ratios, the maximal lift is attained around $\stiff=0.01$--$0.015$ (\emph{c.f.} Fig.~\ref{parametric60}). Maximal lift occurs at this stiffness value due to an optimal balance between two dominant competing effects---efficient advection of LEV and the magnitude of flow velocity in the shear layer above the suction surface of the airfoil---which are analyzed in this section. For lower mass ratios, $\mass<1$, a significant jump in performance as compared to $\mass>1$ is observed in the flap-vortex regime, though the peak performance still corresponds to a similar stiffness of $\stiff=0.015$ (\emph{c.f.} Fig.~\ref{parametric60}). This implies that while the static effects described in this section are applicable for the lower-mass cases, there are additional dynamical effects of the flap-fluid interactions that are beneficial to lift. These flap dynamics and associated lift improvements are described in the next section.

As mentioned earlier, the smaller flap deflection allows the LEV to advect through the region between the shear layer and the flap. This spatial gap can either favorably or adversely influence the LEV-advection process and therefore, the LEV strength, $\lev$. For demonstration, consider the vorticity contours plotted in Fig.~\ref{fv_tlev} for the cases of a relatively highly ($\stiff=0.005$) and weakly ($\stiff=0.015$) deployed flap with $\mass=1.875$ and $\sixty$ location at a representative time instant of $\periodt=0.82$ which corresponds to the LEV advecting above the flap. The $\stiff=0.005$ flap in Fig.~\ref{fv_tlev1} is largely deployed such that the LEV has to maneuver above and around the flap through the relatively smaller region as compared to $\stiff=0.015$ in Fig.~\ref{fv_tlev2}. The lack of flow passage width in the $\stiff=0.005$ case distorts the advecting LEV as seen in Fig.~\ref{fv_tlev1} while a sufficiently large but not excessively wide region in Fig.~\ref{fv_tlev2} for $\stiff=0.015$ effectively squeezes and enhances the rolling-up of the LEV. Therefore, the LEV strength is respectively worsened and enhanced for $\stiff=0.005$ and $\stiff=0.015$. 
Now, the mean $\lev$ for the various cases of stiffness and mass ratio at $\sixty$ location is plotted in Fig.~\ref{fv_static_trends1_1}, where the methodology for quantifying $\lev$ is described in Appendix~\ref{quantvortstrength}. To correlate the effect of $\lev$ on performance, the mean lift generated by the suction surface of the airfoil is also plotted in Fig.~\ref{fv_static_trends1_3}. From Fig.~\ref{fv_static_trends1_1} it can be seen that $\lev$ increases as the stiffness is increased from $\stiff=0.005$ to $\stiff=0.015$ for all mass ratios. This increase in $\lev$ contributes to the initial rise in the suction surface lift in Fig.~\ref{fv_static_trends1_3}. As the stiffness is further increased and yields a qualitatively different  regime than the flap-vortex regime, $\lev$ begins to decrease, as the weakly deployed flap approaches the baseline case. This reduction in $\lev$ and associated drop in lift can be observed for the $\stiff=0.02$, $\mass=18.75$ case (dashed green line) in Fig.~\ref{fv_static_trends1_1}, which according to Fig.~\ref{parametric60} belongs to the non-interactive regime.

\begin{figure}
\begin{adjustwidth}{}{0.3cm} 
\centering
\begin{subfigure}[t]{0.3\textwidth}
\centering
\includegraphics[scale=1]{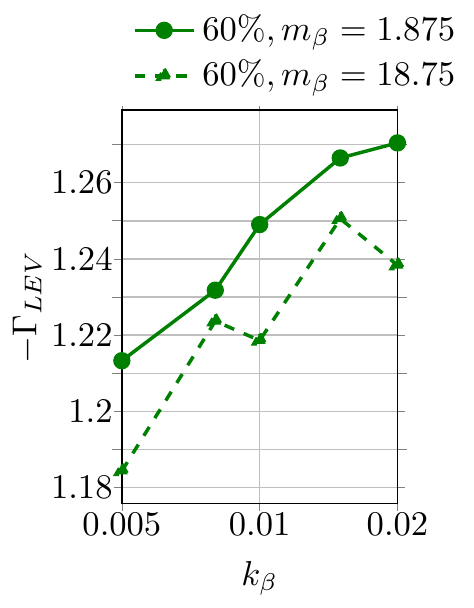}
\caption{\centering Mean negative LEV strength.}
\label{fv_static_trends1_1}
\end{subfigure}
\begin{subfigure}[t]{0.3\textwidth}
\centering
\includegraphics[scale=1]{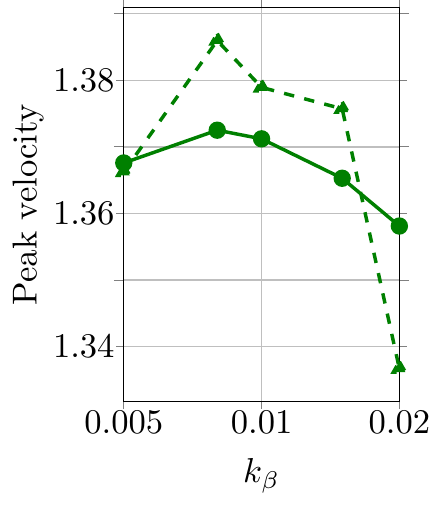}
\caption{\centering Mean peak velocity attained at $x=0.2$.}
\label{fv_static_trends1_2}
\end{subfigure}
\begin{subfigure}[t]{0.3\textwidth}
\centering
\includegraphics[scale=1]{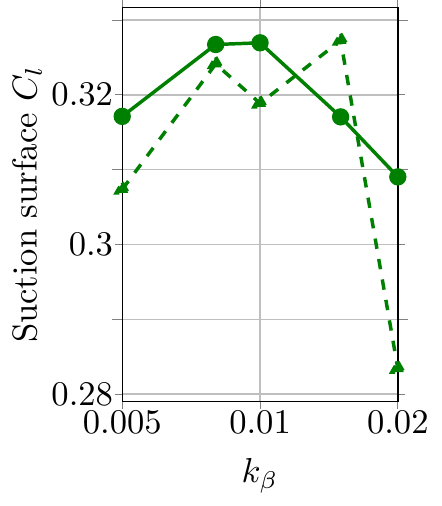}
\caption{\centering Mean suction lift.}
\label{fv_static_trends1_3}
\end{subfigure}
\end{adjustwidth}
\caption{Plots of various physical quantities characteristic of static behavior in the flap-vortex regime averaged over a time period for $\sixty$ location, $\mass>1$ and varying stiffness.}
\label{fv_static_trends1}
\end{figure}

We also note that the increasing LEV strength with increasing stiffness, however, does not always continue to improve performance---the suction surface lift begins to drop after $\stiff=0.01$--$0.015$ (\emph{c.f.} Fig.~\ref{fv_static_trends1_3}). Similar to the static-behavior analysis in Sec.~\ref{fs_static}, this reversal of performance trend is associated with the increasing gap between the flap and the shear layer as the stiffness is increased. Recalling from Sec.~\ref{fs_static}, a large gap resulted in a reduced barrier to upstream-propagating reverse flow, thereby permitting a further growth of the TEV. The resulting reduced flow velocity in the shear layer above the suction surface of the airfoil near the airfoil leading edge is detrimental to maintaining a lower pressure zone and maximizing lift upstream of the flap. As a proxy for the flow velocity within the shear layer, the mean of peak $x$-component of flow velocity attained at a representative location of $x=0.2$ is plotted in Fig.~\ref{fv_static_trends1_2}. The peak velocity is taken as the maximal value across all $y$-locations, at $x=0.2$, which is then averaged over a lift cycle. A reduction in the peak velocity for stiffness values beyond $\stiff=0.008$ can be observed, which in turn contributes to the reduction in the suction surface lift in Fig.~\ref{fv_static_trends1_3}.

\begin{figure}
\centering
\begin{subfigure}[t]{0.55\textwidth}
\centering
\includegraphics[scale=1]{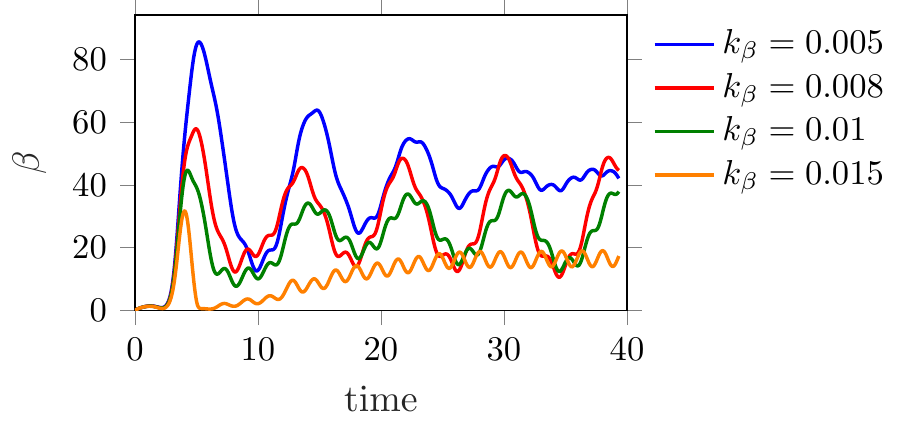}
\caption{Flap deflection.}
\label{fv_static_defl1}
\end{subfigure}
\begin{subfigure}[t]{0.4\textwidth}
\centering
\includegraphics[scale=1]{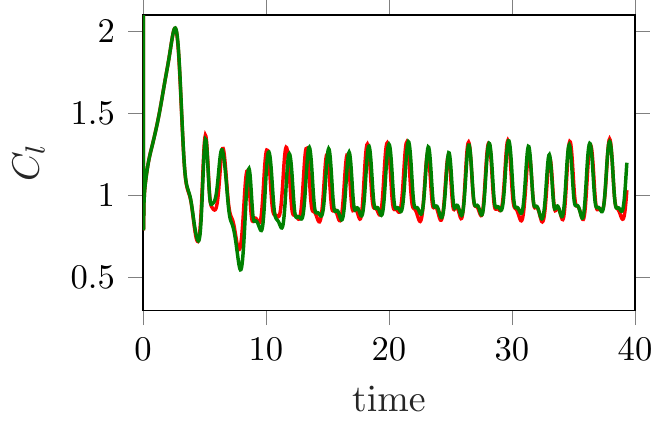}
\caption{Lift coefficient.}
\label{fv_static_defl2}
\end{subfigure}
\caption{Demonstration of the presence of dual frequencies in the flap deflection signal only minimally affecting the frequency of the airfoil lift signal for select stiffness values at the largest mass ratio of $\mass=18.75$ and $\sixty$ location.}
\label{fv_static_defl}
\end{figure}

Finally, we remark that the $\stiff=0.008$ and $\stiff=0.01$ cases at the largest mass case of $\mass=18.75$ exhibit certain inconsistencies, which are manifested as slightly fluctuated variations in $\lev$ and suction surface lift contribution in Fig.~\ref{fv_static_trends1_1} and \ref{fv_static_trends1_3}, respectively. However, these fluctuations are not present in the performance plots of Fig. \ref{parametric60}. In order to understand the cause for these inconsistencies in Fig.~\ref{fv_static_trends1}, consider the plots of flap deflection over many convective time units for the largest mass ratio, $\mass=18.75$, cases in the flap-vortex regime in Fig.~\ref{fv_static_defl1}. It can be observed that the $\stiff=0.008$ and $\stiff=0.01$ cases undergo large-amplitude low-frequency oscillations with superimposed small-amplitude high-frequency oscillations. Here, the high- and low-frequency components correspond to the frequency of vortex shedding and vacuum-scaled natural frequency of the large mass flap, respectively. 
To understand the effect of dual frequencies in flap dynamics on the airfoil lift response, the lift signals for these cases are plotted in Fig.~\ref{fv_static_defl2}. It can be seen that the lift dynamics also exhibit low-frequency (albeit small-amplitude) oscillations enveloping the more dominant high-frequency signal. The presence of this low-frequency envelope makes the high-(vortex-shedding)-frequency lift dynamics notably aperiodic, which results in the conflicting consequences in Fig.~\ref{fv_static_trends1}. Specifically, while the mean quantities in the performance plots of Fig. \ref{parametric60} are evaluated across several lift cycles, in the more detailed analysis presented in Fig.~\ref{fv_static_trends1}, the mean quantities are evaluated in just one lift cycle. For all the cases except $\mass=18.75$, consistent results are attained regardless of whether the mean quantities are evaluated in a single time period of lift or multiple lift cycles, since the lift signal is largely periodic. However, due to the aperiodic nature of $\mass=18.75$ cases, the mean quantities in one lift cycle can non-negligibly vary from the mean evaluated across several cycles. 

We emphasize that the dual frequencies in the deflection and lift signals are present only in the large mass ratio cases of $\mass=18.75$. This is because the large mass of the flap denies the flap to respond quickly to the vortex-shedding process prompting the flap to slowly oscillate at its lower natural frequency. However, not all $\mass=18.75$ flaps exhibit such a nature---dual frequencies are mostly observed only for $\stiff=0.008$ and $\stiff=0.01$ while the adjacent stiffness cases of $\stiff=0.005$ and $\stiff=0.015$ undergo largely small-amplitude, high-frequency oscillations with low-frequency oscillations being decayed as observed from Fig.~\ref{fv_static_defl1}. This is because, in the former cases of $\stiff=0.008$ and $\stiff=0.01$, the mean flap configurations are such that the flaps are strongly positioned within the vortex-shedding region. On the other hand, in the latter cases of $\stiff=0.005$ and $\stiff=0.015$, the flaps oscillate above and below the core vortex-shedding region, respectively, thereby eluding the strong aerodynamic fluctuations of vortex shedding. In any case, these effects are not addressed in detail here, as the single-cycle analysis is appropriate for the smaller mass ratios and, even for the higher masses, sufficient to identify the primary mechanisms driving lift improvements/detriments of focus here. 

In summary, the competing effects of (advantageous) increasing LEV strength and (disadvantageous) decreasing flow velocity in the shear layer above the suction surface near the leading edge as the stiffness is increased, results in maximal lift being attained around a stiffness of $\stiff=0.01$--$0.015$ in the flap-vortex interaction regime. In this section, we analyzed the lift enhancing mechanisms associated to the mean flap deflection through the lens of higher mass cases of $\mass>1$ due to their better approximation towards a quasi-static response. In the next section, we will investigate how flap dynamics, largely encoded through mass ratio, plays a role in favorably or adversely affecting these mechanisms. 

\subsubsection{Dynamic behavior: effect of varying inertia}
\label{fv_dynamic}

\begin{figure}
    \centering
    \includegraphics[scale=1]{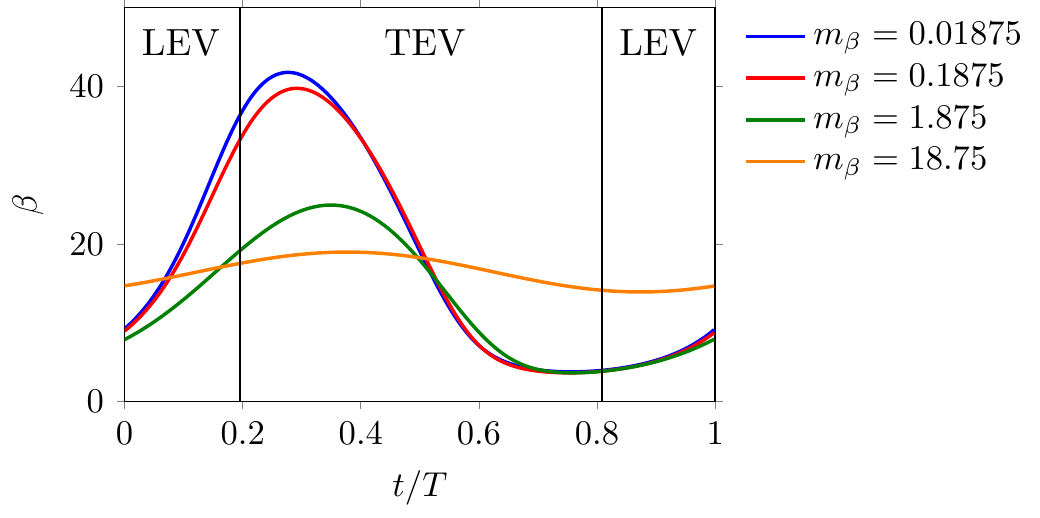}
    \caption{Flap deflection in one time period of the lift cycle for $\sixty$ location and $\stiff=0.015$. The time period is divided into two portions dominated by the LEV and TEV where the time instants of demarcation are $\periodt\approx 0.2$ and $\periodt\approx 0.8$.}
    \label{deflfv}
\end{figure}

Similar to the flap-shear regime, the primary effect of flap mass is to modify the flap amplitude and phase of oscillations in the flap-vortex regime. For demonstration, Fig.~\ref{deflfv} plots flap deflections in one time period of the lift cycle for several mass ratios, $\stiff=0.015$ and $\sixty$ location. Distinct phase behavior (visible by the shifted time instants of peak flap deflection) and a reducing amplitude with increasing flap mass can be clearly observed. In order to understand the effects of mass-dependent varying amplitude and phase on the flap-fluid interactions and associated performance, we divide the time period of lift cycle into two parts---the LEV- and TEV-dominant portions---as described in Sec.~\ref{qualitative}. Also, the phase is defined in the same manner as in Sec.~\ref{fs_dynamic} and Appendix~\ref{quantvortstrength}. 

\begin{figure}
    \centering
    \includegraphics[scale=1]{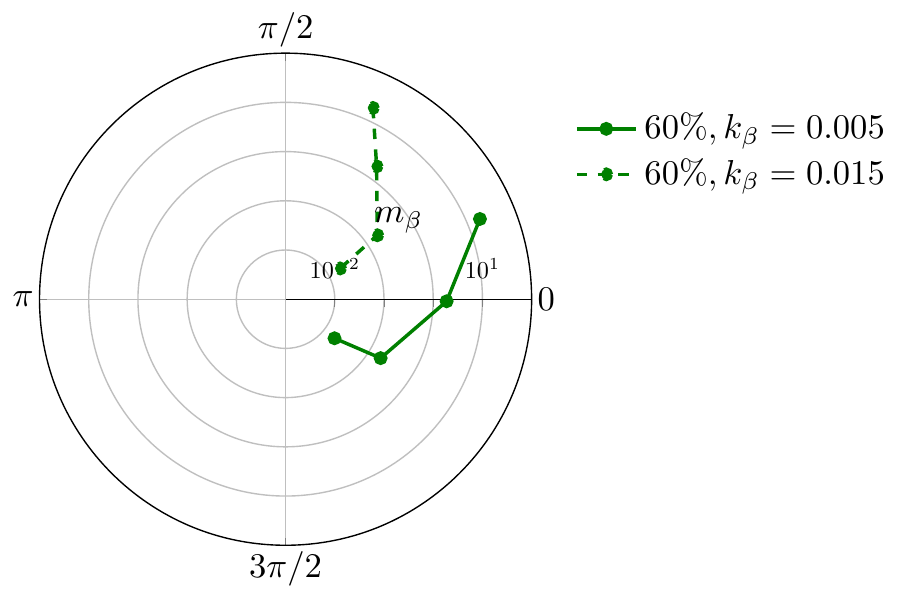}
    \caption{Phase difference of flap oscillations v/s mass ratio (along the radial axis) for
various stiffness at $\sixty$ location.}
    \label{fv_dynamic_phase}
\end{figure}

In the flap-vortex regime, the ability of the flap to constructively interact with the LEV while countering the TEV and TEV-induced reverse flow dictates whether there are performance benefits or detriments. This ability of the flap to enhance the LEV and mitigate the TEV is linked to the timing (or phase) and amplitude of the flap dynamics relative to the TEV-formation and LEV-advection process. To motivate the importance of timing, we focus on the dynamics of the lowest mass flap, $\mass=0.01875$, in Fig.~\ref{deflfv} and associated interactions with the LEV and the TEV in Fig.~\ref{fv}. During the TEV-portion, the flap oscillates downwards as seen from Fig.~\ref{deflfv}. This motion is in response to the downstream advection of a newly shed LEV and can be visualized in going from Fig.~\ref{fv2} to \ref{fv4}. Once the LEV has advected beyond the flap, the close vicinity of the LEV to the flap, its low pressure and the velocity induced by the clockwise rotating flow pulls the flap upwards in the LEV-portion as seen in going from Fig.~\ref{fv4} to \ref{fv1}. This upwards flap motion in the LEV-portion can also be observed in Fig.~\ref{deflfv}. This type of downwards and upwards oscillating flap in the TEV- and LEV-portions, respectively, correspond to a phase of $\phase\approx0$ (based on the previous definition of phase) and enhances the performance in two ways. Firstly, the downward flap motion during the TEV-portion allows the LEV to efficiently advect downstream without hindrance from the flap. Secondly, this downward motion also counters the upstream-propagating reverse flow induced by the TEV and therefore mitigates the growth of the TEV. Motivated by the observed importance of timing of flap oscillations, we now probe the phase for all mass ratios, $\sixty$ location and stiffness values of $\stiff=0.005$ and $\stiff=0.015$ in Fig.~\ref{fv_dynamic_phase}. The figure demonstrates that lower-mass flaps have a favorable phase closer to zero than the larger-mass flaps, specifically when $\stiff=0.015$, implying that the flap oscillations of lower-mass flaps enhance LEV advection while mitigating TEV-induced reverse flow.


\begin{figure}
\centering
\begin{subfigure}[t]{0.45\textwidth}
\centering
\includegraphics[scale=1]{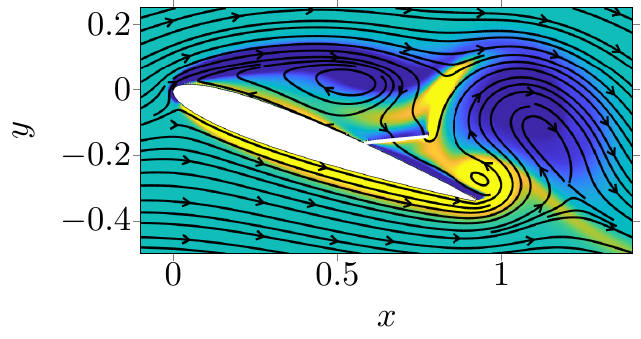}
\caption{Increased roll-up of LEV and TEV mitigation at $\mass=0.01875$.}
\label{fv_flapstream1}
\end{subfigure}
\begin{subfigure}[t]{0.45\textwidth}
\centering
\includegraphics[scale=1]{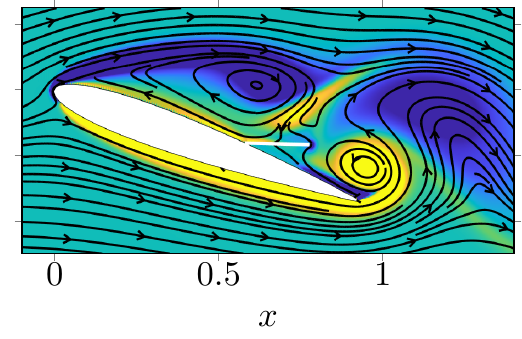}
\caption{Decreased TEV mitigation at $\mass=18.75$.}
\label{fv_flapstream2}
\end{subfigure}
\caption{Demonstration of the mitigation of TEV and roll-up enhancement of LEV facilitated by the downward-oscillating flap via vorticity contours with superimposed streamlines at $\periodt=0.45$ for $\stiff=0.015$ and $\sixty$ location.}
\label{fv_flapstream}
\end{figure}

We also note that in addition to this favorable phase of $\phase\approx0$, a large angular velocity of the flap further augments the LEV while mitigating the TEV. For demonstration, consider the vorticity contours at $\periodt=0.45$ plotted in Fig.~\ref{fv_flapstream} for the lowest ($\mass=0.01875$) and largest ($\mass=18.75$) mass ratios at $\stiff=0.015$ and $\sixty$ location. The lower-mass flap oscillates with a larger angular velocity and therefore during the downward flap motion in the TEV-portion, the flap strongly pulls the flow from the high-vorticity shear layer towards itself (interpreted from the streamlines passing through the flap in Fig.~\ref{fv_flapstream1}) and stimulates the rolling-up of the LEV. At the same time, the large angular velocity also counters the upstream-propagating TEV-induced reverse flow more strongly in Fig.~\ref{fv_flapstream1}. On the other hand, the quasi-static nature of the larger-mass flap in Fig.~\ref{fv_flapstream2} does not counter the TEV-induced reverse flow as much as the higher-velocity lower-mass flap resulting in increased TEV strengths which can be visualized in Fig.~\ref{fv_flapstream2}.

\begin{figure}
\begin{adjustwidth}{}{0.3cm} 
\centering
\begin{subfigure}[t]{0.3\textwidth}
\centering
\includegraphics[scale=1]{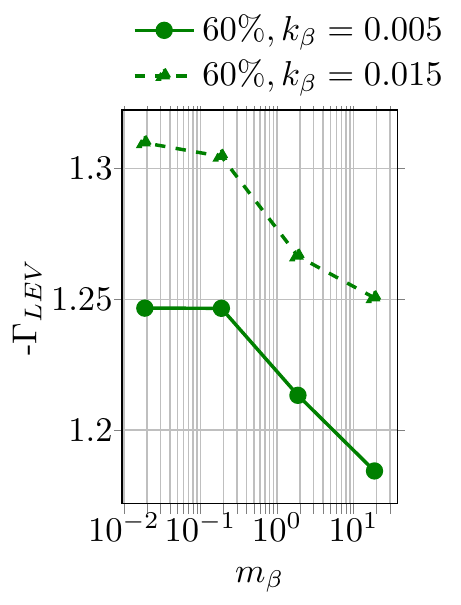}
\caption{\centering Mean negative LEV strength.}
\label{fv_dynamic_trends1_1}
\end{subfigure}
\begin{subfigure}[t]{0.3\textwidth}
\centering
\includegraphics[scale=1]{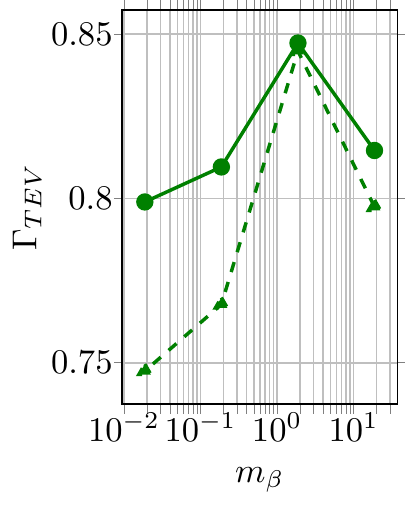}
\caption{\centering Mean TEV strength.}
\label{fv_dynamic_trends1_2}
\end{subfigure}
\begin{subfigure}[t]{0.3\textwidth}
\centering
\includegraphics[scale=1]{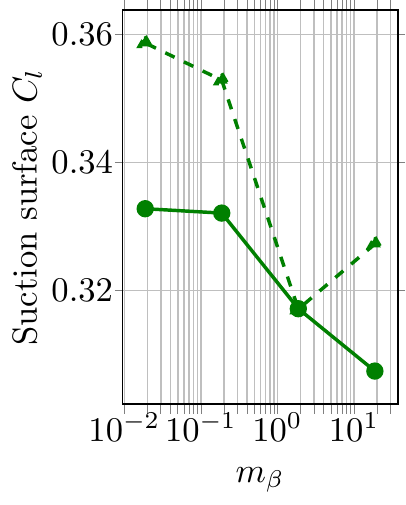}
\caption{\centering Mean suction lift.}
\label{fv_dynamic_trends1_3}
\end{subfigure}
\end{adjustwidth}
\caption{Plots of various physical quantities characteristic of dynamic behavior in the flap-vortex interaction regime averaged over one time period for $\sixty$ location, and varying mass ratio and stiffness.}
\label{fv_dynamic_trends1}
\end{figure}

For quantifying and generalizing the LEV enhancement and TEV mitigation capabilities of  different mass ratio flaps at $\sixty$ location and stiffness values of $\stiff=0.005$ and $\stiff=0.015$, we plot the LEV ($\lev$) and TEV ($\tev$) circulation strengths averaged over a lift cycle in Fig.~\ref{fv_dynamic_trends1_1} and \ref{fv_dynamic_trends1_2}, respectively. Refer to Appendix~\ref{quantvortstrength} for the methodology of evaluating $\lev$ and $\tev$. It can be seen that the lower-mass flaps have the highest $\lev$ and lowest $\tev$ for a range of stiffness values which in turn contribute to the corresponding highest suction surface lift, which is plotted in Fig.~\ref{fv_dynamic_trends1_3}.

In summary, the lowest inertia flap is observed to provide the best performance improvement for a fixed stiffness in the flap-vortex regime. This was primarily due to the downward flap motion during the TEV-dominant portion of the lift cycle coupled with large flap angular velocity which augmented the rolling-up of the LEV and countered the TEV-induced reverse flow the most. 


\subsubsection{Occurrence of resonance and connections to added mass}

\begin{figure}
\centering
\begin{subfigure}[t]{0.45\textwidth}
    \centering
    \includegraphics[scale=1]{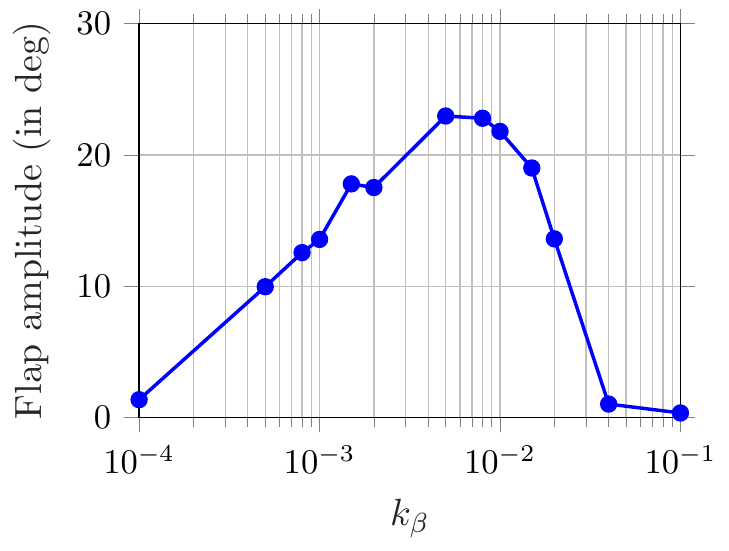}
    \caption{Amplitude of flap oscillations for a flap hinged at $\sixty$ location and $\mass=0.01875$.}
    \label{amp}
\end{subfigure}
\begin{subfigure}[t]{0.45\textwidth}
    \centering
    \includegraphics[scale=1]{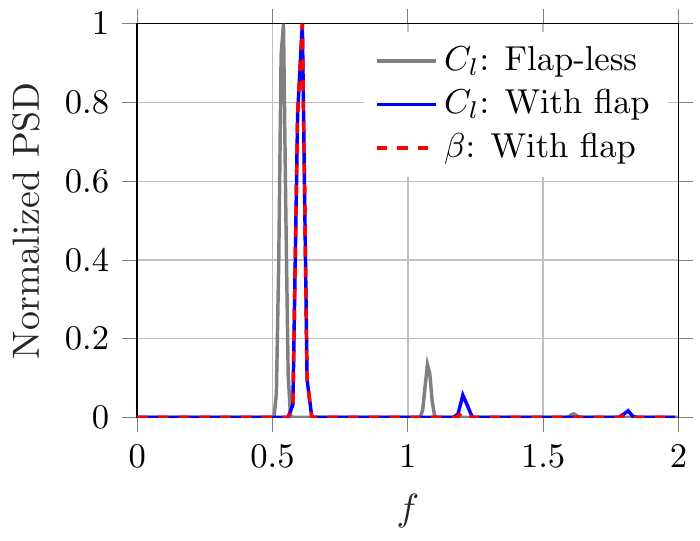}
    \caption{Normalized power spectral density (PSD) of the lift and flap deflection signals for the flap-less case and the case of flap hinged at $\sixty$ location, $\stiff=0.015$ and $\mass=0.01875$.}
    \label{psd}
\end{subfigure}
\caption{Identification of resonance in the flap-vortex regime via plots of amplitude of flap oscillations and frequency spectrum.}
\end{figure}

In this section, we investigate the occurrence of resonance in the flap dynamics and its effect on aerodynamic performance. In Fig.~\ref{amp}, the amplitude of flap oscillations is plotted for the flap at $\sixty$ location and lowest mass, $\mass=0.01875$. The appearance of a peak in amplitude (near a stiffness of $\stiff=0.005$--$0.008$ which belongs to the flap-vortex regime) suggests that resonance may be a feature of improved performance for this regime. To investigate this possibility, we probe the relationship between the flap dynamics in the FSI system and the forcing frequency, which in this case is the frequency of vortex shedding, $f_v$. 
To identify $f_v$, the normalized power spectral density (PSD) of the lift dynamics of the flap-less case is plotted in Fig.~\ref{psd}. From this figure, the frequency of vortex shedding of the flap-less case is found to be $f_v=0.54$. We also plot the PSD of the lift and flap dynamics for parameters near the performance peak hypothesized to be associated with resonance, $\stiff=0.008$, $\mass=0.01875$ and $\sixty$. It can be observed that both the lift and flap dynamics exhibit similar frequency spectra and more importantly, the presence of the flap only slightly increases the frequency of the airfoil-flap system as compared to the flap-less case. Therefore, we consider the baseline vortex-shedding frequency, $f_v=0.54$, for the following theoretical analysis. 

Now, the vacuum-scaled natural frequency of the flap, $f_\beta=1/2\pi\sqrt{\stiff/\inert}$, is $f_\beta = 4.50$. Here, $\inert$ is the moment of inertia of the flap as defined in Sec.~\ref{problemsetup}, which can be interchangeably used with mass ratio as $\inert = \mass (\flap/\chord)^4/3$, where $\flap$ is the dimensional length of the flap. This natural frequency of the flap is, however, greater than $f_v$ by an order of magnitude.  This mismatch of frequencies is because the vacuum-scaled frequency does not account for the effect of added mass in the FSI system. Accordingly, the FSI-scaled natural frequency of the flap is $f'_\beta = 1/2\pi\sqrt{\stiff/(\inert+\addedinert)}$, where $\addedinert$ denotes the non-dimensional added moment of inertia of the displaced fluid.
Recall from Sec.~\ref{problemsetup} that mass ratio of the flap is the ratio of the dimensional moment of inertia of the flap ($\inertdim$) to that of the displaced fluid ($\addedinertdim$), \emph{i.e.} $\mass\approx\inertdim/\addedinertdim$. Here, an approximation symbol is considered because while defining $\mass$ in Sec.~\ref{problemsetup}, an $\mathcal{O}(1)$ approximation was utilized. Therefore, the mass ratio associated to the added inertia of the displaced fluid is $\mass\approx\addedinertdim/\addedinertdim=1$ and based on the inertia-mass-ratio relationship described above, $\addedinert\approx 0.00053$. Now, on incorporating this added mass term, the FSI-scaled natural frequency of the above-mentioned case is $f'_\beta=0.61$, which is much closer to $f_v=0.54$ than $f_\beta$. Therefore, the above-mentioned large amplitude of flap oscillations in addition to the close frequency matching between vortex shedding and FSI-scaled natural frequency strongly suggests that resonance is a driver of improved performance in the flap-vortex regime. Interestingly, the added inertia term also indicates why the performance trends in Fig.~\ref{parametric60} and flap dynamics in Fig.~\ref{deflfv} of the lowest two mass flaps $\mass=0.01875$ and $\mass=0.1875$ are very similar: when accounting for the effects of added mass, which are prominent for low mass ratios, the aggregated masses of \emph{i.e.} $1.01875$ and $1.1875$, respectively, are relatively similar.

Finally, to understand the effect of resonance on performance, recall from Fig.~\ref{parametric} that the global maximum lift across all parameters occurred in the flap-vortex regime at $\stiff=0.015$, $\mass=0.01875$ and $\sixty$ location. Although the amplitude at $\stiff=0.015$ is lower than that at $\stiff=0.008$ from Fig.~\ref{amp}, it may in part be due to the physical limitations of the flap oscillating closer to the airfoil (due to larger stiffness \emph{i.e.} lower mean deflection) and not owed completely to the dynamical theory of the associated spring-mass system. In addition to these flap limitations, the fact that across a wide range of stiffness and inertia that spans over several orders of magnitude, stiffness values corresponding to maximum lift ($\stiff=0.015$) and amplitude ($\stiff=0.008$) are only separated by a factor of two indicates that maximum lift approximately occurs near resonance.

In summary, resonance was found to occur in the flap-vortex regime by noting the large-amplitude flap oscillations and matching of the vortex-shedding frequency and FSI-scaled natural frequency of the flap. Across orders of magnitude variations in stiffness and inertia, maximal lift was found to occur near resonance.

\section{Conclusions}
\label{conclusions}

\begin{sidewaysfigure}
\centering
\begin{forest}
  for tree={
    rounded corners, 
    draw, 
    align=center, 
    l=15mm,
    l sep=1mm,
    parent anchor=south,
    child anchor=north,
    edge path={
      \noexpand\path [\forestoption{edge}, thick]
        (!u.parent anchor) -- +(0,-5pt) -| (.child anchor)\forestoption{edge label};
    },
  }, 
   [FSI physics of the airfoil-flap-flow system
     [Flap-shear regime
        [Flap extended towards the shear layer\\ blocks TEV-induced reverse flow, l=10mm
            [Static behavior, l=20mm
                [SLEV,tier=feature
                    [$\deflmean \downarrow$\\$\rightarrow \slev\uparrow$\\$\rightarrow$Lift$\uparrow$,l=10mm,name=fsstatic1,tier=physics]
                ]
                [Maximal lift  at\\$\stiff\approx0.0015{,}\sixty$,name=fsstatic,before computing xy={s/.average={s}{siblings}},no edge,l=110mm,tier=optimal]
                [Flow velocity\\above\\suction surface,tier=feature
                    [Flap detached\\ from shear layer\\ $\rightarrow$Flow velocity $\downarrow$\\ $\rightarrow$Lift $\downarrow$,l=10mm,name=fsstatic2,tier=physics]
                ]
            ]
            [Dynamic behavior, l=20mm
                [TEV-portion
                    [$\fifty{,}\sixty$ \\location
                        [Upward flap\\ motion, l=20mm,tier=feature
                            [$\phase\approx\pi$\\+Flap velocity$\uparrow$\\ $\rightarrow$Flap aids\\reverse flow\\$\rightarrow$Lift$\downarrow$,name=fsdynamic2,tier=physics]
                        ]
                    ]
                    [$\twenty$ \\location
                        [SLEV, l=20mm,tier=feature
                            [$\phase\approx-\pi/2$\\$\rightarrow\slev\uparrow$\\$\rightarrow$Lift$\uparrow$,l=10mm,name=fsdynamic1,tier=physics]
                        ]
                    ]
                ]
                [Maximal lift  at\\$\mass\approx1.875$,name=fsdynamic,before computing xy={s/.average={s}{siblings}},no edge,tier=optimal]
                [LEV-portion
                    [PLEV,tier=feature
                        [$\phase\approx-\pi/2\rightarrow$Efficient\\SLEV-to-PLEV\\transfer\\+Slowed PLEV\\advection\\$\rightarrow\plev\uparrow\rightarrow$Lift$\uparrow$,l=10mm,name=fsdynamic3,tier=physics]
                    ]
                ]
            ]
       ]
     ]
     [Flap-vortex regime
        [Flap and LEV cumulatively\\ block TEV-induced reverse flow, l=10mm
            [Static behavior, l=20mm
                [LEV,tier=feature
                    [$\deflmean \downarrow$\\$\rightarrow \lev\uparrow$\\$\rightarrow$Lift$\uparrow$,l=10mm,name=fvstatic1,tier=physics]
                ]
                [Maximal lift at\\$\stiff\approx0.015$,no edge,name=fvstatic,before computing xy={s/.average={s}{siblings}},no edge,tier=optimal]
                [Flow velocity\\above\\suction surface,tier=feature
                    [$\deflmean \downarrow\rightarrow$Flow\\velocity $\uparrow$\\ $\rightarrow$Lift $\downarrow$,l=10mm,name=fvstatic2,tier=physics]
                ]
            ]
            [Dynamic behavior, l=20mm
                [LEV\\and\\TEV,tier=feature
                    [$\phase\approx 0$\\+Flap velocity$\uparrow$\\$\rightarrow\lev\uparrow+$\\$\tev\downarrow$\\$\rightarrow$Lift$\uparrow$,name=fvdynamic1,tier=physics]
                ]
                [Maximal lift at\\$\mass\approx0.01875$,no edge,name=fvdynamic,before computing xy={s/.average={s}{siblings}},no edge,tier=optimal]
            ]
        ]
     ]
   ] 
\draw [dashed] (fsstatic1) -- (fsstatic);
\draw [dashed] (fsstatic2) -- (fsstatic);
\draw [dashed] (fsdynamic1) -- (fsdynamic);
\draw [dashed] (fsdynamic2) -- (fsdynamic);
\draw [dashed] (fsdynamic3) -- (fsdynamic);
\draw [dashed] (fvstatic1) -- (fvstatic);
\draw [dashed] (fvstatic2) -- (fvstatic);
\draw [dashed] (fvdynamic1) -- (fvdynamic);
\end{forest}
\caption{Summary of the analysis of the fluid-structure interaction (FSI) physics of flow past a passively deployable flap attached on an airfoil.}
\label{flowchart}
\end{sidewaysfigure}
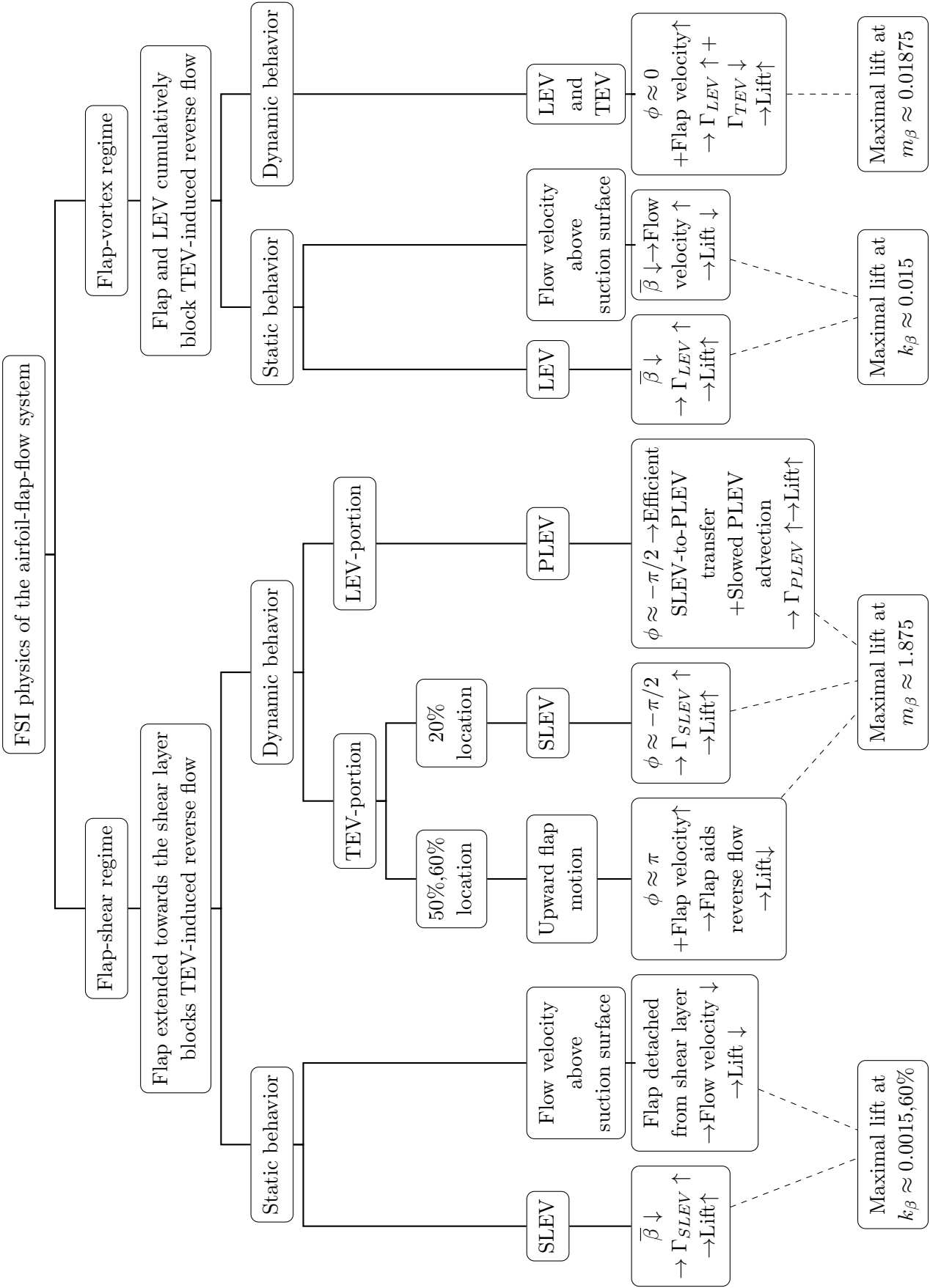

In this manuscript, we numerically modeled covert feathers as a passively deployable torsional flap mounted on the suction surface of a stationary airfoil and investigated the flow physics of this airfoil-flap-flow system via high-fidelity 2D simulations. A low Reynolds number of $\reynolds=1{,}000$ and a post-stall angle of attack of $20^\circ$, where significant flow separation and vortex shedding occurred, were considered. We performed a systematic parametric study of this system by varying the flap location, hinge stiffness and flap inertia.  Lift improvements as high as $27\%$ with respect to the baseline flap-less airfoil were achieved with most and least benefits provided by the flaps located at downstream and upstream flap locations of $\sixty$ and $\twenty$ of the chord from the leading edge, respectively.
The primary role of stiffness was to set a mean flap deflection angle which in turn played a dominant role in setting the overall dynamical regime and qualitative trends in aerodynamic performance. Flap inertia had a secondary role in modifying these trends by inducing time-dependent dynamics. 

The summary of the various flap-fluid interaction mechanisms that modulated airfoil performance and yielded certain parameters to be maximal is provided as a flow chart in Fig.~\ref{flowchart}. Firstly, two flow regimes that provided enhanced performance were identified using a k-means flow classification algorithm---flap-shear and flap-vortex interaction regimes. The flap-shear regime was characterized by the flap extending towards the shear layer while the flap-vortex regime featured significant interactions between the flap and the vortex-shedding process. In the flap-shear regime, the pressure dam effect was found to be the dominant mechanism for enhancing lift where a low pressure region upstream of the flap was maintained due to the blocking of the upstream propagation of reverse flow induced by flow separation and trailing edge vortex (TEV). Flap inertia dictated the phase and amplitude of flap dynamics with respect to large-scale flow features, encoding the effectiveness of this pressure-dam to mean lift. In the flap-vortex regime, there was a similar pressure-dam like blockade caused by the flap and the leading edge vortex (LEV). In addition, the flap dynamics interacted directly with the vortex-shedding process to modulate the strength and advection dynamics of the LEV and TEV.

Next, noting the effects of stiffness and inertia on the flap dynamics, the analysis of each regime was divided into two parts which investigated the static and dynamic behavior of the flap. The static- and dynamic-behavior analysis delineated how the stiffness-dependent mean flap deflection angle and inertia-dependent flap amplitude, velocity and phase affected various flow structures, respectively. The dynamic-behavior analysis of the flap-shear regime was further divided into a detailed analysis of the TEV- and LEV-dominant portions of the lift cycle for the upstream ($\twenty$) and downstream ($\fifty,\sixty$) locations of the flap. 

The reader is referred to Fig.~\ref{flowchart} for an overview of all the different flow features that are affected by the mean flap deflection angle $\deflmean$, phase of flap oscillations $\phase$, and flap angular velocity. However, we emphasize that the general theme of the analysis is that any mechanism that leads to the enhancement of the secondary LEV (SLEV) and primary LEV (PLEV) in the flap-shear regime and LEV in the flap-vortex regime while mitigating the TEV and derived effects such as TEV-induced reverse flow and reduced flow velocity above the suction surface are beneficial for mean lift. The ``$\uparrow$'' and ``$\downarrow$'' symbols in Fig.~\ref{flowchart} denote ``increases'' and ``decreases'', respectively, ``$+$'' symbol denotes ``in addition to'' and ``$\rightarrow$'' symbol denotes ``implies''. The resulting maximal parameters in each regime are provided in the lower-most block in Fig.~\ref{flowchart}. 

This work affirms the lift enhancement capabilities of covert feathers on birds and covert-inspired passive flow control at post-stall angles of attack reported in literature. The numerous physical insights described in this manuscript provide plausible hypothesis for the utility of these feathers in biological flight and can be leveraged to facilitate the use of covert-inspired designs in bio-inspired aerial vehicles. The FSI dynamics and associated lift generating mechanisms at lower $Re$ can be also used to augment the findings of experiments and numerical simulations at higher $Re$ number, which are generally expensive to be performed routinely. Future work could also involve studying the influence of the covert-inspired flap on dynamic pitching of an airfoil that more realistically models the landing and perching dynamics of birds.

\section{Acknowledgements}

We gratefully acknowledge funding through the National Science Foundation under grant CBET 20-29028. We also thank Aimy Wissa, Ahmed Othman and Ernold Thompson for the discussions on exploring the flow physics of the airfoil-flap system.


\appendix

\section{Flow classification}
\label{classification}

\begin{figure}
\centering
\begin{subfigure}[t]{0.45\textwidth}
\centering
\includegraphics[scale=1]{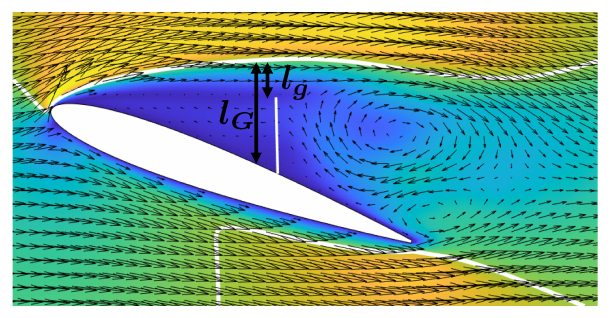}
\caption{Gap ratio in flap-shear interaction regime}
\label{lengths1}
\end{subfigure}
\begin{subfigure}[t]{0.45\textwidth}
\centering
\includegraphics[scale=1]{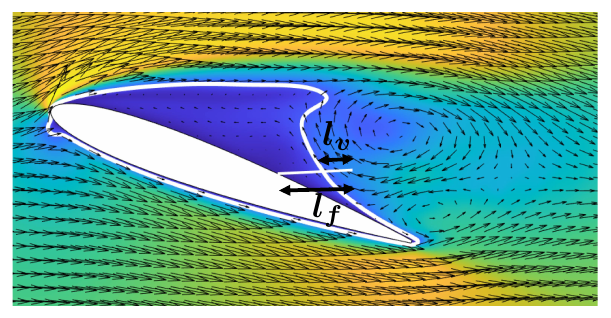}
\caption{Flap ratio in flap-vortex interaction regime}
\label{lengths2}
\end{subfigure}
\caption{Demonstration of length scales in the primary flow regimes. Both contour plots are identical and correspond to the mean velocity magnitude of the flap-less case.}
\label{lengths}
\end{figure}

The flow classification algorithm is derived from \citet{nair2022effects} but applied on a larger data-set consisting of all flap locations of $\twenty$, $\fifty$ and $\sixty$. Due to the significance of stiffness-dependent mean flap configurations described in Sec.~\ref{parametricstudy}, we base the classification algorithm on time-averaged data. First, the mean velocity magnitude contour for the flap-less case is plotted in Fig.~\ref{lengths}. The mean is evaluated using eleven snapshots of velocity magnitude uniformly collected in one time period of the lift cycle. We emphasize that the contour of the flap-less case is considered instead of the flap cases to identify the locations of the baseline flow-field modified by the flap. Next, on top of this contour, we superimpose the mean flap configurations. For demonstration, we show the mean flap configuration superimposed on flap-less velocity contours for the cases of $\stiff=0.0015$ and $\stiff=0.01$, with hinge location at $\sixty$ chord and $\mass=1.875$ in Fig.~\ref{lengths1} and \ref{lengths2}, which correspond to the flap-shear and flap-vortex interaction regimes, respectively.

Next, we define two length scales that will be used as inputs to a k-means clustering algorithm. To define these length scales, we draw two contour lines that denote: (a) the outer boundary of the shear layer defined as the locus of points that satisfy $|\vel|/\velbs=0.99$, as shown in Fig. \ref{lengths1} and (b) the outer boundary of the recirculation zone that excludes the vortex-shedding region defined as the locus of points that satisfy $|\vel|/\velbs=0.25$, as shown in Fig. \ref{lengths2}. Now, the two length scales are defined to be the (a) gap ratio, $\gapratio = \gap/\gaplarge$, where $\gap$ is the distance between the flap tip and the nearest point on the shear layer while $\gaplarge$ is the vertical distance between the corresponding point on the shear layer and the airfoil surface as sketched in Fig.~\ref{lengths1} and (b) $\flapratio = \flapvortex/\flap$, where $\flapvortex$ is the length of the portion of the flap inside the vortex region, while $\flap$ is the total length of the flap, as sketched in Fig.~\ref{lengths2}.

\begin{figure}
\centering
\includegraphics[scale=1]{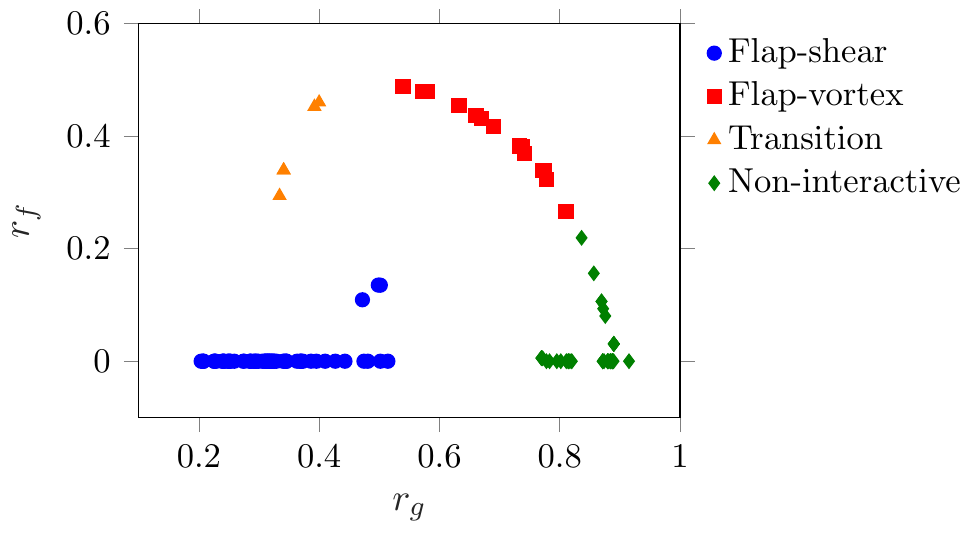}
\caption{Clusters determined by the k-means algorithm with gap ratio, $\gapratio$ and flap ratio, $\flapratio$ as the inputs. The data points include all the cases of mass ratio, stiffness and hinge location considered in the parametric study.}
\label{clusters}
\end{figure}

Next, these length scales are evaluated for all the parametric cases and used as inputs to a k-means clustering algorithm to segregate the cases into different regimes. The number of clusters is set to $k=4$ and the algorithm is repeated 50 times on the same data-set to ensure global convergence. The resulting clusters determined by the algorithm are plotted in Fig.~\ref{clusters}, which contains the data for all values of stiffness, mass ratio and flap locations. 
The cases with by a low gap ratio (the flap is near the shear layer) and flap ratio (the flap does not protrude into the vortex region) are classified into the flap-shear interaction regime. On the other hand the cases with a large gap ratio (the flap is far away from the nominal mean shear layer) and flap ratio (the flap is deflected downwards and into the region where it will interact with vortex-shedding phenomena) are assigned to the flap-vortex interaction regime. The third cluster corresponds to a transition regime between the primary two regimes (the large flap ratio indicates protrusion into the region where vortex shedding will occur but the small gap ratio suggests that the flap remains significantly deployed). Finally, the last cluster corresponds to all the non-interactive cases---the mean lift is essentially unchanged from the baseline (flap-less) case because the flap is hardly deployed or the flap has completely flipped over. The final results of this flow classification algorithm are superimposed on the results of the parametric study in Fig.~\ref{parametric}.



\section{Quantifying vortex circulation strength and phase of flap oscillations}
\label{quantvortstrength}

\begin{figure}
\centering
\includegraphics[scale=1]{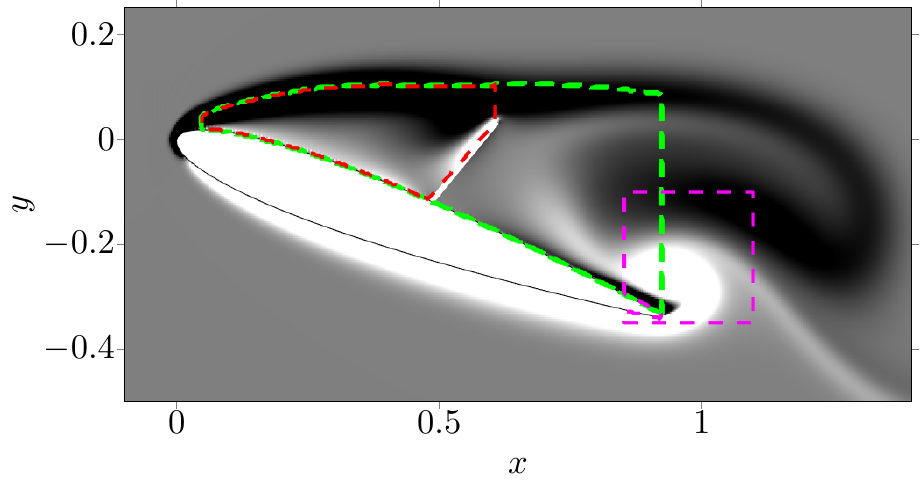}
\put(10,122){
\color{green}\hdashrule[0.5ex][x]{0.05\textwidth}{1.5pt}{3pt} \color{black} \footnotesize LEV
}
\put(10,110){
\color{red}\hdashrule[0.5ex][x]{0.05\textwidth}{1.5pt}{3pt} \color{black} \footnotesize SLEV
}
\put(10,98){
\color{magenta}\hdashrule[0.5ex][x]{0.05\textwidth}{1.5pt}{3pt} \color{black} \footnotesize TEV
}
\caption{Demonstration of areas of integration for quantifying LEV, SLEV and TEV strengths.}
\label{idvort}
\end{figure}

The methodology used to quantify the circulation strengths of the LEV, TEV and SLEV is described in this section first. The vorticity contour of a representative case of $\fifty$ location, $\mass=1.875$ and $\stiff=0.0015$ at $\periodt=0.36$ is plotted in Fig. \ref{idvort} for demonstration. The LEV, SLEV and TEV vortex strengths are quantified as,
\begin{equation}
\lev = \int_{A_{green}} \gamma^- dA, \quad \quad
\slev = \int_{A_{red}} \gamma^- dA, \quad \quad
\tev = \int_{A_{magenta}} \gamma^+ dA,
\label{vortstrength}
\end{equation}
where $\Gamma$ denotes the circulation strength; $\gamma$ is the vorticity; $+$ and $-$ superscripts on $\gamma$ denote anti-clockwise and clockwise rotating vorticity, respectively; and $A_{color}$ is the area enclosed by the surface whose boundary is colored by ``color'' as shown in Fig. \ref{idvort}. The primary LEV strength is evaluated as, $\plev=\lev-\slev$.

The green box used to quantify the LEV strength is defined using three lines: (a) shear layer defined as the locus of points where $|\vel|/\velocityscale=0.99$, bounding from above, (b) a vertical line joining the airfoil trailing edge and the shear layer, bounding from the right, and (c) suction surface of the airfoil, bounding from below. The red box used to quantify the SLEV strength is defined similarly as the green box, except that the lines bounding from the right correspond to the flap and a vertical line connecting the flap tip and the shear layer. Finally the rectangular magenta box used to quantify the TEV strength has dimensions of $[0.85,1.1]\chord\times[-0.35,-0.1]\chord$.

Next, we describe the reason for defining phase in Sec.~\ref{fs_dynamic} as $\phase = 2\pi(\underset{\periodt}{\text{argmax}}(\defl)-0.2)$. We note that the adverse effect of TEV-induced reverse flow can be countered if the flap moves downward against the reverse flow in the TEV-portion of the lift cycle. In other words, the flap is desired to be approximately at its maximum deflection at the start of the TEV-portion. Accordingly, based on the above-mentioned definition of phase, $\phase=0$ is expected to mitigate TEV-induced reverse flow the most. Retrospectively, phase was defined in such a way to denote desired performance when $\phase=0$.



\bibliography{references}

\begin{thebibliography}{36}%
\makeatletter
\providecommand \@ifxundefined [1]{%
 \@ifx{#1\undefined}
}%
\providecommand \@ifnum [1]{%
 \ifnum #1\expandafter \@firstoftwo
 \else \expandafter \@secondoftwo
 \fi
}%
\providecommand \@ifx [1]{%
 \ifx #1\expandafter \@firstoftwo
 \else \expandafter \@secondoftwo
 \fi
}%
\providecommand \natexlab [1]{#1}%
\providecommand \enquote  [1]{``#1''}%
\providecommand \bibnamefont  [1]{#1}%
\providecommand \bibfnamefont [1]{#1}%
\providecommand \citenamefont [1]{#1}%
\providecommand \href@noop [0]{\@secondoftwo}%
\providecommand \href [0]{\begingroup \@sanitize@url \@href}%
\providecommand \@href[1]{\@@startlink{#1}\@@href}%
\providecommand \@@href[1]{\endgroup#1\@@endlink}%
\providecommand \@sanitize@url [0]{\catcode `\\12\catcode `\$12\catcode
  `\&12\catcode `\#12\catcode `\^12\catcode `\_12\catcode `\%12\relax}%
\providecommand \@@startlink[1]{}%
\providecommand \@@endlink[0]{}%
\providecommand \url  [0]{\begingroup\@sanitize@url \@url }%
\providecommand \@url [1]{\endgroup\@href {#1}{\urlprefix }}%
\providecommand \urlprefix  [0]{URL }%
\providecommand \Eprint [0]{\href }%
\providecommand \doibase [0]{https://doi.org/}%
\providecommand \selectlanguage [0]{\@gobble}%
\providecommand \bibinfo  [0]{\@secondoftwo}%
\providecommand \bibfield  [0]{\@secondoftwo}%
\providecommand \translation [1]{[#1]}%
\providecommand \BibitemOpen [0]{}%
\providecommand \bibitemStop [0]{}%
\providecommand \bibitemNoStop [0]{.\EOS\space}%
\providecommand \EOS [0]{\spacefactor3000\relax}%
\providecommand \BibitemShut  [1]{\csname bibitem#1\endcsname}%
\let\auto@bib@innerbib\@empty
\bibitem [{\citenamefont {Gad-el{-}Hak}(2007)}]{gad2007flow}%
  \BibitemOpen
  \bibfield  {author} {\bibinfo {author} {\bibfnamefont {M.}~\bibnamefont
  {Gad-el{-}Hak}},\ }\href@noop {} {\emph {\bibinfo {title} {Flow control:
  passive, active, and reactive flow management}}}\ (\bibinfo  {publisher}
  {Cambridge University Press},\ \bibinfo {year} {2007})\BibitemShut {NoStop}%
\bibitem [{\citenamefont {Lin}(2002)}]{lin2002review}%
  \BibitemOpen
  \bibfield  {author} {\bibinfo {author} {\bibfnamefont {J.~C.}\ \bibnamefont
  {Lin}},\ }\bibfield  {title} {\bibinfo {title} {Review of research on
  low-profile vortex generators to control boundary-layer separation},\
  }\href@noop {} {\bibfield  {journal} {\bibinfo  {journal} {Progress in
  Aerospace Sciences}\ }\textbf {\bibinfo {volume} {38}},\ \bibinfo {pages}
  {389} (\bibinfo {year} {2002})}\BibitemShut {NoStop}%
\bibitem [{\citenamefont {Wang}\ \emph {et~al.}(2008)\citenamefont {Wang},
  \citenamefont {Li},\ and\ \citenamefont {Choi}}]{wang2008gurney}%
  \BibitemOpen
  \bibfield  {author} {\bibinfo {author} {\bibfnamefont {J.}~\bibnamefont
  {Wang}}, \bibinfo {author} {\bibfnamefont {Y.}~\bibnamefont {Li}},\ and\
  \bibinfo {author} {\bibfnamefont {K.-S.}\ \bibnamefont {Choi}},\ }\bibfield
  {title} {\bibinfo {title} {Gurney flap—lift enhancement, mechanisms and
  applications},\ }\href@noop {} {\bibfield  {journal} {\bibinfo  {journal}
  {Progress in Aerospace Sciences}\ }\textbf {\bibinfo {volume} {44}},\
  \bibinfo {pages} {22} (\bibinfo {year} {2008})}\BibitemShut {NoStop}%
\bibitem [{\citenamefont {Saric}\ \emph {et~al.}(2011)\citenamefont {Saric},
  \citenamefont {Carpenter},\ and\ \citenamefont {Reed}}]{saric2011passive}%
  \BibitemOpen
  \bibfield  {author} {\bibinfo {author} {\bibfnamefont {W.~S.}\ \bibnamefont
  {Saric}}, \bibinfo {author} {\bibfnamefont {A.~L.}\ \bibnamefont
  {Carpenter}},\ and\ \bibinfo {author} {\bibfnamefont {H.~L.}\ \bibnamefont
  {Reed}},\ }\bibfield  {title} {\bibinfo {title} {Passive control of
  transition in three-dimensional boundary layers, with emphasis on discrete
  roughness elements},\ }\href@noop {} {\bibfield  {journal} {\bibinfo
  {journal} {Philosophical Transactions of the Royal Society A: Mathematical,
  Physical and Engineering Sciences}\ }\textbf {\bibinfo {volume} {369}},\
  \bibinfo {pages} {1352} (\bibinfo {year} {2011})}\BibitemShut {NoStop}%
\bibitem [{\citenamefont {Choi}\ \emph {et~al.}(2012)\citenamefont {Choi},
  \citenamefont {Park}, \citenamefont {Sagong},\ and\ \citenamefont
  {Lee}}]{choi2012biomimetic}%
  \BibitemOpen
  \bibfield  {author} {\bibinfo {author} {\bibfnamefont {H.}~\bibnamefont
  {Choi}}, \bibinfo {author} {\bibfnamefont {H.}~\bibnamefont {Park}}, \bibinfo
  {author} {\bibfnamefont {W.}~\bibnamefont {Sagong}},\ and\ \bibinfo {author}
  {\bibfnamefont {S.-i.}\ \bibnamefont {Lee}},\ }\bibfield  {title} {\bibinfo
  {title} {Biomimetic flow control based on morphological features of living
  creatures},\ }\href@noop {} {\bibfield  {journal} {\bibinfo  {journal}
  {Physics of Fluids}\ }\textbf {\bibinfo {volume} {24}},\ \bibinfo {pages}
  {121302} (\bibinfo {year} {2012})}\BibitemShut {NoStop}%
\bibitem [{\citenamefont {Gen{\c{c}}}\ \emph {et~al.}(2020)\citenamefont
  {Gen{\c{c}}}, \citenamefont {Koca}, \citenamefont {Demir},\ and\
  \citenamefont {A{\c{c}}{\i}kel}}]{gencc2020traditional}%
  \BibitemOpen
  \bibfield  {author} {\bibinfo {author} {\bibfnamefont {M.~S.}\ \bibnamefont
  {Gen{\c{c}}}}, \bibinfo {author} {\bibfnamefont {K.}~\bibnamefont {Koca}},
  \bibinfo {author} {\bibfnamefont {H.}~\bibnamefont {Demir}},\ and\ \bibinfo
  {author} {\bibfnamefont {H.~H.}\ \bibnamefont {A{\c{c}}{\i}kel}},\ }\bibfield
   {title} {\bibinfo {title} {Traditional and new types of passive flow control
  techniques to pave the way for high maneuverability and low structural weight
  for uavs and mavs},\ }in\ \href@noop {} {\emph {\bibinfo {booktitle}
  {Autonomous Vehicles}}}\ (\bibinfo  {publisher} {IntechOpen},\ \bibinfo
  {year} {2020})\BibitemShut {NoStop}%
\bibitem [{\citenamefont {Rao}\ \emph {et~al.}(2017)\citenamefont {Rao},
  \citenamefont {Ikeda}, \citenamefont {Nakata},\ and\ \citenamefont
  {Liu}}]{rao2017owl}%
  \BibitemOpen
  \bibfield  {author} {\bibinfo {author} {\bibfnamefont {C.}~\bibnamefont
  {Rao}}, \bibinfo {author} {\bibfnamefont {T.}~\bibnamefont {Ikeda}}, \bibinfo
  {author} {\bibfnamefont {T.}~\bibnamefont {Nakata}},\ and\ \bibinfo {author}
  {\bibfnamefont {H.}~\bibnamefont {Liu}},\ }\bibfield  {title} {\bibinfo
  {title} {Owl-inspired leading-edge serrations play a crucial role in
  aerodynamic force production and sound suppression},\ }\href@noop {}
  {\bibfield  {journal} {\bibinfo  {journal} {Bioinspiration \& Biomimetics}\
  }\textbf {\bibinfo {volume} {12}},\ \bibinfo {pages} {046008} (\bibinfo
  {year} {2017})}\BibitemShut {NoStop}%
\bibitem [{\citenamefont {Hu}\ \emph {et~al.}(2008)\citenamefont {Hu},
  \citenamefont {Tamai},\ and\ \citenamefont {Murphy}}]{hu2008flexible}%
  \BibitemOpen
  \bibfield  {author} {\bibinfo {author} {\bibfnamefont {H.}~\bibnamefont
  {Hu}}, \bibinfo {author} {\bibfnamefont {M.}~\bibnamefont {Tamai}},\ and\
  \bibinfo {author} {\bibfnamefont {J.~T.}\ \bibnamefont {Murphy}},\ }\bibfield
   {title} {\bibinfo {title} {Flexible-membrane airfoils at low {R}eynolds
  numbers},\ }\href@noop {} {\bibfield  {journal} {\bibinfo  {journal} {Journal
  of Aircraft}\ }\textbf {\bibinfo {volume} {45}},\ \bibinfo {pages} {1767}
  (\bibinfo {year} {2008})}\BibitemShut {NoStop}%
\bibitem [{\citenamefont {Gamble}\ \emph {et~al.}(2020)\citenamefont {Gamble},
  \citenamefont {Harvey},\ and\ \citenamefont {Inman}}]{gamble2020load}%
  \BibitemOpen
  \bibfield  {author} {\bibinfo {author} {\bibfnamefont {L.~L.}\ \bibnamefont
  {Gamble}}, \bibinfo {author} {\bibfnamefont {C.}~\bibnamefont {Harvey}},\
  and\ \bibinfo {author} {\bibfnamefont {D.~J.}\ \bibnamefont {Inman}},\
  }\bibfield  {title} {\bibinfo {title} {Load alleviation of feather-inspired
  compliant airfoils for instantaneous flow control},\ }\href@noop {}
  {\bibfield  {journal} {\bibinfo  {journal} {Bioinspiration \& Biomimetics}\
  }\textbf {\bibinfo {volume} {15}},\ \bibinfo {pages} {056010} (\bibinfo
  {year} {2020})}\BibitemShut {NoStop}%
\bibitem [{\citenamefont {Kim}\ \emph {et~al.}(2015)\citenamefont {Kim},
  \citenamefont {Park}, \citenamefont {Jab{\l}o{\'n}ski}, \citenamefont {Choi}
  \emph {et~al.}}]{kim2015function}%
  \BibitemOpen
  \bibfield  {author} {\bibinfo {author} {\bibfnamefont {J.}~\bibnamefont
  {Kim}}, \bibinfo {author} {\bibfnamefont {H.}~\bibnamefont {Park}}, \bibinfo
  {author} {\bibfnamefont {P.~G.}\ \bibnamefont {Jab{\l}o{\'n}ski}}, \bibinfo
  {author} {\bibfnamefont {H.}~\bibnamefont {Choi}}, \emph {et~al.},\
  }\bibfield  {title} {\bibinfo {title} {The function of the alula in avian
  flight},\ }\href@noop {} {\bibfield  {journal} {\bibinfo  {journal}
  {Scientific Reports}\ }\textbf {\bibinfo {volume} {5}},\ \bibinfo {pages} {1}
  (\bibinfo {year} {2015})}\BibitemShut {NoStop}%
\bibitem [{\citenamefont {Walker}\ \emph {et~al.}(2010)\citenamefont {Walker},
  \citenamefont {Thomas},\ and\ \citenamefont {Taylor}}]{walker2010deformable}%
  \BibitemOpen
  \bibfield  {author} {\bibinfo {author} {\bibfnamefont {S.~M.}\ \bibnamefont
  {Walker}}, \bibinfo {author} {\bibfnamefont {A.~L.}\ \bibnamefont {Thomas}},\
  and\ \bibinfo {author} {\bibfnamefont {G.~K.}\ \bibnamefont {Taylor}},\
  }\bibfield  {title} {\bibinfo {title} {Deformable wing kinematics in
  free-flying hoverflies},\ }\href@noop {} {\bibfield  {journal} {\bibinfo
  {journal} {Journal of the Royal Society Interface}\ }\textbf {\bibinfo
  {volume} {7}},\ \bibinfo {pages} {131} (\bibinfo {year} {2010})}\BibitemShut
  {NoStop}%
\bibitem [{\citenamefont {Ito}\ \emph {et~al.}(2019)\citenamefont {Ito},
  \citenamefont {Duan},\ and\ \citenamefont {Wissa}}]{ito2019function}%
  \BibitemOpen
  \bibfield  {author} {\bibinfo {author} {\bibfnamefont {M.~R.}\ \bibnamefont
  {Ito}}, \bibinfo {author} {\bibfnamefont {C.}~\bibnamefont {Duan}},\ and\
  \bibinfo {author} {\bibfnamefont {A.~A.}\ \bibnamefont {Wissa}},\ }\bibfield
  {title} {\bibinfo {title} {The function of the alula on engineered wings: a
  detailed experimental investigation of a bioinspired leading-edge device},\
  }\href@noop {} {\bibfield  {journal} {\bibinfo  {journal} {Bioinspiration \&
  Biomimetics}\ }\textbf {\bibinfo {volume} {14}},\ \bibinfo {pages} {056015}
  (\bibinfo {year} {2019})}\BibitemShut {NoStop}%
\bibitem [{\citenamefont {Linehan}\ and\ \citenamefont
  {Mohseni}(2020)}]{linehan2020scaling}%
  \BibitemOpen
  \bibfield  {author} {\bibinfo {author} {\bibfnamefont {T.}~\bibnamefont
  {Linehan}}\ and\ \bibinfo {author} {\bibfnamefont {K.}~\bibnamefont
  {Mohseni}},\ }\bibfield  {title} {\bibinfo {title} {Scaling trends of
  bird’s alular feathers in connection to leading-edge vortex flow over
  hand-wing},\ }\href@noop {} {\bibfield  {journal} {\bibinfo  {journal}
  {Scientific Reports}\ }\textbf {\bibinfo {volume} {10}},\ \bibinfo {pages}
  {1} (\bibinfo {year} {2020})}\BibitemShut {NoStop}%
\bibitem [{\citenamefont {Carruthers}\ \emph {et~al.}(2007)\citenamefont
  {Carruthers}, \citenamefont {Thomas},\ and\ \citenamefont
  {Taylor}}]{carruthers2007automatic}%
  \BibitemOpen
  \bibfield  {author} {\bibinfo {author} {\bibfnamefont {A.~C.}\ \bibnamefont
  {Carruthers}}, \bibinfo {author} {\bibfnamefont {A.~L.}\ \bibnamefont
  {Thomas}},\ and\ \bibinfo {author} {\bibfnamefont {G.~K.}\ \bibnamefont
  {Taylor}},\ }\bibfield  {title} {\bibinfo {title} {Automatic aeroelastic
  devices in the wings of a steppe eagle aquila nipalensis},\ }\href@noop {}
  {\bibfield  {journal} {\bibinfo  {journal} {Journal of Experimental Biology}\
  }\textbf {\bibinfo {volume} {210}},\ \bibinfo {pages} {4136} (\bibinfo {year}
  {2007})}\BibitemShut {NoStop}%
\bibitem [{\citenamefont {Videler}(2006)}]{videler2006avian}%
  \BibitemOpen
  \bibfield  {author} {\bibinfo {author} {\bibfnamefont {J.~J.}\ \bibnamefont
  {Videler}},\ }\href@noop {} {\emph {\bibinfo {title} {Avian flight}}}\
  (\bibinfo  {publisher} {Oxford University Press},\ \bibinfo {year}
  {2006})\BibitemShut {NoStop}%
\bibitem [{\citenamefont {Jiakun}\ \emph {et~al.}(2020)\citenamefont {Jiakun},
  \citenamefont {Zhe}, \citenamefont {Fangbao},\ and\ \citenamefont
  {Gang}}]{jiakun2020review}%
  \BibitemOpen
  \bibfield  {author} {\bibinfo {author} {\bibfnamefont {H.}~\bibnamefont
  {Jiakun}}, \bibinfo {author} {\bibfnamefont {H.}~\bibnamefont {Zhe}},
  \bibinfo {author} {\bibfnamefont {T.}~\bibnamefont {Fangbao}},\ and\ \bibinfo
  {author} {\bibfnamefont {C.}~\bibnamefont {Gang}},\ }\bibfield  {title}
  {\bibinfo {title} {Review on bio-inspired flight systems and bionic
  aerodynamics},\ }\href@noop {} {\bibfield  {journal} {\bibinfo  {journal}
  {Chinese Journal of Aeronautics}\ } (\bibinfo {year} {2020})}\BibitemShut
  {NoStop}%
\bibitem [{\citenamefont {Bechert}\ \emph {et~al.}(2000)\citenamefont
  {Bechert}, \citenamefont {Bruse}, \citenamefont {Hage},\ and\ \citenamefont
  {Meyer}}]{bechert2000fluid}%
  \BibitemOpen
  \bibfield  {author} {\bibinfo {author} {\bibfnamefont {D.}~\bibnamefont
  {Bechert}}, \bibinfo {author} {\bibfnamefont {M.}~\bibnamefont {Bruse}},
  \bibinfo {author} {\bibfnamefont {W.}~\bibnamefont {Hage}},\ and\ \bibinfo
  {author} {\bibfnamefont {R.}~\bibnamefont {Meyer}},\ }\bibfield  {title}
  {\bibinfo {title} {Fluid mechanics of biological surfaces and their
  technological application},\ }\href@noop {} {\bibfield  {journal} {\bibinfo
  {journal} {Naturwissenschaften}\ }\textbf {\bibinfo {volume} {87}},\ \bibinfo
  {pages} {157} (\bibinfo {year} {2000})}\BibitemShut {NoStop}%
\bibitem [{\citenamefont {Duan}\ \emph {et~al.}(2018)\citenamefont {Duan},
  \citenamefont {Waite},\ and\ \citenamefont {Wissa}}]{duan2018design}%
  \BibitemOpen
  \bibfield  {author} {\bibinfo {author} {\bibfnamefont {C.}~\bibnamefont
  {Duan}}, \bibinfo {author} {\bibfnamefont {J.}~\bibnamefont {Waite}},\ and\
  \bibinfo {author} {\bibfnamefont {A.}~\bibnamefont {Wissa}},\ }\bibfield
  {title} {\bibinfo {title} {Design optimization of a covert feather-inspired
  deployable structure for increased lift},\ }in\ \href@noop {} {\emph
  {\bibinfo {booktitle} {2018 Applied Aerodynamics Conference}}}\ (\bibinfo
  {year} {2018})\ p.\ \bibinfo {pages} {3174}\BibitemShut {NoStop}%
\bibitem [{\citenamefont {Kernstine}\ \emph {et~al.}(2008)\citenamefont
  {Kernstine}, \citenamefont {Moore}, \citenamefont {Cutler},\ and\
  \citenamefont {Mittal}}]{kernstine2008initial}%
  \BibitemOpen
  \bibfield  {author} {\bibinfo {author} {\bibfnamefont {K.}~\bibnamefont
  {Kernstine}}, \bibinfo {author} {\bibfnamefont {C.}~\bibnamefont {Moore}},
  \bibinfo {author} {\bibfnamefont {A.}~\bibnamefont {Cutler}},\ and\ \bibinfo
  {author} {\bibfnamefont {R.}~\bibnamefont {Mittal}},\ }\bibfield  {title}
  {\bibinfo {title} {Initial characterization of self-activated movable flaps,"
  pop-up feathers"},\ }in\ \href@noop {} {\emph {\bibinfo {booktitle} {46th
  AIAA Aerospace Sciences Meeting and Exhibit}}}\ (\bibinfo {year} {2008})\ p.\
  \bibinfo {pages} {369}\BibitemShut {NoStop}%
\bibitem [{\citenamefont {Schluter}(2010)}]{schluter2010lift}%
  \BibitemOpen
  \bibfield  {author} {\bibinfo {author} {\bibfnamefont {J.~U.}\ \bibnamefont
  {Schluter}},\ }\bibfield  {title} {\bibinfo {title} {Lift enhancement at low
  {R}eynolds numbers using self-activated movable flaps},\ }\href@noop {}
  {\bibfield  {journal} {\bibinfo  {journal} {Journal of aircraft}\ }\textbf
  {\bibinfo {volume} {47}},\ \bibinfo {pages} {348} (\bibinfo {year}
  {2010})}\BibitemShut {NoStop}%
\bibitem [{\citenamefont {Duan}\ and\ \citenamefont
  {Wissa}(2021)}]{duan2021covert}%
  \BibitemOpen
  \bibfield  {author} {\bibinfo {author} {\bibfnamefont {C.}~\bibnamefont
  {Duan}}\ and\ \bibinfo {author} {\bibfnamefont {A.}~\bibnamefont {Wissa}},\
  }\bibfield  {title} {\bibinfo {title} {Covert-inspired flaps for lift
  enhancement and stall mitigation},\ }\href@noop {} {\bibfield  {journal}
  {\bibinfo  {journal} {Bioinspiration \& Biomimetics}\ } (\bibinfo {year}
  {2021})}\BibitemShut {NoStop}%
\bibitem [{\citenamefont {Johnston}\ and\ \citenamefont
  {Gopalarathnam}(2012)}]{johnston2012investigation}%
  \BibitemOpen
  \bibfield  {author} {\bibinfo {author} {\bibfnamefont {J.}~\bibnamefont
  {Johnston}}\ and\ \bibinfo {author} {\bibfnamefont {A.}~\bibnamefont
  {Gopalarathnam}},\ }\bibfield  {title} {\bibinfo {title} {Investigation of a
  bio-inspired lift-enhancing effector on a 2d airfoil},\ }\href@noop {}
  {\bibfield  {journal} {\bibinfo  {journal} {Bioinspiration \& Biomimetics}\
  }\textbf {\bibinfo {volume} {7}},\ \bibinfo {pages} {036003} (\bibinfo {year}
  {2012})}\BibitemShut {NoStop}%
\bibitem [{\citenamefont {Altman}\ and\ \citenamefont
  {Allemand}(2016)}]{altman2016post}%
  \BibitemOpen
  \bibfield  {author} {\bibinfo {author} {\bibfnamefont {A.}~\bibnamefont
  {Altman}}\ and\ \bibinfo {author} {\bibfnamefont {G.}~\bibnamefont
  {Allemand}},\ }\bibfield  {title} {\bibinfo {title} {Post-stall performance
  improvement through bio-inspired passive covert feathers},\ }in\ \href@noop
  {} {\emph {\bibinfo {booktitle} {54th AIAA Aerospace Sciences Meeting}}}\
  (\bibinfo {year} {2016})\ p.\ \bibinfo {pages} {2042}\BibitemShut {NoStop}%
\bibitem [{\citenamefont {Wang}\ \emph {et~al.}(2019)\citenamefont {Wang},
  \citenamefont {Alam},\ and\ \citenamefont {Zhou}}]{wang2019experimental}%
  \BibitemOpen
  \bibfield  {author} {\bibinfo {author} {\bibfnamefont {L.}~\bibnamefont
  {Wang}}, \bibinfo {author} {\bibfnamefont {M.~M.}\ \bibnamefont {Alam}},\
  and\ \bibinfo {author} {\bibfnamefont {Y.}~\bibnamefont {Zhou}},\ }\bibfield
  {title} {\bibinfo {title} {Experimental study of a passive control of airfoil
  lift using bioinspired feather flap},\ }\href@noop {} {\bibfield  {journal}
  {\bibinfo  {journal} {Bioinspiration \& Biomimetics}\ }\textbf {\bibinfo
  {volume} {14}},\ \bibinfo {pages} {066005} (\bibinfo {year}
  {2019})}\BibitemShut {NoStop}%
\bibitem [{\citenamefont {Fang}\ \emph {et~al.}(2019)\citenamefont {Fang},
  \citenamefont {Gong}, \citenamefont {Revell}, \citenamefont {Chen},
  \citenamefont {Harwood},\ and\ \citenamefont {O’Connor}}]{fang2019passive}%
  \BibitemOpen
  \bibfield  {author} {\bibinfo {author} {\bibfnamefont {Z.}~\bibnamefont
  {Fang}}, \bibinfo {author} {\bibfnamefont {C.}~\bibnamefont {Gong}}, \bibinfo
  {author} {\bibfnamefont {A.}~\bibnamefont {Revell}}, \bibinfo {author}
  {\bibfnamefont {G.}~\bibnamefont {Chen}}, \bibinfo {author} {\bibfnamefont
  {A.}~\bibnamefont {Harwood}},\ and\ \bibinfo {author} {\bibfnamefont
  {J.}~\bibnamefont {O’Connor}},\ }\bibfield  {title} {\bibinfo {title}
  {Passive separation control of a {NACA}0012 airfoil via a flexible flap},\
  }\href@noop {} {\bibfield  {journal} {\bibinfo  {journal} {Physics of
  Fluids}\ }\textbf {\bibinfo {volume} {31}},\ \bibinfo {pages} {101904}
  (\bibinfo {year} {2019})}\BibitemShut {NoStop}%
\bibitem [{\citenamefont {Bramesfeld}\ and\ \citenamefont
  {Maughmer}(2002)}]{bramesfeld2002experimental}%
  \BibitemOpen
  \bibfield  {author} {\bibinfo {author} {\bibfnamefont {G.}~\bibnamefont
  {Bramesfeld}}\ and\ \bibinfo {author} {\bibfnamefont {M.~D.}\ \bibnamefont
  {Maughmer}},\ }\bibfield  {title} {\bibinfo {title} {Experimental
  investigation of self-actuating, upper-surface, high-lift-enhancing
  effectors},\ }\href@noop {} {\bibfield  {journal} {\bibinfo  {journal}
  {Journal of Aircraft}\ }\textbf {\bibinfo {volume} {39}},\ \bibinfo {pages}
  {120} (\bibinfo {year} {2002})}\BibitemShut {NoStop}%
\bibitem [{\citenamefont {Meyer}\ \emph {et~al.}(2007)\citenamefont {Meyer},
  \citenamefont {Hage}, \citenamefont {Bechert}, \citenamefont {Schatz},
  \citenamefont {Knacke},\ and\ \citenamefont {Thiele}}]{meyer2007separation}%
  \BibitemOpen
  \bibfield  {author} {\bibinfo {author} {\bibfnamefont {R.}~\bibnamefont
  {Meyer}}, \bibinfo {author} {\bibfnamefont {W.}~\bibnamefont {Hage}},
  \bibinfo {author} {\bibfnamefont {D.~W.}\ \bibnamefont {Bechert}}, \bibinfo
  {author} {\bibfnamefont {M.}~\bibnamefont {Schatz}}, \bibinfo {author}
  {\bibfnamefont {T.}~\bibnamefont {Knacke}},\ and\ \bibinfo {author}
  {\bibfnamefont {F.}~\bibnamefont {Thiele}},\ }\bibfield  {title} {\bibinfo
  {title} {Separation control by self-activated movable flaps},\ }\href@noop {}
  {\bibfield  {journal} {\bibinfo  {journal} {AIAA Journal}\ }\textbf {\bibinfo
  {volume} {45}},\ \bibinfo {pages} {191} (\bibinfo {year} {2007})}\BibitemShut
  {NoStop}%
\bibitem [{\citenamefont {Izquierdo}\ and\ \citenamefont
  {Marques}(2021)}]{izquierdo2021experimental}%
  \BibitemOpen
  \bibfield  {author} {\bibinfo {author} {\bibfnamefont {D.~O.}\ \bibnamefont
  {Izquierdo}}\ and\ \bibinfo {author} {\bibfnamefont {F.~D.}\ \bibnamefont
  {Marques}},\ }\bibfield  {title} {\bibinfo {title} {Experimental analysis of
  passive bio-inspired covert feathers for stall and post-stall performance
  enhancement},\ }\href@noop {} {\bibfield  {journal} {\bibinfo  {journal}
  {Meccanica}\ }\textbf {\bibinfo {volume} {56}},\ \bibinfo {pages} {2671}
  (\bibinfo {year} {2021})}\BibitemShut {NoStop}%
\bibitem [{\citenamefont {Rosti}\ \emph
  {et~al.}(2018{\natexlab{a}})\citenamefont {Rosti}, \citenamefont
  {Omidyeganeh},\ and\ \citenamefont {Pinelli}}]{rosti2018numerical}%
  \BibitemOpen
  \bibfield  {author} {\bibinfo {author} {\bibfnamefont {M.~E.}\ \bibnamefont
  {Rosti}}, \bibinfo {author} {\bibfnamefont {M.}~\bibnamefont {Omidyeganeh}},\
  and\ \bibinfo {author} {\bibfnamefont {A.}~\bibnamefont {Pinelli}},\
  }\bibfield  {title} {\bibinfo {title} {Numerical simulation of a passive
  control of the flow around an aerofoil using a flexible, self adaptive
  flaplet},\ }\href@noop {} {\bibfield  {journal} {\bibinfo  {journal} {Flow,
  Turbulence and Combustion}\ }\textbf {\bibinfo {volume} {100}},\ \bibinfo
  {pages} {1111} (\bibinfo {year} {2018}{\natexlab{a}})}\BibitemShut {NoStop}%
\bibitem [{\citenamefont {Rosti}\ \emph
  {et~al.}(2018{\natexlab{b}})\citenamefont {Rosti}, \citenamefont
  {Omidyeganeh},\ and\ \citenamefont {Pinelli}}]{rosti2018passive}%
  \BibitemOpen
  \bibfield  {author} {\bibinfo {author} {\bibfnamefont {M.~E.}\ \bibnamefont
  {Rosti}}, \bibinfo {author} {\bibfnamefont {M.}~\bibnamefont {Omidyeganeh}},\
  and\ \bibinfo {author} {\bibfnamefont {A.}~\bibnamefont {Pinelli}},\
  }\bibfield  {title} {\bibinfo {title} {Passive control of the flow around
  unsteady aerofoils using a self-activated deployable flap},\ }\href@noop {}
  {\bibfield  {journal} {\bibinfo  {journal} {Journal of Turbulence}\ }\textbf
  {\bibinfo {volume} {19}},\ \bibinfo {pages} {204} (\bibinfo {year}
  {2018}{\natexlab{b}})}\BibitemShut {NoStop}%
\bibitem [{\citenamefont {Nair}\ and\ \citenamefont
  {Goza}(2022{\natexlab{a}})}]{nair2022effects}%
  \BibitemOpen
  \bibfield  {author} {\bibinfo {author} {\bibfnamefont {N.~J.}\ \bibnamefont
  {Nair}}\ and\ \bibinfo {author} {\bibfnamefont {A.}~\bibnamefont {Goza}},\
  }\bibfield  {title} {\bibinfo {title} {Effects of torsional stiffness and
  inertia on a passively deployable flap for aerodynamic lift enhancement},\
  }in\ \href@noop {} {\emph {\bibinfo {booktitle} {AIAA SCITECH 2022 Forum}}}\
  (\bibinfo {year} {2022})\ p.\ \bibinfo {pages} {1968}\BibitemShut {NoStop}%
\bibitem [{\citenamefont {Nair}\ \emph {et~al.}(2022)\citenamefont {Nair},
  \citenamefont {Flynn},\ and\ \citenamefont {Goza}}]{nair2022numerical}%
  \BibitemOpen
  \bibfield  {author} {\bibinfo {author} {\bibfnamefont {N.~J.}\ \bibnamefont
  {Nair}}, \bibinfo {author} {\bibfnamefont {Z.}~\bibnamefont {Flynn}},\ and\
  \bibinfo {author} {\bibfnamefont {A.}~\bibnamefont {Goza}},\ }\bibfield
  {title} {\bibinfo {title} {Numerical study of multiple bio-inspired
  torsionally hinged flaps for passive flow control},\ }\href@noop {}
  {\bibfield  {journal} {\bibinfo  {journal} {Fluids}\ }\textbf {\bibinfo
  {volume} {7}},\ \bibinfo {pages} {44} (\bibinfo {year} {2022})}\BibitemShut
  {NoStop}%
\bibitem [{\citenamefont {Gracey}(1941)}]{gracey1941additional}%
  \BibitemOpen
  \bibfield  {author} {\bibinfo {author} {\bibfnamefont {W.}~\bibnamefont
  {Gracey}},\ }\bibfield  {title} {\bibinfo {title} {The additional-mass effect
  of plates as determined by experiments},\ }\href@noop {} {\bibfield
  {journal} {\bibinfo  {journal} {NACA Report}\ }\textbf {\bibinfo {volume}
  {707}},\ \bibinfo {pages} {81} (\bibinfo {year} {1941})}\BibitemShut
  {NoStop}%
\bibitem [{\citenamefont {Goza}\ and\ \citenamefont
  {Colonius}(2017)}]{goza2017strongly}%
  \BibitemOpen
  \bibfield  {author} {\bibinfo {author} {\bibfnamefont {A.}~\bibnamefont
  {Goza}}\ and\ \bibinfo {author} {\bibfnamefont {T.}~\bibnamefont
  {Colonius}},\ }\bibfield  {title} {\bibinfo {title} {A strongly-coupled
  immersed-boundary formulation for thin elastic structures},\ }\href@noop {}
  {\bibfield  {journal} {\bibinfo  {journal} {Journal of Computational
  Physics}\ }\textbf {\bibinfo {volume} {336}},\ \bibinfo {pages} {401}
  (\bibinfo {year} {2017})}\BibitemShut {NoStop}%
\bibitem [{\citenamefont {Nair}\ and\ \citenamefont
  {Goza}(2022{\natexlab{b}})}]{nair2022strongly}%
  \BibitemOpen
  \bibfield  {author} {\bibinfo {author} {\bibfnamefont {N.~J.}\ \bibnamefont
  {Nair}}\ and\ \bibinfo {author} {\bibfnamefont {A.}~\bibnamefont {Goza}},\
  }\bibfield  {title} {\bibinfo {title} {A strongly coupled immersed boundary
  method for fluid-structure interaction that mimics the efficiency of
  stationary body methods},\ }\href@noop {} {\bibfield  {journal} {\bibinfo
  {journal} {Journal of Computational Physics}\ ,\ \bibinfo {pages} {110897}}
  (\bibinfo {year} {2022}{\natexlab{b}})}\BibitemShut {NoStop}%
\bibitem [{\citenamefont {Colonius}\ and\ \citenamefont
  {Taira}(2008)}]{colonius2008fast}%
  \BibitemOpen
  \bibfield  {author} {\bibinfo {author} {\bibfnamefont {T.}~\bibnamefont
  {Colonius}}\ and\ \bibinfo {author} {\bibfnamefont {K.}~\bibnamefont
  {Taira}},\ }\bibfield  {title} {\bibinfo {title} {A fast immersed boundary
  method using a nullspace approach and multi-domain far-field boundary
  conditions},\ }\href@noop {} {\bibfield  {journal} {\bibinfo  {journal}
  {Computer Methods in Applied Mechanics and Engineering}\ }\textbf {\bibinfo
  {volume} {197}},\ \bibinfo {pages} {2131} (\bibinfo {year}
  {2008})}\BibitemShut {NoStop}%
\end{thebibliography}%

\end{document}